\documentclass[longauth]{aa}

\usepackage{graphicx}
\usepackage{natbib}

\usepackage[table]{xcolor}

\usepackage{comment}

\usepackage{txfonts}
\usepackage{booktabs}
\usepackage{ulem}
\usepackage{subcaption}

\newcommand{\Mstarsun}{\logten(M_{\ast}/M_\odot)}

\usepackage[pdfencoding=auto,psdextra]{hyperref}
\hypersetup{
    colorlinks=true,
    linkcolor=blue,
    filecolor=magenta,      
    urlcolor=blue,
    citecolor=blue
}
\makeatletter
\newcommand{\citenopar}[1]{%
  \begingroup
  \@tempcnta\z@
  \@for\@II:=#1\do{\advance\@tempcnta\@ne}%
  \@tempcntb\z@
  \@for\@II:=#1\do{%
    \advance\@tempcntb\@ne
    \citeauthor{\@II} \citeyear{\@II}%
    \ifnum\@tempcntb<\@tempcnta, \fi%
  }%
  \endgroup
}
\makeatother

\makeatletter
\renewcommand*\aa@pageof{, page \thepage{} of \pageref*{LastPage}}
\makeatother

\usepackage[utf8]{inputenc}

\usepackage[switch, modulo]{lineno}

\usepackage{euclid}

\begin{document}
%
%

\title{Euclid Quick Data Release (Q1)} 
\subtitle{Exploring galaxy properties with a multi-modal foundation model}

   \titlerunning{\Euclid\/: {\tt AstroPT} foundation model}
   \authorrunning{Euclid Collaboration: M. Siudek et al.}
   
%
\newcommand{\orcid}[1]{} 
\author{Euclid Collaboration: M.~Siudek\orcid{0000-0002-2949-2155}\thanks{\email{msiudek@iac.es}}\inst{\ref{aff1},\ref{aff2},\ref{aff3}}
\and M.~Huertas-Company\orcid{0000-0002-1416-8483}\inst{\ref{aff3},\ref{aff1},\ref{aff4},\ref{aff5}}
\and M.~Smith\orcid{0000-0003-0220-5125}\inst{\ref{aff6},\ref{aff7}}
\and G.~Martinez-Solaeche\inst{\ref{aff8}}
\and F.~Lanusse\orcid{0000-0001-7956-0542}\inst{\ref{aff9}}
\and S.~Ho\orcid{0000-0002-1068-160X}\inst{\ref{aff10},\ref{aff11},\ref{aff12}}
\and E.~Angeloudi\orcid{0009-0002-9769-0388}\inst{\ref{aff3},\ref{aff13}}
\and P.~A.~C.~Cunha\orcid{0000-0002-9454-859X}\inst{\ref{aff14},\ref{aff15}}
\and H.~Dom\'inguez~S\'anchez\orcid{0000-0002-9013-1316}\inst{\ref{aff16}}
\and M.~Dunn\orcid{0000-0001-5374-1644}\inst{\ref{aff3},\ref{aff1},\ref{aff13}}
\and Y.~Fu\orcid{0000-0002-0759-0504}\inst{\ref{aff17},\ref{aff18}}
\and P.~Iglesias-Navarro\orcid{0009-0009-8959-2404}\inst{\ref{aff3},\ref{aff13}}
\and J.~Junais\orcid{0000-0002-7016-4532}\inst{\ref{aff3},\ref{aff13}}
\and J.~H.~Knapen\orcid{0000-0003-1643-0024}\inst{\ref{aff3},\ref{aff13}}
\and B.~Laloux\orcid{0000-0001-9996-9732}\inst{\ref{aff19},\ref{aff20}}
\and M.~Mezcua\orcid{0000-0003-4440-259X}\inst{\ref{aff2},\ref{aff21}}
\and W.~Roster\orcid{0000-0002-9149-6528}\inst{\ref{aff20}}
\and G.~Stevens\orcid{0000-0002-8885-4443}\inst{\ref{aff22}}
\and J.~Vega-Ferrero\orcid{0000-0003-2338-5567}\inst{\ref{aff23}}
\and N.~Aghanim\orcid{0000-0002-6688-8992}\inst{\ref{aff24}}
\and B.~Altieri\orcid{0000-0003-3936-0284}\inst{\ref{aff25}}
\and A.~Amara\inst{\ref{aff26}}
\and S.~Andreon\orcid{0000-0002-2041-8784}\inst{\ref{aff27}}
\and N.~Auricchio\orcid{0000-0003-4444-8651}\inst{\ref{aff28}}
\and H.~Aussel\orcid{0000-0002-1371-5705}\inst{\ref{aff9}}
\and C.~Baccigalupi\orcid{0000-0002-8211-1630}\inst{\ref{aff29},\ref{aff30},\ref{aff31},\ref{aff32}}
\and M.~Baldi\orcid{0000-0003-4145-1943}\inst{\ref{aff33},\ref{aff28},\ref{aff34}}
\and S.~Bardelli\orcid{0000-0002-8900-0298}\inst{\ref{aff28}}
\and P.~Battaglia\orcid{0000-0002-7337-5909}\inst{\ref{aff28}}
\and A.~Biviano\orcid{0000-0002-0857-0732}\inst{\ref{aff30},\ref{aff29}}
\and A.~Bonchi\orcid{0000-0002-2667-5482}\inst{\ref{aff35}}
\and E.~Branchini\orcid{0000-0002-0808-6908}\inst{\ref{aff36},\ref{aff37},\ref{aff27}}
\and M.~Brescia\orcid{0000-0001-9506-5680}\inst{\ref{aff38},\ref{aff19}}
\and J.~Brinchmann\orcid{0000-0003-4359-8797}\inst{\ref{aff15},\ref{aff14}}
\and S.~Camera\orcid{0000-0003-3399-3574}\inst{\ref{aff39},\ref{aff40},\ref{aff41}}
\and G.~Ca\~nas-Herrera\orcid{0000-0003-2796-2149}\inst{\ref{aff42},\ref{aff43},\ref{aff17}}
\and V.~Capobianco\orcid{0000-0002-3309-7692}\inst{\ref{aff41}}
\and C.~Carbone\orcid{0000-0003-0125-3563}\inst{\ref{aff44}}
\and J.~Carretero\orcid{0000-0002-3130-0204}\inst{\ref{aff45},\ref{aff46}}
\and S.~Casas\orcid{0000-0002-4751-5138}\inst{\ref{aff47}}
\and F.~J.~Castander\orcid{0000-0001-7316-4573}\inst{\ref{aff2},\ref{aff21}}
\and M.~Castellano\orcid{0000-0001-9875-8263}\inst{\ref{aff48}}
\and G.~Castignani\orcid{0000-0001-6831-0687}\inst{\ref{aff28}}
\and S.~Cavuoti\orcid{0000-0002-3787-4196}\inst{\ref{aff19},\ref{aff49}}
\and K.~C.~Chambers\orcid{0000-0001-6965-7789}\inst{\ref{aff50}}
\and A.~Cimatti\inst{\ref{aff51}}
\and C.~Colodro-Conde\inst{\ref{aff3}}
\and G.~Congedo\orcid{0000-0003-2508-0046}\inst{\ref{aff52}}
\and C.~J.~Conselice\orcid{0000-0003-1949-7638}\inst{\ref{aff53}}
\and L.~Conversi\orcid{0000-0002-6710-8476}\inst{\ref{aff54},\ref{aff25}}
\and Y.~Copin\orcid{0000-0002-5317-7518}\inst{\ref{aff55}}
\and F.~Courbin\orcid{0000-0003-0758-6510}\inst{\ref{aff56},\ref{aff57}}
\and H.~M.~Courtois\orcid{0000-0003-0509-1776}\inst{\ref{aff58}}
\and M.~Cropper\orcid{0000-0003-4571-9468}\inst{\ref{aff59}}
\and A.~Da~Silva\orcid{0000-0002-6385-1609}\inst{\ref{aff60},\ref{aff61}}
\and H.~Degaudenzi\orcid{0000-0002-5887-6799}\inst{\ref{aff62}}
\and G.~De~Lucia\orcid{0000-0002-6220-9104}\inst{\ref{aff30}}
\and A.~M.~Di~Giorgio\orcid{0000-0002-4767-2360}\inst{\ref{aff63}}
\and J.~Dinis\orcid{0000-0001-5075-1601}\inst{\ref{aff60},\ref{aff61}}
\and C.~Dolding\orcid{0009-0003-7199-6108}\inst{\ref{aff59}}
\and H.~Dole\orcid{0000-0002-9767-3839}\inst{\ref{aff24}}
\and F.~Dubath\orcid{0000-0002-6533-2810}\inst{\ref{aff62}}
\and C.~A.~J.~Duncan\orcid{0009-0003-3573-0791}\inst{\ref{aff53}}
\and X.~Dupac\inst{\ref{aff25}}
\and S.~Dusini\orcid{0000-0002-1128-0664}\inst{\ref{aff64}}
\and S.~Escoffier\orcid{0000-0002-2847-7498}\inst{\ref{aff65}}
\and M.~Farina\orcid{0000-0002-3089-7846}\inst{\ref{aff63}}
\and R.~Farinelli\inst{\ref{aff28}}
\and F.~Faustini\orcid{0000-0001-6274-5145}\inst{\ref{aff35},\ref{aff48}}
\and S.~Ferriol\inst{\ref{aff55}}
\and F.~Finelli\orcid{0000-0002-6694-3269}\inst{\ref{aff28},\ref{aff66}}
\and S.~Fotopoulou\orcid{0000-0002-9686-254X}\inst{\ref{aff22}}
\and M.~Frailis\orcid{0000-0002-7400-2135}\inst{\ref{aff30}}
\and E.~Franceschi\orcid{0000-0002-0585-6591}\inst{\ref{aff28}}
\and S.~Galeotta\orcid{0000-0002-3748-5115}\inst{\ref{aff30}}
\and K.~George\orcid{0000-0002-1734-8455}\inst{\ref{aff67}}
\and B.~Gillis\orcid{0000-0002-4478-1270}\inst{\ref{aff52}}
\and C.~Giocoli\orcid{0000-0002-9590-7961}\inst{\ref{aff28},\ref{aff34}}
\and J.~Gracia-Carpio\inst{\ref{aff20}}
\and B.~R.~Granett\orcid{0000-0003-2694-9284}\inst{\ref{aff27}}
\and A.~Grazian\orcid{0000-0002-5688-0663}\inst{\ref{aff68}}
\and F.~Grupp\inst{\ref{aff20},\ref{aff67}}
\and S.~Gwyn\orcid{0000-0001-8221-8406}\inst{\ref{aff69}}
\and S.~V.~H.~Haugan\orcid{0000-0001-9648-7260}\inst{\ref{aff70}}
\and W.~Holmes\inst{\ref{aff71}}
\and I.~M.~Hook\orcid{0000-0002-2960-978X}\inst{\ref{aff72}}
\and F.~Hormuth\inst{\ref{aff73}}
\and A.~Hornstrup\orcid{0000-0002-3363-0936}\inst{\ref{aff74},\ref{aff75}}
\and K.~Jahnke\orcid{0000-0003-3804-2137}\inst{\ref{aff76}}
\and M.~Jhabvala\inst{\ref{aff77}}
\and E.~Keih\"anen\orcid{0000-0003-1804-7715}\inst{\ref{aff78}}
\and S.~Kermiche\orcid{0000-0002-0302-5735}\inst{\ref{aff65}}
\and A.~Kiessling\orcid{0000-0002-2590-1273}\inst{\ref{aff71}}
\and B.~Kubik\orcid{0009-0006-5823-4880}\inst{\ref{aff55}}
\and M.~K\"ummel\orcid{0000-0003-2791-2117}\inst{\ref{aff67}}
\and M.~Kunz\orcid{0000-0002-3052-7394}\inst{\ref{aff79}}
\and H.~Kurki-Suonio\orcid{0000-0002-4618-3063}\inst{\ref{aff80},\ref{aff81}}
\and Q.~Le~Boulc'h\inst{\ref{aff82}}
\and A.~M.~C.~Le~Brun\orcid{0000-0002-0936-4594}\inst{\ref{aff83}}
\and D.~Le~Mignant\orcid{0000-0002-5339-5515}\inst{\ref{aff84}}
\and S.~Ligori\orcid{0000-0003-4172-4606}\inst{\ref{aff41}}
\and P.~B.~Lilje\orcid{0000-0003-4324-7794}\inst{\ref{aff70}}
\and V.~Lindholm\orcid{0000-0003-2317-5471}\inst{\ref{aff80},\ref{aff81}}
\and I.~Lloro\orcid{0000-0001-5966-1434}\inst{\ref{aff85}}
\and G.~Mainetti\orcid{0000-0003-2384-2377}\inst{\ref{aff82}}
\and D.~Maino\inst{\ref{aff86},\ref{aff44},\ref{aff87}}
\and E.~Maiorano\orcid{0000-0003-2593-4355}\inst{\ref{aff28}}
\and O.~Mansutti\orcid{0000-0001-5758-4658}\inst{\ref{aff30}}
\and S.~Marcin\inst{\ref{aff88}}
\and O.~Marggraf\orcid{0000-0001-7242-3852}\inst{\ref{aff89}}
\and M.~Martinelli\orcid{0000-0002-6943-7732}\inst{\ref{aff48},\ref{aff90}}
\and N.~Martinet\orcid{0000-0003-2786-7790}\inst{\ref{aff84}}
\and F.~Marulli\orcid{0000-0002-8850-0303}\inst{\ref{aff91},\ref{aff28},\ref{aff34}}
\and R.~Massey\orcid{0000-0002-6085-3780}\inst{\ref{aff92}}
\and S.~Maurogordato\inst{\ref{aff93}}
\and H.~J.~McCracken\orcid{0000-0002-9489-7765}\inst{\ref{aff94}}
\and E.~Medinaceli\orcid{0000-0002-4040-7783}\inst{\ref{aff28}}
\and S.~Mei\orcid{0000-0002-2849-559X}\inst{\ref{aff95},\ref{aff96}}
\and M.~Melchior\inst{\ref{aff97}}
\and Y.~Mellier\inst{\ref{aff98},\ref{aff94}}
\and M.~Meneghetti\orcid{0000-0003-1225-7084}\inst{\ref{aff28},\ref{aff34}}
\and E.~Merlin\orcid{0000-0001-6870-8900}\inst{\ref{aff48}}
\and G.~Meylan\inst{\ref{aff99}}
\and A.~Mora\orcid{0000-0002-1922-8529}\inst{\ref{aff100}}
\and M.~Moresco\orcid{0000-0002-7616-7136}\inst{\ref{aff91},\ref{aff28}}
\and L.~Moscardini\orcid{0000-0002-3473-6716}\inst{\ref{aff91},\ref{aff28},\ref{aff34}}
\and R.~Nakajima\orcid{0009-0009-1213-7040}\inst{\ref{aff89}}
\and C.~Neissner\orcid{0000-0001-8524-4968}\inst{\ref{aff101},\ref{aff46}}
\and S.-M.~Niemi\inst{\ref{aff42}}
\and J.~W.~Nightingale\orcid{0000-0002-8987-7401}\inst{\ref{aff102}}
\and C.~Padilla\orcid{0000-0001-7951-0166}\inst{\ref{aff101}}
\and S.~Paltani\orcid{0000-0002-8108-9179}\inst{\ref{aff62}}
\and F.~Pasian\orcid{0000-0002-4869-3227}\inst{\ref{aff30}}
\and K.~Pedersen\inst{\ref{aff103}}
\and W.~J.~Percival\orcid{0000-0002-0644-5727}\inst{\ref{aff104},\ref{aff105},\ref{aff106}}
\and V.~Pettorino\inst{\ref{aff42}}
\and S.~Pires\orcid{0000-0002-0249-2104}\inst{\ref{aff9}}
\and G.~Polenta\orcid{0000-0003-4067-9196}\inst{\ref{aff35}}
\and M.~Poncet\inst{\ref{aff107}}
\and L.~A.~Popa\inst{\ref{aff108}}
\and L.~Pozzetti\orcid{0000-0001-7085-0412}\inst{\ref{aff28}}
\and F.~Raison\orcid{0000-0002-7819-6918}\inst{\ref{aff20}}
\and A.~Renzi\orcid{0000-0001-9856-1970}\inst{\ref{aff109},\ref{aff64}}
\and J.~Rhodes\orcid{0000-0002-4485-8549}\inst{\ref{aff71}}
\and G.~Riccio\inst{\ref{aff19}}
\and E.~Romelli\orcid{0000-0003-3069-9222}\inst{\ref{aff30}}
\and M.~Roncarelli\orcid{0000-0001-9587-7822}\inst{\ref{aff28}}
\and R.~Saglia\orcid{0000-0003-0378-7032}\inst{\ref{aff67},\ref{aff20}}
\and Z.~Sakr\orcid{0000-0002-4823-3757}\inst{\ref{aff110},\ref{aff111},\ref{aff112}}
\and A.~G.~S\'anchez\orcid{0000-0003-1198-831X}\inst{\ref{aff20}}
\and D.~Sapone\orcid{0000-0001-7089-4503}\inst{\ref{aff113}}
\and B.~Sartoris\orcid{0000-0003-1337-5269}\inst{\ref{aff67},\ref{aff30}}
\and J.~A.~Schewtschenko\orcid{0000-0002-4913-6393}\inst{\ref{aff52}}
\and P.~Schneider\orcid{0000-0001-8561-2679}\inst{\ref{aff89}}
\and T.~Schrabback\orcid{0000-0002-6987-7834}\inst{\ref{aff114}}
\and M.~Scodeggio\inst{\ref{aff44}}
\and A.~Secroun\orcid{0000-0003-0505-3710}\inst{\ref{aff65}}
\and G.~Seidel\orcid{0000-0003-2907-353X}\inst{\ref{aff76}}
\and M.~Seiffert\orcid{0000-0002-7536-9393}\inst{\ref{aff71}}
\and S.~Serrano\orcid{0000-0002-0211-2861}\inst{\ref{aff21},\ref{aff115},\ref{aff2}}
\and P.~Simon\inst{\ref{aff89}}
\and C.~Sirignano\orcid{0000-0002-0995-7146}\inst{\ref{aff109},\ref{aff64}}
\and G.~Sirri\orcid{0000-0003-2626-2853}\inst{\ref{aff34}}
\and L.~Stanco\orcid{0000-0002-9706-5104}\inst{\ref{aff64}}
\and J.~Steinwagner\orcid{0000-0001-7443-1047}\inst{\ref{aff20}}
\and P.~Tallada-Cresp\'{i}\orcid{0000-0002-1336-8328}\inst{\ref{aff45},\ref{aff46}}
\and A.~N.~Taylor\inst{\ref{aff52}}
\and I.~Tereno\inst{\ref{aff60},\ref{aff116}}
\and S.~Toft\orcid{0000-0003-3631-7176}\inst{\ref{aff117},\ref{aff118}}
\and R.~Toledo-Moreo\orcid{0000-0002-2997-4859}\inst{\ref{aff119}}
\and F.~Torradeflot\orcid{0000-0003-1160-1517}\inst{\ref{aff46},\ref{aff45}}
\and I.~Tutusaus\orcid{0000-0002-3199-0399}\inst{\ref{aff111}}
\and L.~Valenziano\orcid{0000-0002-1170-0104}\inst{\ref{aff28},\ref{aff66}}
\and J.~Valiviita\orcid{0000-0001-6225-3693}\inst{\ref{aff80},\ref{aff81}}
\and T.~Vassallo\orcid{0000-0001-6512-6358}\inst{\ref{aff67},\ref{aff30}}
\and G.~Verdoes~Kleijn\orcid{0000-0001-5803-2580}\inst{\ref{aff18}}
\and A.~Veropalumbo\orcid{0000-0003-2387-1194}\inst{\ref{aff27},\ref{aff37},\ref{aff36}}
\and Y.~Wang\orcid{0000-0002-4749-2984}\inst{\ref{aff120}}
\and J.~Weller\orcid{0000-0002-8282-2010}\inst{\ref{aff67},\ref{aff20}}
\and A.~Zacchei\orcid{0000-0003-0396-1192}\inst{\ref{aff30},\ref{aff29}}
\and G.~Zamorani\orcid{0000-0002-2318-301X}\inst{\ref{aff28}}
\and F.~M.~Zerbi\inst{\ref{aff27}}
\and I.~A.~Zinchenko\orcid{0000-0002-2944-2449}\inst{\ref{aff67}}
\and E.~Zucca\orcid{0000-0002-5845-8132}\inst{\ref{aff28}}
\and V.~Allevato\orcid{0000-0001-7232-5152}\inst{\ref{aff19}}
\and M.~Ballardini\orcid{0000-0003-4481-3559}\inst{\ref{aff121},\ref{aff122},\ref{aff28}}
\and M.~Bolzonella\orcid{0000-0003-3278-4607}\inst{\ref{aff28}}
\and E.~Bozzo\orcid{0000-0002-8201-1525}\inst{\ref{aff62}}
\and C.~Burigana\orcid{0000-0002-3005-5796}\inst{\ref{aff123},\ref{aff66}}
\and R.~Cabanac\orcid{0000-0001-6679-2600}\inst{\ref{aff111}}
\and A.~Cappi\inst{\ref{aff28},\ref{aff93}}
\and D.~Di~Ferdinando\inst{\ref{aff34}}
\and J.~A.~Escartin~Vigo\inst{\ref{aff20}}
\and L.~Gabarra\orcid{0000-0002-8486-8856}\inst{\ref{aff124}}
\and J.~Mart\'{i}n-Fleitas\orcid{0000-0002-8594-569X}\inst{\ref{aff100}}
\and S.~Matthew\orcid{0000-0001-8448-1697}\inst{\ref{aff52}}
\and N.~Mauri\orcid{0000-0001-8196-1548}\inst{\ref{aff51},\ref{aff34}}
\and R.~B.~Metcalf\orcid{0000-0003-3167-2574}\inst{\ref{aff91},\ref{aff28}}
\and A.~Pezzotta\orcid{0000-0003-0726-2268}\inst{\ref{aff125},\ref{aff20}}
\and M.~P\"ontinen\orcid{0000-0001-5442-2530}\inst{\ref{aff80}}
\and C.~Porciani\orcid{0000-0002-7797-2508}\inst{\ref{aff89}}
\and I.~Risso\orcid{0000-0003-2525-7761}\inst{\ref{aff126}}
\and V.~Scottez\inst{\ref{aff98},\ref{aff127}}
\and M.~Sereno\orcid{0000-0003-0302-0325}\inst{\ref{aff28},\ref{aff34}}
\and M.~Tenti\orcid{0000-0002-4254-5901}\inst{\ref{aff34}}
\and M.~Viel\orcid{0000-0002-2642-5707}\inst{\ref{aff29},\ref{aff30},\ref{aff32},\ref{aff31},\ref{aff128}}
\and M.~Wiesmann\orcid{0009-0000-8199-5860}\inst{\ref{aff70}}
\and Y.~Akrami\orcid{0000-0002-2407-7956}\inst{\ref{aff129},\ref{aff130}}
\and I.~T.~Andika\orcid{0000-0001-6102-9526}\inst{\ref{aff131},\ref{aff132}}
\and S.~Anselmi\orcid{0000-0002-3579-9583}\inst{\ref{aff64},\ref{aff109},\ref{aff133}}
\and M.~Archidiacono\orcid{0000-0003-4952-9012}\inst{\ref{aff86},\ref{aff87}}
\and F.~Atrio-Barandela\orcid{0000-0002-2130-2513}\inst{\ref{aff134}}
\and C.~Benoist\inst{\ref{aff93}}
\and K.~Benson\inst{\ref{aff59}}
\and D.~Bertacca\orcid{0000-0002-2490-7139}\inst{\ref{aff109},\ref{aff68},\ref{aff64}}
\and M.~Bethermin\orcid{0000-0002-3915-2015}\inst{\ref{aff135}}
\and L.~Bisigello\orcid{0000-0003-0492-4924}\inst{\ref{aff68}}
\and A.~Blanchard\orcid{0000-0001-8555-9003}\inst{\ref{aff111}}
\and L.~Blot\orcid{0000-0002-9622-7167}\inst{\ref{aff136},\ref{aff133}}
\and M.~L.~Brown\orcid{0000-0002-0370-8077}\inst{\ref{aff53}}
\and S.~Bruton\orcid{0000-0002-6503-5218}\inst{\ref{aff137}}
\and A.~Calabro\orcid{0000-0003-2536-1614}\inst{\ref{aff48}}
\and B.~Camacho~Quevedo\orcid{0000-0002-8789-4232}\inst{\ref{aff21},\ref{aff2}}
\and F.~Caro\inst{\ref{aff48}}
\and C.~S.~Carvalho\inst{\ref{aff116}}
\and T.~Castro\orcid{0000-0002-6292-3228}\inst{\ref{aff30},\ref{aff31},\ref{aff29},\ref{aff128}}
\and Y.~Charles\inst{\ref{aff84}}
\and F.~Cogato\orcid{0000-0003-4632-6113}\inst{\ref{aff91},\ref{aff28}}
\and A.~R.~Cooray\orcid{0000-0002-3892-0190}\inst{\ref{aff138}}
\and O.~Cucciati\orcid{0000-0002-9336-7551}\inst{\ref{aff28}}
\and S.~Davini\orcid{0000-0003-3269-1718}\inst{\ref{aff37}}
\and F.~De~Paolis\orcid{0000-0001-6460-7563}\inst{\ref{aff139},\ref{aff140},\ref{aff141}}
\and G.~Desprez\orcid{0000-0001-8325-1742}\inst{\ref{aff18}}
\and A.~D\'iaz-S\'anchez\orcid{0000-0003-0748-4768}\inst{\ref{aff142}}
\and J.~J.~Diaz\inst{\ref{aff3}}
\and S.~Di~Domizio\orcid{0000-0003-2863-5895}\inst{\ref{aff36},\ref{aff37}}
\and J.~M.~Diego\orcid{0000-0001-9065-3926}\inst{\ref{aff16}}
\and P.-A.~Duc\orcid{0000-0003-3343-6284}\inst{\ref{aff135}}
\and A.~Enia\orcid{0000-0002-0200-2857}\inst{\ref{aff33},\ref{aff28}}
\and Y.~Fang\inst{\ref{aff67}}
\and A.~G.~Ferrari\orcid{0009-0005-5266-4110}\inst{\ref{aff34}}
\and P.~G.~Ferreira\orcid{0000-0002-3021-2851}\inst{\ref{aff124}}
\and A.~Finoguenov\orcid{0000-0002-4606-5403}\inst{\ref{aff80}}
\and A.~Fontana\orcid{0000-0003-3820-2823}\inst{\ref{aff48}}
\and A.~Franco\orcid{0000-0002-4761-366X}\inst{\ref{aff140},\ref{aff139},\ref{aff141}}
\and K.~Ganga\orcid{0000-0001-8159-8208}\inst{\ref{aff95}}
\and J.~Garc\'ia-Bellido\orcid{0000-0002-9370-8360}\inst{\ref{aff129}}
\and T.~Gasparetto\orcid{0000-0002-7913-4866}\inst{\ref{aff30}}
\and V.~Gautard\inst{\ref{aff143}}
\and E.~Gaztanaga\orcid{0000-0001-9632-0815}\inst{\ref{aff2},\ref{aff21},\ref{aff144}}
\and F.~Giacomini\orcid{0000-0002-3129-2814}\inst{\ref{aff34}}
\and F.~Gianotti\orcid{0000-0003-4666-119X}\inst{\ref{aff28}}
\and G.~Gozaliasl\orcid{0000-0002-0236-919X}\inst{\ref{aff145},\ref{aff80}}
\and M.~Guidi\orcid{0000-0001-9408-1101}\inst{\ref{aff33},\ref{aff28}}
\and C.~M.~Gutierrez\orcid{0000-0001-7854-783X}\inst{\ref{aff146}}
\and A.~Hall\orcid{0000-0002-3139-8651}\inst{\ref{aff52}}
\and W.~G.~Hartley\inst{\ref{aff62}}
\and S.~Hemmati\orcid{0000-0003-2226-5395}\inst{\ref{aff147}}
\and C.~Hern\'andez-Monteagudo\orcid{0000-0001-5471-9166}\inst{\ref{aff13},\ref{aff3}}
\and H.~Hildebrandt\orcid{0000-0002-9814-3338}\inst{\ref{aff148}}
\and J.~Hjorth\orcid{0000-0002-4571-2306}\inst{\ref{aff103}}
\and J.~J.~E.~Kajava\orcid{0000-0002-3010-8333}\inst{\ref{aff149},\ref{aff150}}
\and Y.~Kang\orcid{0009-0000-8588-7250}\inst{\ref{aff62}}
\and V.~Kansal\orcid{0000-0002-4008-6078}\inst{\ref{aff151},\ref{aff152}}
\and D.~Karagiannis\orcid{0000-0002-4927-0816}\inst{\ref{aff121},\ref{aff153}}
\and K.~Kiiveri\inst{\ref{aff78}}
\and C.~C.~Kirkpatrick\inst{\ref{aff78}}
\and S.~Kruk\orcid{0000-0001-8010-8879}\inst{\ref{aff25}}
\and J.~Le~Graet\orcid{0000-0001-6523-7971}\inst{\ref{aff65}}
\and L.~Legrand\orcid{0000-0003-0610-5252}\inst{\ref{aff154},\ref{aff155}}
\and M.~Lembo\orcid{0000-0002-5271-5070}\inst{\ref{aff121},\ref{aff122}}
\and F.~Lepori\orcid{0009-0000-5061-7138}\inst{\ref{aff156}}
\and G.~Leroy\orcid{0009-0004-2523-4425}\inst{\ref{aff157},\ref{aff92}}
\and G.~F.~Lesci\orcid{0000-0002-4607-2830}\inst{\ref{aff91},\ref{aff28}}
\and J.~Lesgourgues\orcid{0000-0001-7627-353X}\inst{\ref{aff47}}
\and L.~Leuzzi\orcid{0009-0006-4479-7017}\inst{\ref{aff91},\ref{aff28}}
\and T.~I.~Liaudat\orcid{0000-0002-9104-314X}\inst{\ref{aff158}}
\and A.~Loureiro\orcid{0000-0002-4371-0876}\inst{\ref{aff159},\ref{aff160}}
\and J.~Macias-Perez\orcid{0000-0002-5385-2763}\inst{\ref{aff161}}
\and G.~Maggio\orcid{0000-0003-4020-4836}\inst{\ref{aff30}}
\and M.~Magliocchetti\orcid{0000-0001-9158-4838}\inst{\ref{aff63}}
\and E.~A.~Magnier\orcid{0000-0002-7965-2815}\inst{\ref{aff50}}
\and F.~Mannucci\orcid{0000-0002-4803-2381}\inst{\ref{aff162}}
\and R.~Maoli\orcid{0000-0002-6065-3025}\inst{\ref{aff163},\ref{aff48}}
\and C.~J.~A.~P.~Martins\orcid{0000-0002-4886-9261}\inst{\ref{aff164},\ref{aff15}}
\and L.~Maurin\orcid{0000-0002-8406-0857}\inst{\ref{aff24}}
\and M.~Miluzio\inst{\ref{aff25},\ref{aff165}}
\and P.~Monaco\orcid{0000-0003-2083-7564}\inst{\ref{aff166},\ref{aff30},\ref{aff31},\ref{aff29}}
\and C.~Moretti\orcid{0000-0003-3314-8936}\inst{\ref{aff32},\ref{aff128},\ref{aff30},\ref{aff29},\ref{aff31}}
\and G.~Morgante\inst{\ref{aff28}}
\and C.~Murray\inst{\ref{aff95}}
\and K.~Naidoo\orcid{0000-0002-9182-1802}\inst{\ref{aff144}}
\and A.~Navarro-Alsina\orcid{0000-0002-3173-2592}\inst{\ref{aff89}}
\and S.~Nesseris\orcid{0000-0002-0567-0324}\inst{\ref{aff129}}
\and F.~Passalacqua\orcid{0000-0002-8606-4093}\inst{\ref{aff109},\ref{aff64}}
\and K.~Paterson\orcid{0000-0001-8340-3486}\inst{\ref{aff76}}
\and L.~Patrizii\inst{\ref{aff34}}
\and A.~Pisani\orcid{0000-0002-6146-4437}\inst{\ref{aff65},\ref{aff167}}
\and D.~Potter\orcid{0000-0002-0757-5195}\inst{\ref{aff156}}
\and S.~Quai\orcid{0000-0002-0449-8163}\inst{\ref{aff91},\ref{aff28}}
\and M.~Radovich\orcid{0000-0002-3585-866X}\inst{\ref{aff68}}
\and S.~Sacquegna\orcid{0000-0002-8433-6630}\inst{\ref{aff139},\ref{aff140},\ref{aff141}}
\and M.~Sahl\'en\orcid{0000-0003-0973-4804}\inst{\ref{aff168}}
\and D.~B.~Sanders\orcid{0000-0002-1233-9998}\inst{\ref{aff50}}
\and E.~Sarpa\orcid{0000-0002-1256-655X}\inst{\ref{aff32},\ref{aff128},\ref{aff31}}
\and A.~Schneider\orcid{0000-0001-7055-8104}\inst{\ref{aff156}}
\and D.~Sciotti\orcid{0009-0008-4519-2620}\inst{\ref{aff48},\ref{aff90}}
\and D.~Scognamiglio\orcid{0000-0001-8450-7885}\inst{\ref{aff71}}
\and E.~Sellentin\inst{\ref{aff169},\ref{aff17}}
\and L.~C.~Smith\orcid{0000-0002-3259-2771}\inst{\ref{aff170}}
\and K.~Tanidis\orcid{0000-0001-9843-5130}\inst{\ref{aff124}}
\and G.~Testera\inst{\ref{aff37}}
\and R.~Teyssier\orcid{0000-0001-7689-0933}\inst{\ref{aff167}}
\and S.~Tosi\orcid{0000-0002-7275-9193}\inst{\ref{aff36},\ref{aff126}}
\and A.~Troja\orcid{0000-0003-0239-4595}\inst{\ref{aff109},\ref{aff64}}
\and M.~Tucci\inst{\ref{aff62}}
\and C.~Valieri\inst{\ref{aff34}}
\and A.~Venhola\orcid{0000-0001-6071-4564}\inst{\ref{aff171}}
\and D.~Vergani\orcid{0000-0003-0898-2216}\inst{\ref{aff28}}
\and G.~Verza\orcid{0000-0002-1886-8348}\inst{\ref{aff10}}
\and P.~Vielzeuf\orcid{0000-0003-2035-9339}\inst{\ref{aff65}}
\and N.~A.~Walton\orcid{0000-0003-3983-8778}\inst{\ref{aff170}}
\and J.~G.~Sorce\orcid{0000-0002-2307-2432}\inst{\ref{aff172},\ref{aff24}}}
										   
\institute{Instituto de Astrof\'isica de Canarias (IAC); Departamento de Astrof\'isica, Universidad de La Laguna (ULL), 38200, La Laguna, Tenerife, Spain\label{aff1}
\and
Institute of Space Sciences (ICE, CSIC), Campus UAB, Carrer de Can Magrans, s/n, 08193 Barcelona, Spain\label{aff2}
\and
Instituto de Astrof\'{\i}sica de Canarias, V\'{\i}a L\'actea, 38205 La Laguna, Tenerife, Spain\label{aff3}
\and
Universit\'e PSL, Observatoire de Paris, Sorbonne Universit\'e, CNRS, LERMA, 75014, Paris, France\label{aff4}
\and
Universit\'e Paris-Cit\'e, 5 Rue Thomas Mann, 75013, Paris, France\label{aff5}
\and
School of Physics, Astronomy and Mathematics, University of Hertfordshire, College Lane, Hatfield AL10 9AB, UK\label{aff6}
\and
Aspia Space, Falmouth, TR10 9TA, UK\label{aff7}
\and
Instituto de Astrof\'isica de Andaluc\'ia, CSIC, Glorieta de la Astronom\'\i a, 18080, Granada, Spain\label{aff8}
\and
Universit\'e Paris-Saclay, Universit\'e Paris Cit\'e, CEA, CNRS, AIM, 91191, Gif-sur-Yvette, France\label{aff9}
\and
Center for Computational Astrophysics, Flatiron Institute, 162 5th Avenue, 10010, New York, NY, USA\label{aff10}
\and
Center for Cosmology and Particle Physics, Department of Physics, New York University, New York, NY 10003, USA\label{aff11}
\and
Princeton University, Princeton, NJ 08540, USA\label{aff12}
\and
Universidad de La Laguna, Departamento de Astrof\'{\i}sica, 38206 La Laguna, Tenerife, Spain\label{aff13}
\and
Faculdade de Ci\^encias da Universidade do Porto, Rua do Campo de Alegre, 4150-007 Porto, Portugal\label{aff14}
\and
Instituto de Astrof\'isica e Ci\^encias do Espa\c{c}o, Universidade do Porto, CAUP, Rua das Estrelas, PT4150-762 Porto, Portugal\label{aff15}
\and
Instituto de F\'isica de Cantabria, Edificio Juan Jord\'a, Avenida de los Castros, 39005 Santander, Spain\label{aff16}
\and
Leiden Observatory, Leiden University, Einsteinweg 55, 2333 CC Leiden, The Netherlands\label{aff17}
\and
Kapteyn Astronomical Institute, University of Groningen, PO Box 800, 9700 AV Groningen, The Netherlands\label{aff18}
\and
INAF-Osservatorio Astronomico di Capodimonte, Via Moiariello 16, 80131 Napoli, Italy\label{aff19}
\and
Max Planck Institute for Extraterrestrial Physics, Giessenbachstr. 1, 85748 Garching, Germany\label{aff20}
\and
Institut d'Estudis Espacials de Catalunya (IEEC),  Edifici RDIT, Campus UPC, 08860 Castelldefels, Barcelona, Spain\label{aff21}
\and
School of Physics, HH Wills Physics Laboratory, University of Bristol, Tyndall Avenue, Bristol, BS8 1TL, UK\label{aff22}
\and
Centro de Estudios de F\'isica del Cosmos de Arag\'on (CEFCA), Plaza San Juan, 1, planta 2, 44001, Teruel, Spain\label{aff23}
\and
Universit\'e Paris-Saclay, CNRS, Institut d'astrophysique spatiale, 91405, Orsay, France\label{aff24}
\and
ESAC/ESA, Camino Bajo del Castillo, s/n., Urb. Villafranca del Castillo, 28692 Villanueva de la Ca\~nada, Madrid, Spain\label{aff25}
\and
School of Mathematics and Physics, University of Surrey, Guildford, Surrey, GU2 7XH, UK\label{aff26}
\and
INAF-Osservatorio Astronomico di Brera, Via Brera 28, 20122 Milano, Italy\label{aff27}
\and
INAF-Osservatorio di Astrofisica e Scienza dello Spazio di Bologna, Via Piero Gobetti 93/3, 40129 Bologna, Italy\label{aff28}
\and
IFPU, Institute for Fundamental Physics of the Universe, via Beirut 2, 34151 Trieste, Italy\label{aff29}
\and
INAF-Osservatorio Astronomico di Trieste, Via G. B. Tiepolo 11, 34143 Trieste, Italy\label{aff30}
\and
INFN, Sezione di Trieste, Via Valerio 2, 34127 Trieste TS, Italy\label{aff31}
\and
SISSA, International School for Advanced Studies, Via Bonomea 265, 34136 Trieste TS, Italy\label{aff32}
\and
Dipartimento di Fisica e Astronomia, Universit\`a di Bologna, Via Gobetti 93/2, 40129 Bologna, Italy\label{aff33}
\and
INFN-Sezione di Bologna, Viale Berti Pichat 6/2, 40127 Bologna, Italy\label{aff34}
\and
Space Science Data Center, Italian Space Agency, via del Politecnico snc, 00133 Roma, Italy\label{aff35}
\and
Dipartimento di Fisica, Universit\`a di Genova, Via Dodecaneso 33, 16146, Genova, Italy\label{aff36}
\and
INFN-Sezione di Genova, Via Dodecaneso 33, 16146, Genova, Italy\label{aff37}
\and
Department of Physics "E. Pancini", University Federico II, Via Cinthia 6, 80126, Napoli, Italy\label{aff38}
\and
Dipartimento di Fisica, Universit\`a degli Studi di Torino, Via P. Giuria 1, 10125 Torino, Italy\label{aff39}
\and
INFN-Sezione di Torino, Via P. Giuria 1, 10125 Torino, Italy\label{aff40}
\and
INAF-Osservatorio Astrofisico di Torino, Via Osservatorio 20, 10025 Pino Torinese (TO), Italy\label{aff41}
\and
European Space Agency/ESTEC, Keplerlaan 1, 2201 AZ Noordwijk, The Netherlands\label{aff42}
\and
Institute Lorentz, Leiden University, Niels Bohrweg 2, 2333 CA Leiden, The Netherlands\label{aff43}
\and
INAF-IASF Milano, Via Alfonso Corti 12, 20133 Milano, Italy\label{aff44}
\and
Centro de Investigaciones Energ\'eticas, Medioambientales y Tecnol\'ogicas (CIEMAT), Avenida Complutense 40, 28040 Madrid, Spain\label{aff45}
\and
Port d'Informaci\'{o} Cient\'{i}fica, Campus UAB, C. Albareda s/n, 08193 Bellaterra (Barcelona), Spain\label{aff46}
\and
Institute for Theoretical Particle Physics and Cosmology (TTK), RWTH Aachen University, 52056 Aachen, Germany\label{aff47}
\and
INAF-Osservatorio Astronomico di Roma, Via Frascati 33, 00078 Monteporzio Catone, Italy\label{aff48}
\and
INFN section of Naples, Via Cinthia 6, 80126, Napoli, Italy\label{aff49}
\and
Institute for Astronomy, University of Hawaii, 2680 Woodlawn Drive, Honolulu, HI 96822, USA\label{aff50}
\and
Dipartimento di Fisica e Astronomia "Augusto Righi" - Alma Mater Studiorum Universit\`a di Bologna, Viale Berti Pichat 6/2, 40127 Bologna, Italy\label{aff51}
\and
Institute for Astronomy, University of Edinburgh, Royal Observatory, Blackford Hill, Edinburgh EH9 3HJ, UK\label{aff52}
\and
Jodrell Bank Centre for Astrophysics, Department of Physics and Astronomy, University of Manchester, Oxford Road, Manchester M13 9PL, UK\label{aff53}
\and
European Space Agency/ESRIN, Largo Galileo Galilei 1, 00044 Frascati, Roma, Italy\label{aff54}
\and
Universit\'e Claude Bernard Lyon 1, CNRS/IN2P3, IP2I Lyon, UMR 5822, Villeurbanne, F-69100, France\label{aff55}
\and
Institut de Ci\`{e}ncies del Cosmos (ICCUB), Universitat de Barcelona (IEEC-UB), Mart\'{i} i Franqu\`{e}s 1, 08028 Barcelona, Spain\label{aff56}
\and
Instituci\'o Catalana de Recerca i Estudis Avan\c{c}ats (ICREA), Passeig de Llu\'{\i}s Companys 23, 08010 Barcelona, Spain\label{aff57}
\and
UCB Lyon 1, CNRS/IN2P3, IUF, IP2I Lyon, 4 rue Enrico Fermi, 69622 Villeurbanne, France\label{aff58}
\and
Mullard Space Science Laboratory, University College London, Holmbury St Mary, Dorking, Surrey RH5 6NT, UK\label{aff59}
\and
Departamento de F\'isica, Faculdade de Ci\^encias, Universidade de Lisboa, Edif\'icio C8, Campo Grande, PT1749-016 Lisboa, Portugal\label{aff60}
\and
Instituto de Astrof\'isica e Ci\^encias do Espa\c{c}o, Faculdade de Ci\^encias, Universidade de Lisboa, Campo Grande, 1749-016 Lisboa, Portugal\label{aff61}
\and
Department of Astronomy, University of Geneva, ch. d'Ecogia 16, 1290 Versoix, Switzerland\label{aff62}
\and
INAF-Istituto di Astrofisica e Planetologia Spaziali, via del Fosso del Cavaliere, 100, 00100 Roma, Italy\label{aff63}
\and
INFN-Padova, Via Marzolo 8, 35131 Padova, Italy\label{aff64}
\and
Aix-Marseille Universit\'e, CNRS/IN2P3, CPPM, Marseille, France\label{aff65}
\and
INFN-Bologna, Via Irnerio 46, 40126 Bologna, Italy\label{aff66}
\and
Universit\"ats-Sternwarte M\"unchen, Fakult\"at f\"ur Physik, Ludwig-Maximilians-Universit\"at M\"unchen, Scheinerstrasse 1, 81679 M\"unchen, Germany\label{aff67}
\and
INAF-Osservatorio Astronomico di Padova, Via dell'Osservatorio 5, 35122 Padova, Italy\label{aff68}
\and
NRC Herzberg, 5071 West Saanich Rd, Victoria, BC V9E 2E7, Canada\label{aff69}
\and
Institute of Theoretical Astrophysics, University of Oslo, P.O. Box 1029 Blindern, 0315 Oslo, Norway\label{aff70}
\and
Jet Propulsion Laboratory, California Institute of Technology, 4800 Oak Grove Drive, Pasadena, CA, 91109, USA\label{aff71}
\and
Department of Physics, Lancaster University, Lancaster, LA1 4YB, UK\label{aff72}
\and
Felix Hormuth Engineering, Goethestr. 17, 69181 Leimen, Germany\label{aff73}
\and
Technical University of Denmark, Elektrovej 327, 2800 Kgs. Lyngby, Denmark\label{aff74}
\and
Cosmic Dawn Center (DAWN), Denmark\label{aff75}
\and
Max-Planck-Institut f\"ur Astronomie, K\"onigstuhl 17, 69117 Heidelberg, Germany\label{aff76}
\and
NASA Goddard Space Flight Center, Greenbelt, MD 20771, USA\label{aff77}
\and
Department of Physics and Helsinki Institute of Physics, Gustaf H\"allstr\"omin katu 2, 00014 University of Helsinki, Finland\label{aff78}
\and
Universit\'e de Gen\`eve, D\'epartement de Physique Th\'eorique and Centre for Astroparticle Physics, 24 quai Ernest-Ansermet, CH-1211 Gen\`eve 4, Switzerland\label{aff79}
\and
Department of Physics, P.O. Box 64, 00014 University of Helsinki, Finland\label{aff80}
\and
Helsinki Institute of Physics, Gustaf H{\"a}llstr{\"o}min katu 2, University of Helsinki, Helsinki, Finland\label{aff81}
\and
Centre de Calcul de l'IN2P3/CNRS, 21 avenue Pierre de Coubertin 69627 Villeurbanne Cedex, France\label{aff82}
\and
Laboratoire d'etude de l'Univers et des phenomenes eXtremes, Observatoire de Paris, Universit\'e PSL, Sorbonne Universit\'e, CNRS, 92190 Meudon, France\label{aff83}
\and
Aix-Marseille Universit\'e, CNRS, CNES, LAM, Marseille, France\label{aff84}
\and
SKA Observatory, Jodrell Bank, Lower Withington, Macclesfield, Cheshire SK11 9FT, UK\label{aff85}
\and
Dipartimento di Fisica "Aldo Pontremoli", Universit\`a degli Studi di Milano, Via Celoria 16, 20133 Milano, Italy\label{aff86}
\and
INFN-Sezione di Milano, Via Celoria 16, 20133 Milano, Italy\label{aff87}
\and
University of Applied Sciences and Arts of Northwestern Switzerland, School of Computer Science, 5210 Windisch, Switzerland\label{aff88}
\and
Universit\"at Bonn, Argelander-Institut f\"ur Astronomie, Auf dem H\"ugel 71, 53121 Bonn, Germany\label{aff89}
\and
INFN-Sezione di Roma, Piazzale Aldo Moro, 2 - c/o Dipartimento di Fisica, Edificio G. Marconi, 00185 Roma, Italy\label{aff90}
\and
Dipartimento di Fisica e Astronomia "Augusto Righi" - Alma Mater Studiorum Universit\`a di Bologna, via Piero Gobetti 93/2, 40129 Bologna, Italy\label{aff91}
\and
Department of Physics, Institute for Computational Cosmology, Durham University, South Road, Durham, DH1 3LE, UK\label{aff92}
\and
Universit\'e C\^{o}te d'Azur, Observatoire de la C\^{o}te d'Azur, CNRS, Laboratoire Lagrange, Bd de l'Observatoire, CS 34229, 06304 Nice cedex 4, France\label{aff93}
\and
Institut d'Astrophysique de Paris, UMR 7095, CNRS, and Sorbonne Universit\'e, 98 bis boulevard Arago, 75014 Paris, France\label{aff94}
\and
Universit\'e Paris Cit\'e, CNRS, Astroparticule et Cosmologie, 75013 Paris, France\label{aff95}
\and
CNRS-UCB International Research Laboratory, Centre Pierre Binetruy, IRL2007, CPB-IN2P3, Berkeley, USA\label{aff96}
\and
University of Applied Sciences and Arts of Northwestern Switzerland, School of Engineering, 5210 Windisch, Switzerland\label{aff97}
\and
Institut d'Astrophysique de Paris, 98bis Boulevard Arago, 75014, Paris, France\label{aff98}
\and
Institute of Physics, Laboratory of Astrophysics, Ecole Polytechnique F\'ed\'erale de Lausanne (EPFL), Observatoire de Sauverny, 1290 Versoix, Switzerland\label{aff99}
\and
Aurora Technology for European Space Agency (ESA), Camino bajo del Castillo, s/n, Urbanizacion Villafranca del Castillo, Villanueva de la Ca\~nada, 28692 Madrid, Spain\label{aff100}
\and
Institut de F\'{i}sica d'Altes Energies (IFAE), The Barcelona Institute of Science and Technology, Campus UAB, 08193 Bellaterra (Barcelona), Spain\label{aff101}
\and
School of Mathematics, Statistics and Physics, Newcastle University, Herschel Building, Newcastle-upon-Tyne, NE1 7RU, UK\label{aff102}
\and
DARK, Niels Bohr Institute, University of Copenhagen, Jagtvej 155, 2200 Copenhagen, Denmark\label{aff103}
\and
Waterloo Centre for Astrophysics, University of Waterloo, Waterloo, Ontario N2L 3G1, Canada\label{aff104}
\and
Department of Physics and Astronomy, University of Waterloo, Waterloo, Ontario N2L 3G1, Canada\label{aff105}
\and
Perimeter Institute for Theoretical Physics, Waterloo, Ontario N2L 2Y5, Canada\label{aff106}
\and
Centre National d'Etudes Spatiales -- Centre spatial de Toulouse, 18 avenue Edouard Belin, 31401 Toulouse Cedex 9, France\label{aff107}
\and
Institute of Space Science, Str. Atomistilor, nr. 409 M\u{a}gurele, Ilfov, 077125, Romania\label{aff108}
\and
Dipartimento di Fisica e Astronomia "G. Galilei", Universit\`a di Padova, Via Marzolo 8, 35131 Padova, Italy\label{aff109}
\and
Institut f\"ur Theoretische Physik, University of Heidelberg, Philosophenweg 16, 69120 Heidelberg, Germany\label{aff110}
\and
Institut de Recherche en Astrophysique et Plan\'etologie (IRAP), Universit\'e de Toulouse, CNRS, UPS, CNES, 14 Av. Edouard Belin, 31400 Toulouse, France\label{aff111}
\and
Universit\'e St Joseph; Faculty of Sciences, Beirut, Lebanon\label{aff112}
\and
Departamento de F\'isica, FCFM, Universidad de Chile, Blanco Encalada 2008, Santiago, Chile\label{aff113}
\and
Universit\"at Innsbruck, Institut f\"ur Astro- und Teilchenphysik, Technikerstr. 25/8, 6020 Innsbruck, Austria\label{aff114}
\and
Satlantis, University Science Park, Sede Bld 48940, Leioa-Bilbao, Spain\label{aff115}
\and
Instituto de Astrof\'isica e Ci\^encias do Espa\c{c}o, Faculdade de Ci\^encias, Universidade de Lisboa, Tapada da Ajuda, 1349-018 Lisboa, Portugal\label{aff116}
\and
Cosmic Dawn Center (DAWN)\label{aff117}
\and
Niels Bohr Institute, University of Copenhagen, Jagtvej 128, 2200 Copenhagen, Denmark\label{aff118}
\and
Universidad Polit\'ecnica de Cartagena, Departamento de Electr\'onica y Tecnolog\'ia de Computadoras,  Plaza del Hospital 1, 30202 Cartagena, Spain\label{aff119}
\and
Infrared Processing and Analysis Center, California Institute of Technology, Pasadena, CA 91125, USA\label{aff120}
\and
Dipartimento di Fisica e Scienze della Terra, Universit\`a degli Studi di Ferrara, Via Giuseppe Saragat 1, 44122 Ferrara, Italy\label{aff121}
\and
Istituto Nazionale di Fisica Nucleare, Sezione di Ferrara, Via Giuseppe Saragat 1, 44122 Ferrara, Italy\label{aff122}
\and
INAF, Istituto di Radioastronomia, Via Piero Gobetti 101, 40129 Bologna, Italy\label{aff123}
\and
Department of Physics, Oxford University, Keble Road, Oxford OX1 3RH, UK\label{aff124}
\and
INAF - Osservatorio Astronomico di Brera, via Emilio Bianchi 46, 23807 Merate, Italy\label{aff125}
\and
INAF-Osservatorio Astronomico di Brera, Via Brera 28, 20122 Milano, Italy, and INFN-Sezione di Genova, Via Dodecaneso 33, 16146, Genova, Italy\label{aff126}
\and
ICL, Junia, Universit\'e Catholique de Lille, LITL, 59000 Lille, France\label{aff127}
\and
ICSC - Centro Nazionale di Ricerca in High Performance Computing, Big Data e Quantum Computing, Via Magnanelli 2, Bologna, Italy\label{aff128}
\and
Instituto de F\'isica Te\'orica UAM-CSIC, Campus de Cantoblanco, 28049 Madrid, Spain\label{aff129}
\and
CERCA/ISO, Department of Physics, Case Western Reserve University, 10900 Euclid Avenue, Cleveland, OH 44106, USA\label{aff130}
\and
Technical University of Munich, TUM School of Natural Sciences, Physics Department, James-Franck-Str.~1, 85748 Garching, Germany\label{aff131}
\and
Max-Planck-Institut f\"ur Astrophysik, Karl-Schwarzschild-Str.~1, 85748 Garching, Germany\label{aff132}
\and
Laboratoire Univers et Th\'eorie, Observatoire de Paris, Universit\'e PSL, Universit\'e Paris Cit\'e, CNRS, 92190 Meudon, France\label{aff133}
\and
Departamento de F{\'\i}sica Fundamental. Universidad de Salamanca. Plaza de la Merced s/n. 37008 Salamanca, Spain\label{aff134}
\and
Universit\'e de Strasbourg, CNRS, Observatoire astronomique de Strasbourg, UMR 7550, 67000 Strasbourg, France\label{aff135}
\and
Center for Data-Driven Discovery, Kavli IPMU (WPI), UTIAS, The University of Tokyo, Kashiwa, Chiba 277-8583, Japan\label{aff136}
\and
California Institute of Technology, 1200 E California Blvd, Pasadena, CA 91125, USA\label{aff137}
\and
Department of Physics \& Astronomy, University of California Irvine, Irvine CA 92697, USA\label{aff138}
\and
Department of Mathematics and Physics E. De Giorgi, University of Salento, Via per Arnesano, CP-I93, 73100, Lecce, Italy\label{aff139}
\and
INFN, Sezione di Lecce, Via per Arnesano, CP-193, 73100, Lecce, Italy\label{aff140}
\and
INAF-Sezione di Lecce, c/o Dipartimento Matematica e Fisica, Via per Arnesano, 73100, Lecce, Italy\label{aff141}
\and
Departamento F\'isica Aplicada, Universidad Polit\'ecnica de Cartagena, Campus Muralla del Mar, 30202 Cartagena, Murcia, Spain\label{aff142}
\and
CEA Saclay, DFR/IRFU, Service d'Astrophysique, Bat. 709, 91191 Gif-sur-Yvette, France\label{aff143}
\and
Institute of Cosmology and Gravitation, University of Portsmouth, Portsmouth PO1 3FX, UK\label{aff144}
\and
Department of Computer Science, Aalto University, PO Box 15400, Espoo, FI-00 076, Finland\label{aff145}
\and
Instituto de Astrof\'\i sica de Canarias, c/ Via Lactea s/n, La Laguna 38200, Spain. Departamento de Astrof\'\i sica de la Universidad de La Laguna, Avda. Francisco Sanchez, La Laguna, 38200, Spain\label{aff146}
\and
Caltech/IPAC, 1200 E. California Blvd., Pasadena, CA 91125, USA\label{aff147}
\and
Ruhr University Bochum, Faculty of Physics and Astronomy, Astronomical Institute (AIRUB), German Centre for Cosmological Lensing (GCCL), 44780 Bochum, Germany\label{aff148}
\and
Department of Physics and Astronomy, Vesilinnantie 5, 20014 University of Turku, Finland\label{aff149}
\and
Serco for European Space Agency (ESA), Camino bajo del Castillo, s/n, Urbanizacion Villafranca del Castillo, Villanueva de la Ca\~nada, 28692 Madrid, Spain\label{aff150}
\and
ARC Centre of Excellence for Dark Matter Particle Physics, Melbourne, Australia\label{aff151}
\and
Centre for Astrophysics \& Supercomputing, Swinburne University of Technology,  Hawthorn, Victoria 3122, Australia\label{aff152}
\and
Department of Physics and Astronomy, University of the Western Cape, Bellville, Cape Town, 7535, South Africa\label{aff153}
\and
DAMTP, Centre for Mathematical Sciences, Wilberforce Road, Cambridge CB3 0WA, UK\label{aff154}
\and
Kavli Institute for Cosmology Cambridge, Madingley Road, Cambridge, CB3 0HA, UK\label{aff155}
\and
Department of Astrophysics, University of Zurich, Winterthurerstrasse 190, 8057 Zurich, Switzerland\label{aff156}
\and
Department of Physics, Centre for Extragalactic Astronomy, Durham University, South Road, Durham, DH1 3LE, UK\label{aff157}
\and
IRFU, CEA, Universit\'e Paris-Saclay 91191 Gif-sur-Yvette Cedex, France\label{aff158}
\and
Oskar Klein Centre for Cosmoparticle Physics, Department of Physics, Stockholm University, Stockholm, SE-106 91, Sweden\label{aff159}
\and
Astrophysics Group, Blackett Laboratory, Imperial College London, London SW7 2AZ, UK\label{aff160}
\and
Univ. Grenoble Alpes, CNRS, Grenoble INP, LPSC-IN2P3, 53, Avenue des Martyrs, 38000, Grenoble, France\label{aff161}
\and
INAF-Osservatorio Astrofisico di Arcetri, Largo E. Fermi 5, 50125, Firenze, Italy\label{aff162}
\and
Dipartimento di Fisica, Sapienza Universit\`a di Roma, Piazzale Aldo Moro 2, 00185 Roma, Italy\label{aff163}
\and
Centro de Astrof\'{\i}sica da Universidade do Porto, Rua das Estrelas, 4150-762 Porto, Portugal\label{aff164}
\and
HE Space for European Space Agency (ESA), Camino bajo del Castillo, s/n, Urbanizacion Villafranca del Castillo, Villanueva de la Ca\~nada, 28692 Madrid, Spain\label{aff165}
\and
Dipartimento di Fisica - Sezione di Astronomia, Universit\`a di Trieste, Via Tiepolo 11, 34131 Trieste, Italy\label{aff166}
\and
Department of Astrophysical Sciences, Peyton Hall, Princeton University, Princeton, NJ 08544, USA\label{aff167}
\and
Theoretical astrophysics, Department of Physics and Astronomy, Uppsala University, Box 515, 751 20 Uppsala, Sweden\label{aff168}
\and
Mathematical Institute, University of Leiden, Einsteinweg 55, 2333 CA Leiden, The Netherlands\label{aff169}
\and
Institute of Astronomy, University of Cambridge, Madingley Road, Cambridge CB3 0HA, UK\label{aff170}
\and
Space physics and astronomy research unit, University of Oulu, Pentti Kaiteran katu 1, FI-90014 Oulu, Finland\label{aff171}
\and
Univ. Lille, CNRS, Centrale Lille, UMR 9189 CRIStAL, 59000 Lille, France\label{aff172}}    

\abstract{
Modern astronomical surveys, such as the \Euclid mission, produce high-dimensional, multi-modal data sets that include imaging and spectroscopic information for millions of galaxies. These data serve as an ideal benchmark for large, pre-trained multi-modal models, which can leverage vast amounts of unlabelled data. In this work, we present the first exploration of \Euclid data with \texttt{AstroPT}, an autoregressive multi-modal foundation model trained on approximately \(300\,000\) optical and infrared \Euclid images and spectral energy distributions (SEDs) from the first \Euclid Quick Data Release. We compare self-supervised pre-training with baseline fully supervised training across several tasks: galaxy morphology classification; redshift estimation; similarity searches; and outlier detection. Our results show that: (a) \texttt{AstroPT} embeddings are highly informative, correlating with morphology and effectively isolating outliers; (b) including infrared data helps to isolate stars, but degrades the identification of edge-on galaxies, which are better captured by optical images; (c) simple fine-tuning of these embeddings for photometric redshift and stellar mass estimation outperforms a fully supervised approach, even when using only \(1\%\) of the training labels; and (d) incorporating SED data into \texttt{AstroPT} via a straightforward multi-modal token-chaining method improves photo-\(z\) predictions, and allow us to identify potentially more interesting anomalies (such as ringed or interacting galaxies) compared to a model pre-trained solely on imaging data.
}
%
%
%
%

   \maketitle
%
%
%
%
   
\section{\label{sc:Intro}Introduction}
 Astronomy's abundance of data requires appropriately scalable, automated frameworks to mine it effectively. This demand has already driven the adoption of advanced machine-learning and deep-learning techniques for a variety of purposes (see for example \citenopar{Dominguez-Sanchez2018, Cheng2020, Cheng2021, Vega-Ferrero2021, Walmsley2022, Walmsley2023, Huertas-Company2024}, or see \citenopar{Huertas-Company2023, Smith2023} for general reviews). However -- in most prior work -- a small subset of labelled examples are used to train supervised models. This means that only a small fraction of available data is used, and that the trained models are very sensitive to 'domain drift' \citep{DS2019} thus limiting their full deployment for scientific analysis \citep{Huertas-Company2023}.

Recent works have demonstrated the transformative impact of models pre-trained on large amounts of unlabelled data (that is, 'foundation models'), which can then be fine-tuned to perform a variety of tasks. This is particularly evident for language processing, where the best-performing `frontier' models now contain many billions of parameters and require `web-scale' data of the order of terabytes to petabytes to be sufficiently pre-trained \citep[e.g.,][]{Grattafiori2024}. Furthermore, the recent rediscovery of neural scaling laws -- empirical relationships that demonstrate how model performance improves predictably with increases in model size, data set size, or computational resources -- has led to great growth in the size of deep-learning models \citep{Cortes1993,Kaplan2020,Henighan2020}.

Astronomical data sets are also growing rapidly in size and complexity, creating the need for models that can scale with data set size and perform multiple tasks effectively~\citep{Smith2024,Pan2024,Walmsley2024,Merz2024,MMU2024}. The advent of the next generation of large-scale sky surveys such as \Euclid~\citep{Laureijs11}, the Dark Energy Spectroscopic Instrument Legacy Survey \citep[DESI;][]{DESI_overview_2022, DESI_Validation_2024}, and the Vera C.\ Rubin Legacy Surveys of Space and Time \citep[LSST;][]{Ivezic2019} opens new opportunities for developing a new generation of astronomical foundation models. These data sets offer multi-wavelength imaging and spectroscopy of millions of galaxies, with relatively few labelled data, and thus provide an ideal sandbox for developing the next generation of machine-learning models. \Euclid's mission is especially valuable for its combination of high-resolution imaging with spectroscopy over a wide field of view, which creates a unique opportunity for advancing the next generation of large-scale machine-learning models. This synergy between cutting-edge surveys and artificial intelligence-driven methods will be instrumental in accelerating adoption of machine learning in astrophysics~\citep{Humphrey-EP22, Leuzzi-TBD, EP-Aussel, EP-Enia, EuclidSkyOverview}.

Several works have started exploring large pre-trained models in astronomy~\citep[e.g.,][]{Vega-Ferrero2024}. \cite{Hayat2021}  and~\cite{2021ApJ...921..177S} first demonstrated the quality of the embedding spaces on Sloan Digital Sky Survey~\citep[SDSS;][]{Gunn1998,Gunn2006} and Mapping Nearby Galaxies
at APO~\citep[MaNGA;][]{Bundy2015} data using contrastive learning, a self-supervised technique where a model is trained to bring similar samples closer in an embedding space while pushing dissimilar ones apart~\citep{Chen2020, Radford2021}. \cite{Smith2024} showed that autoregressive galaxy imaging models follow similar scaling laws to those reported for text. Complementing these imaging methods, {\tt AstroCLIP} \citep{Parker2024} pioneered a multi-modal strategy by aligning spectra and imaging via contrastive learning. 
Following a similar approach, the {\tt PAPERCLIP} framework \citep{Mishra-Sharma2024} explores the intersection of astronomical imaging and natural language by fine-tuning a contrastive language–image pretraining~\citep[CLIP;][]{Radford2021} CLIP model to align observations (e.g., galaxy images) and textual descriptions. This allows users to search for images based on descriptive language, and vice versa. \cite{Zhang2024} and \cite{Rizhko2024} present the first attempts to incorporate time-series data with CLIP, constructing an astronomical multi-modal data set from time-series photometry data, spectra, and astrophysical metadata. These advancements illustrate the expanding frontier of multi-modal approaches, which combine data modalities to provide a more comprehensive understanding of astronomical phenomena.
 As just one example of a downstream use-case, researchers can use these models to gain insight into both the structural properties and the dynamical evolution of galaxies, reaching state-of-the-art performance on classification and redshift estimation without any explicit optimisations towards these tasks. 

\begin{figure}[ht] 
\centering 
\includegraphics[width=0.49\textwidth]{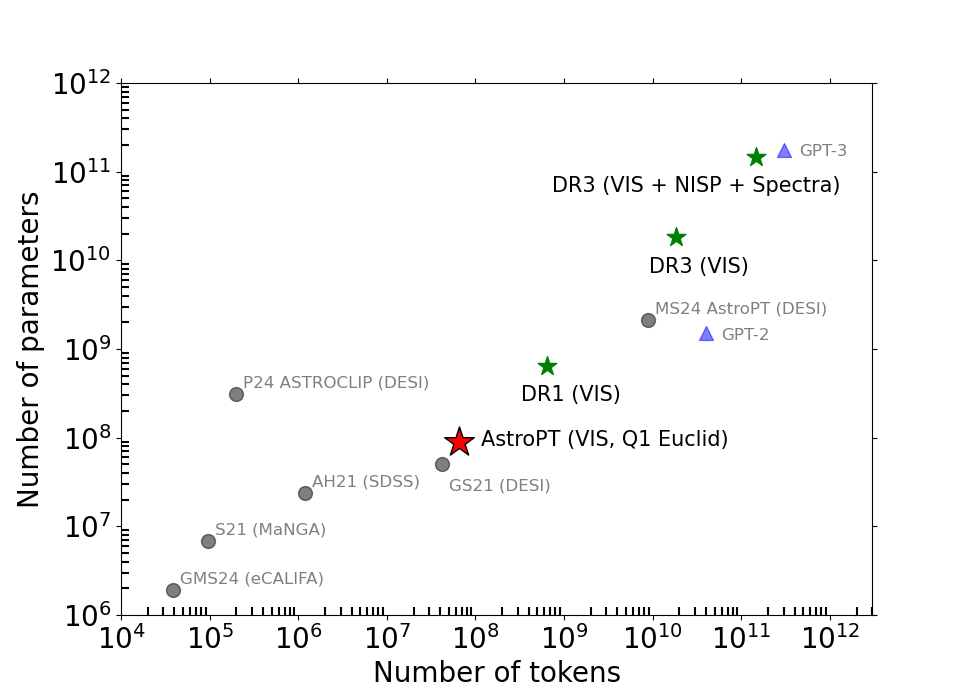} 
\caption{Scaling laws for foundation models in astronomy compared to the GPT series. The number of tokens is defined as subword units in text (GPT models); pairs of augmented images (contrastive learning models); or a number of patches ({\tt AstroPT} model). The projected scaling of future models incorporating \Euclid data in upcoming data releases are marked with green stars. Our current model, trained on Q1 \Euclid data, is represented by a red star. The models from the literature are referenced as follows: AH21~\citep{Hayat2021}; S21~\citep{2021ApJ...921..177S}, GS21~\citep{Stein2021}; GMS24~\citep{2024A&A...688A.160M}; P24~\citep{Parker2024}; and MS24~\citep{Smith2024}.} 
\label{fig:scaling_laws} 
\end{figure}

Following these developments, in this work we pre-train and use a multi-modal autoregressive foundation model (\texttt{AstroPT}) to explore the optical and infrared imaging components, and the spectral energy distributions (SEDs) of \Euclid Q1 galaxies~\citep{Q1cite}. Multi-modal models are capable of integrating and learning from different types of data (such as images, spectra, and text) simultaneously, while autoregressive models predict data sequentially by learning dependencies across inputs, such as pixel sequences in images or wavelength features in spectra or SEDs. Furthermore, autoregressive models can operate on the union of available observations without being restricted solely to cross-matched data sets, which typically represent only a small fraction of the available data.
These properties make models like \texttt{AstroPT} well-suited to analysing the diverse, high-dimensional data produced by modern surveys.
In Fig.~\ref{fig:scaling_laws} we illustrate the rapid growth of foundation models in astronomy, demonstrating an increasing trend in model size as astronomical data sets expand. 
With each new data release -- assuming that galaxy scaling laws hold in the same way as previously found in textual models -- models are projected to scale closer to the size of GPT-2 and GPT-3,  large-scale transformer-based language models developed by OpenAI, representing significant advancements in natural language processing~\citep{brown2020}. 
Additionally, incorporating spectral data into multi-modal models significantly enhances their complexity, potentially requiring a further increase in model size. 

We evaluate the performance of an {\tt AstroPT} model on a variety of downstream tasks -- specific, practical applications of a pre-trained model -- such as galaxy morphology classification, redshift prediction, or stellar mass estimations. By comparing these results to those achieved by a baseline fully supervised model, we assess the advantages of pre-training and multi-modal learning for astronomical data analysis.
Unsupervised and self-supervised learning methods are particularly appealing for downstream tasks because they can generate semantically meaningful representations without requiring extensive labelled data sets. Such methods were successfully applied in astronomy for galaxy morphology studies~\citep[e.g.,][]{Hocking2018,Slijepcevic2024}, estimations of photometric redshift~\citep[photo-$z$; e.g.,][]{Merz2024,Pattnaik2025}, and physical properties~\citep[e.g.,][]{Bisigello-EP23, Chu2024, EP-Enia, Gai2024}, as well as for anomaly detection~\citep[see][for a review]{Fotopoulou2024}, and similarity searches~\citep[e.g.,][]{Stein2021, Walmsley2022Practical, Parker2024}.
Recent advancements in self-supervised learning have significantly closed the performance gap between unsupervised and supervised learning~\citep[e.g.,][]{Cheng2020, Mohale2024}, making these methods viable alternatives for downstream tasks.

This paper is structured as follows. In Sect.~\ref{sc:data}, we introduce the \Euclid mission and the Q1 sample selection. The \texttt{AstroPT} framework is presented in Sect.~\ref{sc:Methodology}, together with other approaches to predict galaxy properties. In Sect.~\ref{sc:Results} we discuss the performance of \texttt{AstroPT} on \Euclid imaging data, in comparison to other methods.  Section~\ref{sc:Summary} summarises the main results. 
In App.~\ref{app:hyperparameters} the hyperparameters of the model are given, App.~\ref{app:VIS_NISP_visualization} presents the embeddings for VIS, and NISP imaging. The star--galaxy separation is discussed in App.~\ref{app:star_galaxy_qso_separation}, and physical properties in App.~\ref{app:quality_of_emb_PhysProp}. Finally, the accuracy of morphological classification, and predictions of stellar masses and photo-$z$ is detailed in App.~\ref{app:statistics}.

\section{\label{sc:data}Data}
The \Euclid mission, launched in 2023, is a mission from the European Space Agency (ESA), dedicated to investigating the nature of dark energy, and dark matter by measuring the accelerated expansion of the Universe. \Euclid primarily focuses on gathering high-resolution imaging and spectroscopic data of around $10^9$ galaxies out to $z \approx 3$, and millions of galaxies out to $z \approx 6$ over 14\,000 $\rm deg^2$~\citep{EuclidSkyOverview}. \Euclid provides broadband optical imaging through the visible imaging channel~\citep[VIS;][]{Q1-TP002}, and multi-band near-infrared spectroscopy and imaging with the near-infrared spectrometer and photometer~\citep[NISP;][]{EuclidSkyNISP}. 
The VIS instrument captures optical data in the visible range corresponding to $r$, $i$, and $z$ filters (\IE covers $\lambda = 0.53$--0.92\,\micron; \citealt{EuclidSkyVIS}), with a spatial resolution of approximately 0.\arcsecond18~\citep{EuclidSkyVIS}. This high-resolution imaging provides an excellent foundation for morphological studies, allowing the extraction of detailed structural properties of galaxies. 
The NISP captures infrared data \citep[\YE, \JE, and \HE cover $\lambda = 0.95$--2.02\,\micron; ][]{Schirmer-EP18}  and is designed for photometric redshift (photo-$z$) determination and galaxy clustering measurements, which complement the VIS data by providing additional depth and accuracy for cosmic surveys.

\subsection{\label{sc:Q1data}Q1 sample}

As part of the \Euclid mission, Q1 represents the first major public data release. Q1 provides a snapshot of early observations taken during the initial phase of the mission, covering $\sim$29.7 million sources over $\rm 63~deg^2$ of the Euclid Deep Fields (EDFs).  These fields are the Euclid Deep Field North, (EDF-N), the Euclid Deep Field (EDF-S), and Euclid Deep Field Fornax (EDF-F). The Q1 release includes raw and processed images (calibrated VIS, NISP frames, and background-subtracted mosaics, and segmentation maps), along with the MER photometric catalogue~\citep{Q1-TP004}. A detailed description of the Q1 data release is presented in \citet{Q1-TP001}, \citet{Q1-TP002}, \citet{Q1-TP003}, and \citet{Q1-TP004}. 

The MER catalogue includes multi-wavelength photometry, combining data from \Euclid VIS and NISP photometry with external photometric surveys spanning the ultraviolet to the near-infrared wavelengths. In particular, the catalogue includes fluxes in the \IE, \HE, \JE, and \YE \Euclid bands, complemented by optical bands ($u$, $g$, $r$, $i$, and $z$) from the Dark Energy Survey~\citep{DES2016}, and the ultraviolet near-infrared optical northern survey (UNIONS; Gwyn et al. in prep.).  UNIONS imaging includes CFHT $u$, $r$, HSC $g$, $z$, and Pan-STARRS $i$, $z$, and is an independent collaboration aimed at enhancing the scientific yield of large and deep surveys of the northern sky. To ensure robust cross-ground-based survey comparisons, the photometry is measured using {\tt A-PHOT}~\citep{Merlin2019}, employing fixed circular apertures with diameters of 2 full width at half maximum (FWHM) of the worst-seeing band (typically around $2^{\prime\prime}$), computed individually for each source based on PSF-matched images. 
The bands used to construct SEDs (i.e., the collection of photometric measurements), and their corresponding fields are listed in Table~\ref{tab:photometric_bands}. 

\begin{figure}[ht] 
\centering 
\includegraphics[width=0.49\textwidth]{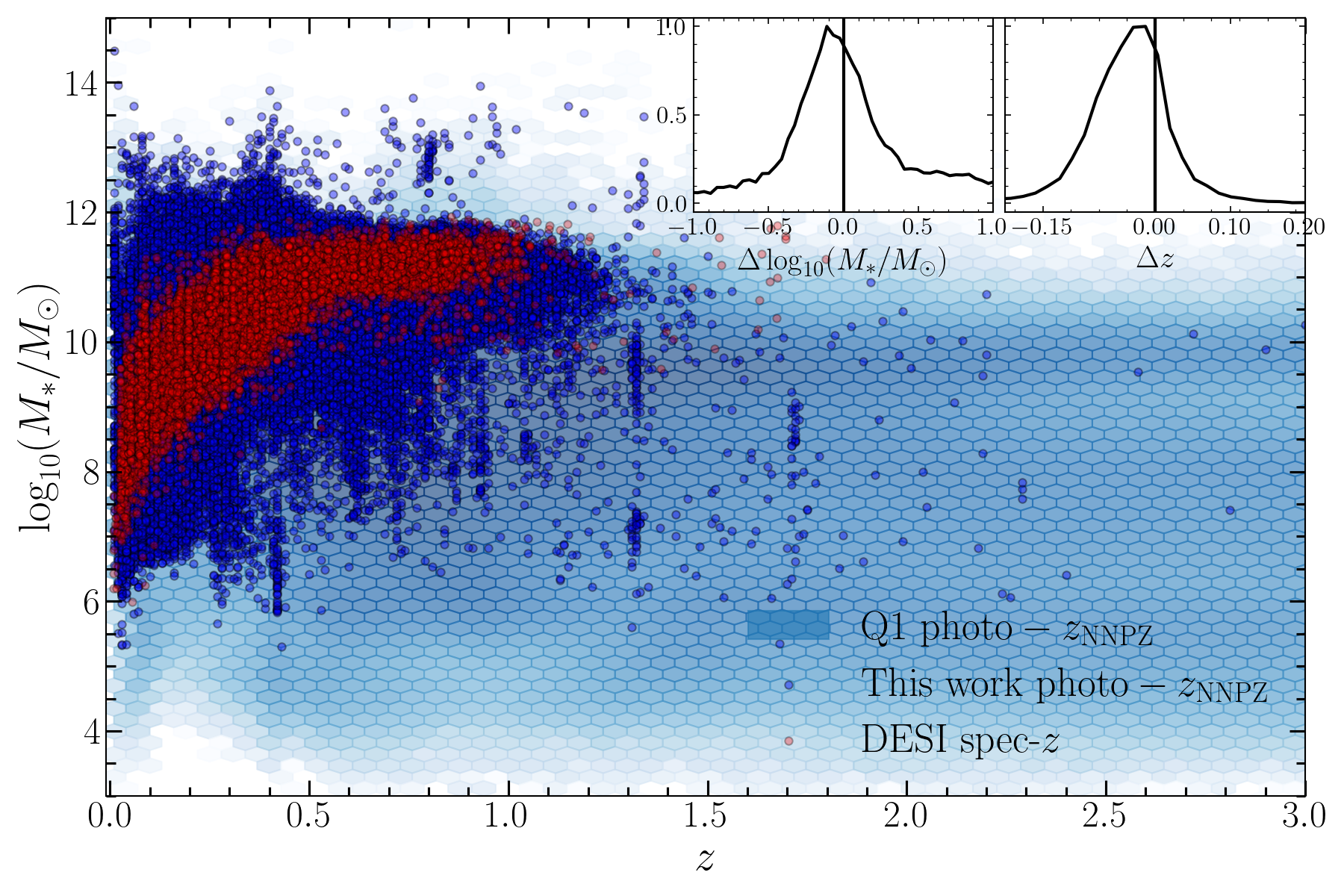} 
\caption{Stellar mass-redshift relation for the Q1 sample (in gray), the sample used in this work (in blue), and our sample with DESI spectroscopic information (in red).  The inset plots show the residuals of redshift and stellar mass differences between the \Euclid pipeline and DESI, respectively. } 
\label{fig:stellarmass_redshift_relation} 
\end{figure}

This data set is further complemented by a \texttt{\{star, galaxy, quasar (QSO)\}} trinary classification, the photo-$z$, and galaxy physical properties estimated with the $k$-nearest-neighbour algorithm~\citep[{\tt NNPZ};][]{Tanaka2018,Desprez-EP10, EP-Enia, Q1-SP031, Q1-TP005}, among others. 
While these estimates are useful for certain downstream tasks (see Sect.~
\ref{sec:downstream_tasks}), spectroscopic redshifts~\citep{DESI2024_EDR} and stellar masses~\citep{Siudek2024} from DESI are used for validation to ensure reliability. The DESI sample constitutes only a small fraction of the entire Q1 data set, and does not cover the full panoply of \Euclid Q1 galaxies (see Fig.~\ref{fig:stellarmass_redshift_relation}).  

\begin{table*}
\caption{Photometric bands, instruments, observed depths, and fields included in the \Euclid photometric catalogue used to build the SEDs of Q1 sources. }\label{tab:photometric_bands}
\centering                                   
\begin{tabular}{ccccc}         
\hline\hline 
Band &  Instrument &  $\lambda$ [nm] &  Depth [mag] & Field\\
\hline
$u$  & CFHT/MegaCam     & 372 & 23.5 & EDF-N \\
$g$  & HSC         & 480 & 25.3 & EDF-N \\
$r$  & CFHT/MegaCam     & 640 & 24.1 & EDF-N \\
$i$  & Pan-STARRS  & 755 & 23.3 & EDF-N \\
$z$  & HSC         & 891 & 23.5 & EDF-N \\     
$g$  & DECam       & 473 & 24.6 & EDF-S, EDF-F \\
$r$  & DECam       & 642 & 24.3 & EDF-S, EDF-F \\
$i$  & DECam       & 784 & 23.7 & EDF-S, EDF-F \\
$z$  & DECam       & 926 & 22.9 & EDF-S, EDF-F \\
\IE  & VIS         & 715 & 24.7 & EDF-N, EDF-S, EDF-F  \\
\YE  & NISP        & 1085 & 23.1 & EDF-N, EDF-S, EDF-F \\     
\JE  & NISP        & 1375 & 23.2 & EDF-N, EDF-S, EDF-F \\   
\HE  & NISP        & 1773 & 23.2 & EDF-N, EDF-S, EDF-F \\ 
\hline
\end{tabular} 
\end{table*}

\begin{figure}[ht] 
\centering 
\includegraphics[width=0.49\textwidth]{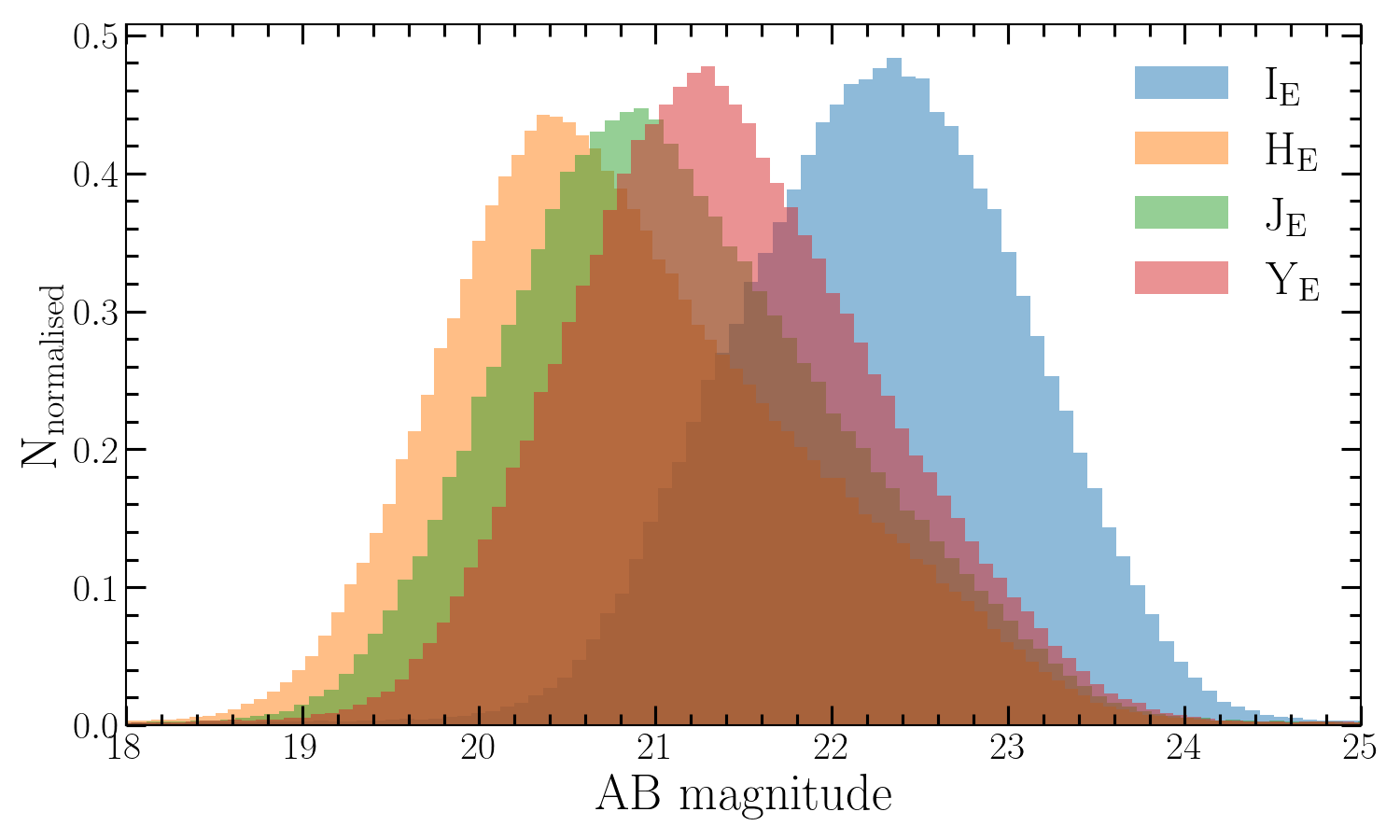} 
\caption{Distribution of \Euclid photometry for the sample used in this paper.} 
\label{fig:mag_distribution} 
\end{figure}

\begin{figure}[ht] 
\centering 
\includegraphics[width=0.49\textwidth]{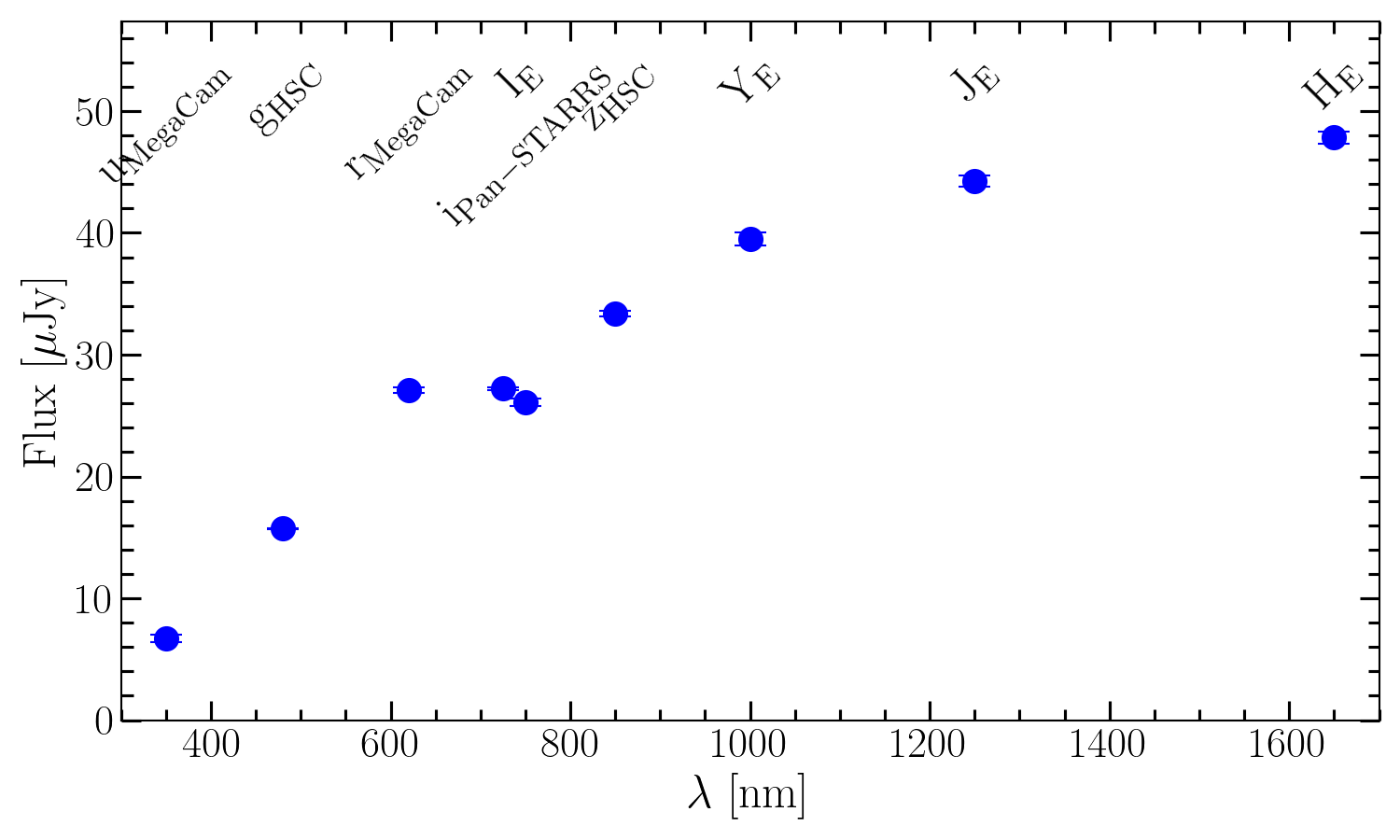} 
\includegraphics[width=0.49\textwidth]{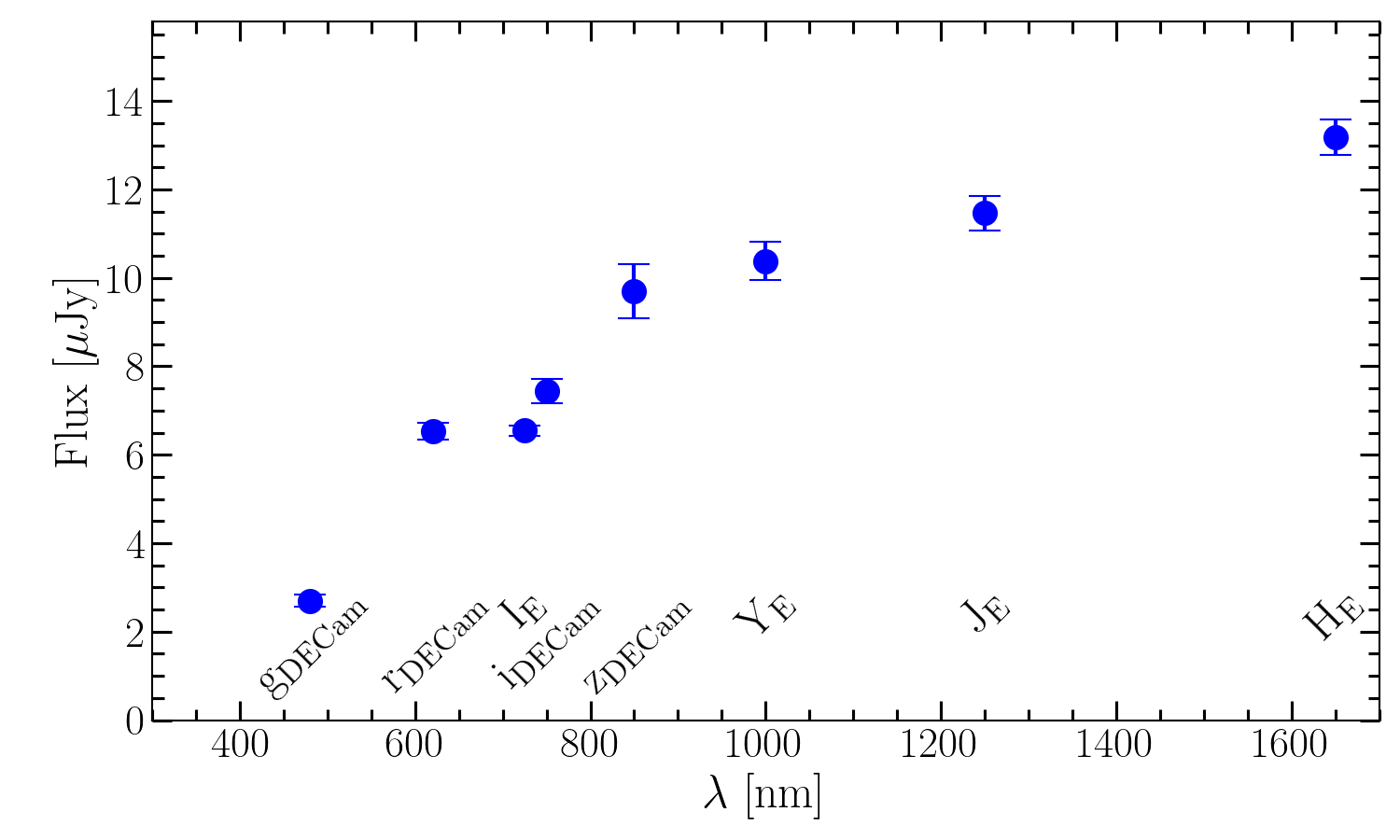} 
\caption{Representative SEDs of galaxies from the Q1 EDF-N (top) and EDF-S fields (bottom). The SED includes multi-wavelength photometric data spanning the ultraviolet to the near-infrared listed in Table~\ref{tab:photometric_bands}. } 
\label{fig:example_SED} 
\end{figure}

In this work, we select sources with a segmentation area greater than 800 pixels in the VIS  data, which corresponds to galaxies that are sufficiently large to have accurate morphological classifications~\citep{Q1-SP048}. The \Euclid survey operates with a pixel scale of approximately 0.\arcsecond1 per pixel, corresponding to a physical resolution of approximately 0.5 kpc at $z\approx0.5$. This criterion ensures that the sample includes only well-resolved objects, allowing for accurate morphological characterisation. We additionally restrict our sample to galaxies with \HE $<22.5$. This magnitude limit ensures the inclusion of objects that are bright enough to be detectable in the near-infrared bands, which will be crucial for obtaining their spectra.  Figure~\ref{fig:mag_distribution} shows the distribution of \Euclid magnitudes for sources in our sample, broken down by the \IE, \YE, \JE, and \HE bands. 
For the selected sample the images are complemented with the SEDs of sources covering the optical through to infrared fluxes.  An example SED constructed from \Euclid and auxiliary survey data is shown in Figure~\ref{fig:example_SED}. This work solely builds on the imaging and photometric data, while we leave the addition of the spectra for future work. 

Our final set comprises approximately 330\,000 galaxies spread across the  Q1 fields, shown in blue in Fig.~\ref{fig:stellarmass_redshift_relation}. We split our
galaxy data set using an 80/10/10 training-test-validation split. This selection introduces constraints in both redshift and stellar mass, which may influence the representation of our data and the performance of downstream tasks. In particular, our sample is mostly limited to $z \lesssim 1$ and $\Mstarsun \gtrsim 7$. These cuts ensure higher-quality images for our analysis but could introduce biases, particularly in studies of low-mass and high-redshift galaxies. The spectroscopic subset from DESI further refines the selection, potentially impacting our ability to characterise the most massive systems with precise redshift information.

\section{\label{sc:Methodology}Methodology}

In this section, we outline the machine learning methodologies employed to analyse the galaxy data set, starting with {\tt AstroPT} in Sect.~\ref{sc:AstroPT}. In Sect.~\ref{sec:Downstreamtaks:methodology} we introduce three separate models used to evaluate the performance of different techniques in downstream tasks (Sect.~\ref{sec:downstream_tasks}).  

\subsection{\label{sc:AstroPT}AstroPT}

\begin{figure*}[h]
\centering
\begin{subfigure}[t]{0.45\textwidth}
\centering
\raisebox{2.5em}{\includegraphics[width=0.5\textwidth]{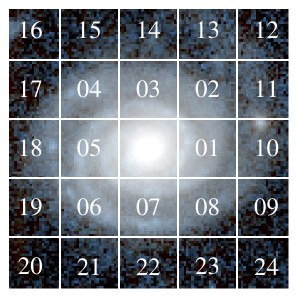}}
\subcaption{The sequence order of galaxy image patches used in this work.}
\end{subfigure}
\hfill
\begin{subfigure}[t]{0.54\textwidth}
\includegraphics[width=\textwidth]{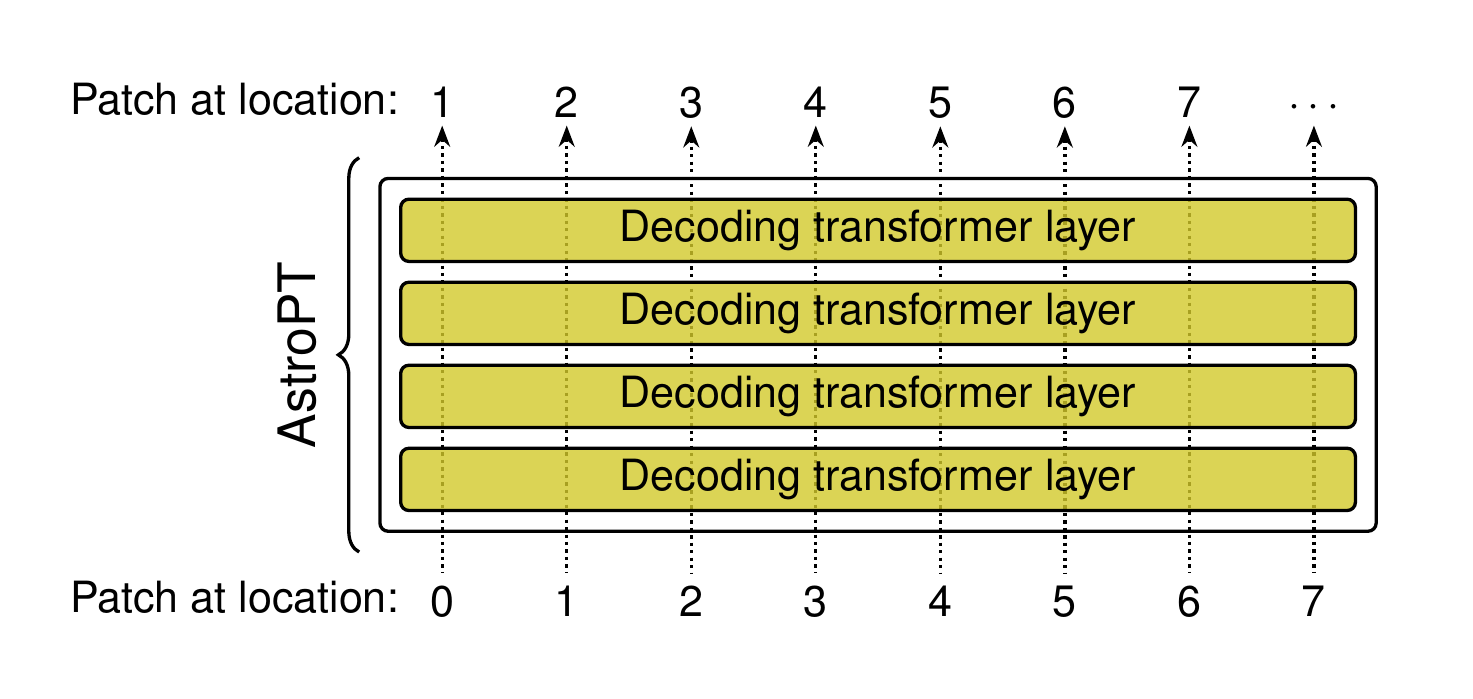}
\subcaption{A schematic of {\tt AstroPT}'s architecture.}
\end{subfigure}
\caption{{\tt AstroPT's} training objective is predicting the next patch in a sequence of galaxy image (and other modalities') patches. On the left we show the order of galaxy patches used in this work, and on the right we show how each patch is processed by {\tt AstroPT}.}
\label{fig_galaxyastropt}
\end{figure*}

{\tt AstroPT}\footnote{\url{https://github.com/Smith42/astroPT}} is an autoregressive transformer designed to extract scientifically useful embeddings from high-resolution galaxy images. 
A visualisation of the {\tt AstroPT} architecture is shown in the right panel in Fig.~\ref{fig_galaxyastropt}. 
Due to space constraints, we do not go into detail about the transformer architecture in this paper. We instead recommend Sect.~4(d) of \citet{Smith2023} for an astronomy-friendly introduction of the transformer model, and Sects. 7 and 9 of the same publication for a further discussion of the ideas behind {\texttt AstroPT} and other foundation models. 

{\tt AstroPT} is initially trained on the task of predicting the next segment of a stream of astronomical data. This training process incentivises the model to learn a compact representation of the data, which captures the essential patterns and relationships inherent in the astronomical observations it is pre-trained on. 
Our data set consists of 224$\times$224 pixel FITS cutouts in the VIS and NISP bands fixed to the same angular size from mosaics, i.e.,  MER background-subtracted calibrated frames. The cutout size is an arbitrary choice, corresponding to a commonly used input size for deep-learning models due to their compatibility with standard convolutional neural network architectures and {\tt AstroCLIP}~\citep{Parker2024}. The fixed size allows for efficient training on GPUs while maintaining a balance between spatial resolution and computational cost. However, this choice has implications for the representation of galaxy morphology: (i) smaller galaxies may be dominated by background noise or multiple objects might be in the frame; and (ii) larger galaxies might not fit entirely within the cutout, leading to truncation of their outer regions. These effects introduce potential biases into the embeddings, as their descriptive power may depend on the relative size of the galaxy within the cutout. 
For images, the training process involves optimising a Huber loss~\citep{Huber1964} objective, with the model attempting to predict the next $16\times16$ pixel patch in a sequence of galaxy image patches (as illustrated in the left panel of Fig.~\ref{fig_galaxyastropt}).
 In this study, we use the same model architecture as the 90 million parameter model described in \cite{Smith2024}.
 To ensure consistency across inputs, each image patch is normalised by taking its $z$-score, defined as
\begin{equation}
    z = \frac{x - \mu}{\sigma},
\end{equation}
 where $x$ is the pixel value, $\mu$, is the mean and $\sigma$ is the standard deviation of the pixel values across the patch. This normalisation centres the data around zero and scales it to have unit variance, improving stability during pre-training~\citep[e.g.,][]{El-Nouby2024}. 
   However, we acknowledge that global normalisation may have an impact on information, like a contrast reduction in faint structures, particularly in the outskirts of galaxies. The verification of alternative normalisation strategies, such as local normalisation or 'min-max' scaling, is left to future work.
 To incorporate SEDs, we use a simple token-chaining mechanism. This process involves concatenating the SED data as a token sequence, to be fed in after the imaging tokens, thus allowing the model to process both modalities  (imaging, and SED information) simultaneously. The SEDs are divided into patches of a length of one, and the SEDs are normalised using the $z$-score over the entire SED sequence. While the $z$-score normalisation preserves the relative shape of SEDs, it could, in principle, reduce the distinctiveness of extreme SEDs (e.g., very blue or very red galaxies). Testing alternative approaches, such as converting fluxes to magnitudes before normalisation, is left for future work. 
We use separate encoder and decoder layers for the images and SEDs. 
 The encoders take as input the raw image or SED patch data, and process these data so that they produce embedding patches that are of a fixed size.
 These embedding patches are then passed into the model `trunk', shown on the right of Fig.~\ref{fig_galaxyastropt}.
 After the model trunk processes these patches, the output is then projected back into a prediction of the next image or SED patch by the decoder head,  which consists of the final layers of the model.

In our case, {\tt AstroPT} was pre-trained three times:
firstly on solely optical (VIS) images;
secondly, on optical and NISP (VIS+NISP) images;
and thirdly on a chained modality stream of optical and NISP images and pre-calculated SEDs (VIS+NISP+SED).
A 90-million parameter model was used in all cases due to its efficiency, as larger architectures did not show significant performance improvements.
We state our full hyperparameters in Appendix~\ref{app:hyperparameters}.

\subsection{Models and evaluation metrics}\label{sec:Downstreamtaks:methodology}

We investigate three methodologies for predicting galaxy properties, utilising supervised and self-supervised techniques. These approaches leverage the same data set, and are applied to both regression tasks (such as photo-$z$ and stellar mass estimation) and classification tasks (such as morphological classification of late/early-type galaxies).
\begin{enumerate}
    \item {Supervised convolutional neural network (CNN)}: This approach involves training a CNN~\citep{Fukushima1980} using the same cutouts as used for {\tt AstroPT}, ensuring consistency in the input data.  The implementation of SEDs is beyond the scope of this paper, and so is left for future work. Our CNN consists of convolutional layers followed by dense layers, which are trained end-to-end using the mean squared error loss (MSE) function for regression tasks, and the binary cross-entropy (BCE) loss function for classification tasks. This method is a fully supervised approach where the network relies entirely on labelled data sets to learn the mapping from input features to output predictions.  
    \item Linear model: Here we evaluate the quality of our representations derived from the {\tt AstroPT} framework via a simple linear or logistic regression model. This is a minimalistic and interpretable approach, where we train our regressor directly on the embeddings generated by the {\tt AstroPT} framework. This approach leverages the representations learned by the self-supervised model and evaluates their utility without additional modifications to the embeddings. For regression tasks, the linear regression model outputs continuous values, while for classification tasks,  the logistic regression model outputs class probabilities.
    \item MLP model: This method also uses self-supervised embeddings from the {\tt AstroPT} model as fixed input features. A multi-layer perceptron~\citep[MLP;][]{rosenblatt1958} is then trained on labelled data for either regression or classification tasks. The MLP consists of four fully connected layers. To ensure stable, efficient training, batch normalisation is applied after each layer and dropout regularisation (with a 30\% probability) is used to prevent overfitting. We use the leaky rectified linear unit~\citep[ReLU;][]{nair2010, Xu2015} as our activation function. The model is trained using the AdamW optimiser~\citep{Loshchilov2017} with weight decay, and the Huber\citep{Huber1964} or BCE loss is used as the loss function for regression or classification tasks, respectively.
\end{enumerate}
 All source code and full implementation details for our downstream tasks are available in the {\tt AstroPT} GitHub repository\footnote{\url{https://github.com/Smith42/astroPT/tree/euclid_scripts/scripts/euclid}}, which provides further insight into the underlying methods. Table~\ref{tab:model_comparison} summarises the key details about the methodologies used for downstream tasks. 

\begin{table*}
    \centering
    \caption{Comparison of different methodologies for regression and classification downstream tasks. }
    \label{tab:model_comparison}
    \begin{tabular}{l l l}
        \hline\hline
         Model & Loss function & Key features \\
        \hline
        Supervised CNN & MSE / BCE & Non-linear, deep feature extraction; trained on labelled data. \\
        Linear Model & MSE / Log Loss & Simple regression, assumes linear relationships, trained on embeddings. \\
        MLP Model &  Huber / BCE & Deep architecture, trained on embeddings. \\
        \hline
    \end{tabular}
\end{table*}

The morphological labels for all our sample come from the morphology catalogue\citep[GZ;][]{Q1-SP047}, for $\approx 5\%$ of our sample the stellar mass labels, and the spectroscopic redshift are taken from from DESI Early Data Release~\citep{DESI2024_EDR}, and associated Value Added Catalogue of physical properties of DESI galaxies~\citep{Siudek2024}. 

To assess the robustness of the models, a Monte Carlo simulation is conducted with 50 (reduced to 10 for supervised CNN due to time limitations) runs for different percentage of data used for training, each time evaluating key metrics given below. 
For evaluating our models' performance, we use different sets of metrics depending on whether the tasks are regression or classification.

For predicting continuous values, such as photo-$z$, or stellar mass estimates, we rely on the following metrics commonly used in regression tasks

\begin{itemize}
    \item Bias (\(\Delta x\)): Indicates systematic offsets in predictions. 
    We apply the normalisation to account for the scale dependence of redshift errors that scale with \(1 + x_{\text{true}}\), we keep it as well for stellar mass estimates:
\begin{equation}
        \Delta x = \frac{x_{\text{pred}} - x_{\text{true}}}{1 + x_{\text{true}}},
\end{equation}
    where \(x_{\text{true}}\) and \(x_{\text{pred}}\) correspond to the true and predicted values, respectively. 
    \item Normalised median absolute deviation (NMAD): A robust statistic to measure scatter that is less sensitive to outliers. It is defined as
\begin{equation}
    \text{NMAD} = 1.4826 \times \text{median}\left( \left| \Delta x - \text{median}(\Delta x) \right| \right).\end{equation}

    \item Outlier fraction: The fraction of predictions with $|\Delta x|$ exceeding 0.15, and 0.25 for photo-$z$, and stellar mass, respectively, quantifying catastrophic prediction errors:
\begin{equation}
    \text{Outlier fraction} = \frac{\text{Number of galaxies with } |\Delta x| > 0.15(0.25)}{N}.
\end{equation}

     The choice of the threshold of 0.15 for photo-$z$ is commonly used to define catastrophic outliers. For stellar mass, while no strict convention exists, a threshold of 0.25 corresponds to standard stellar mass error~\citep[][]{Siudek2024}.

\end{itemize}

 For binary classification tasks such as morphological classification (such as early-type versus late-type galaxies), we use the following metrics.

\begin{itemize}
    \item {Accuracy:} The proportion of correct predictions (both true positives and true negatives) among all predictions:
\begin{equation}
    \text{Accuracy} = \frac{N_{\rm TP} + N_{\rm TN}}{N_{\rm TP} + N_{\rm FP} + N_{\rm TN} + N_{\rm FN}},
\end{equation}

    where \( N_{\rm TP} \), \( N_{\rm TN} \), \( N_{\rm FP} \), and \( N_{\rm FN} \) are the number of true positives, true negatives, false positives, and false negatives, respectively.

    \item {Precision:} The proportion of true positives among all instances predicted as positive:
\begin{equation}
    \text{Precision} = \frac{N_{\rm TP}}{N_{\rm TP} + N_{\rm FP}}.
\end{equation}

    \item {Recall (true positive rate, TPR):} The proportion of true positives among all actual positive instances:
\begin{equation}
    \text{Recall} = \frac{N_{\rm TP}}{N_{\rm TP} + N_{\rm FN}}.
\end{equation}

    \item {False positive rate (FPR):} The proportion of false positives among all actual negatives:
\begin{equation}
    \text{FPR} = \frac{N_{\rm FP}}{N_{\rm FP} + N_{\rm TN}}.
\end{equation}

    \item {F1 score:} The harmonic mean of precision and recall, balancing the trade-off between them:
\begin{equation}
    F1 = 2 \times \frac{\text{precision} \times \text{recall}}{\text{precision} + \text{recall}}.
\end{equation}
\end{itemize}

\section{\label{sc:Results}Results}

In this work, we explore the latent representations of galaxies derived from our pre-trained {\tt AstroPT} models applied to three distinct data sets. 
The complementary nature of VIS, VIS+NISP, and VIS+NISP+SED data facilitates a systematic evaluation of the information content in the latent spaces derived from our models. 
This comparison also allows us to assess how well each wavelength constrains a galaxy's physical parameters and to understand the limitations of relying solely on optical imaging.

\subsection{\label{sc:embeddings}{\tt AstroPT}: Quality of self-supervised embeddings}

In this section, we assess the quality of self-supervised embeddings obtained from the \Euclid VIS, VIS+NISP, and VIS+NISP+SED data via {\tt AstroPT}.
 We do this by analysing their correlation with source properties.
 In all downstream tasks (including anomaly and similarity search, see Sect.~\ref{sec:downstream_tasks}), we use the raw 768-dimensional neural network embeddings as input. For visualisation purposes, we project this high-dimensional embedding space onto two dimensions using uniform manifold approximation and projection~\citep[UMAP;][]{McInnes2018}.
 We initialise UMAP with a consistent set of hyperparameters for all visualisations ($\tt n\_neighbours =15$, $\tt min\_dist = 0.1$, $\tt n\_components=2$, $\tt random\_state=42$) to ensure consistency across tasks.  We validate different UMAP parametrisations, to: (i) determine whether the randomness in UMAP initialisation impacts the visualisations; (ii) to emphasize local structures that can potentially create tighter clusters; and (iii) to highlight the global relationships. While the shapes of the projections varied slightly across these settings, the trends related to galaxy morphological and physical properties remained consistent. Thus, we stick to the default values that provide a good trade-off between the local, and global structures. The UMAP projection facilitates an intuitive exploration of clustering patterns and structures within the data. 
UMAP is particularly effective in preserving local and global structures,  often outperforming methods such as principal component analysis (PCA) in revealing underlying relationships in the data~\citep[e.g.,][]{McInnes2018}. 
 However, we note that the dimensionality reduction algorithm ultimately influences the patterns and separations observed in these visualisations. While the UMAP projections suggest that relevant morphological and physical information is encoded in the {\tt AstroPT} embeddings, the degree of separation achieved depends on the algorithm's ability to preserve the global structure.


To more directly evaluate the utility of these embeddings, we also examine their performance in downstream tasks through approaches like the linear and MLP models (see Sect.~\ref{sec:downstream_tasks}). These analyses provide a complementary and quantitative assessment of the embeddings' capacity to capture source properties, independent of the visualisation technique used.

\begin{figure*}[ht] 
\centering 
\includegraphics[width=0.99\textwidth]{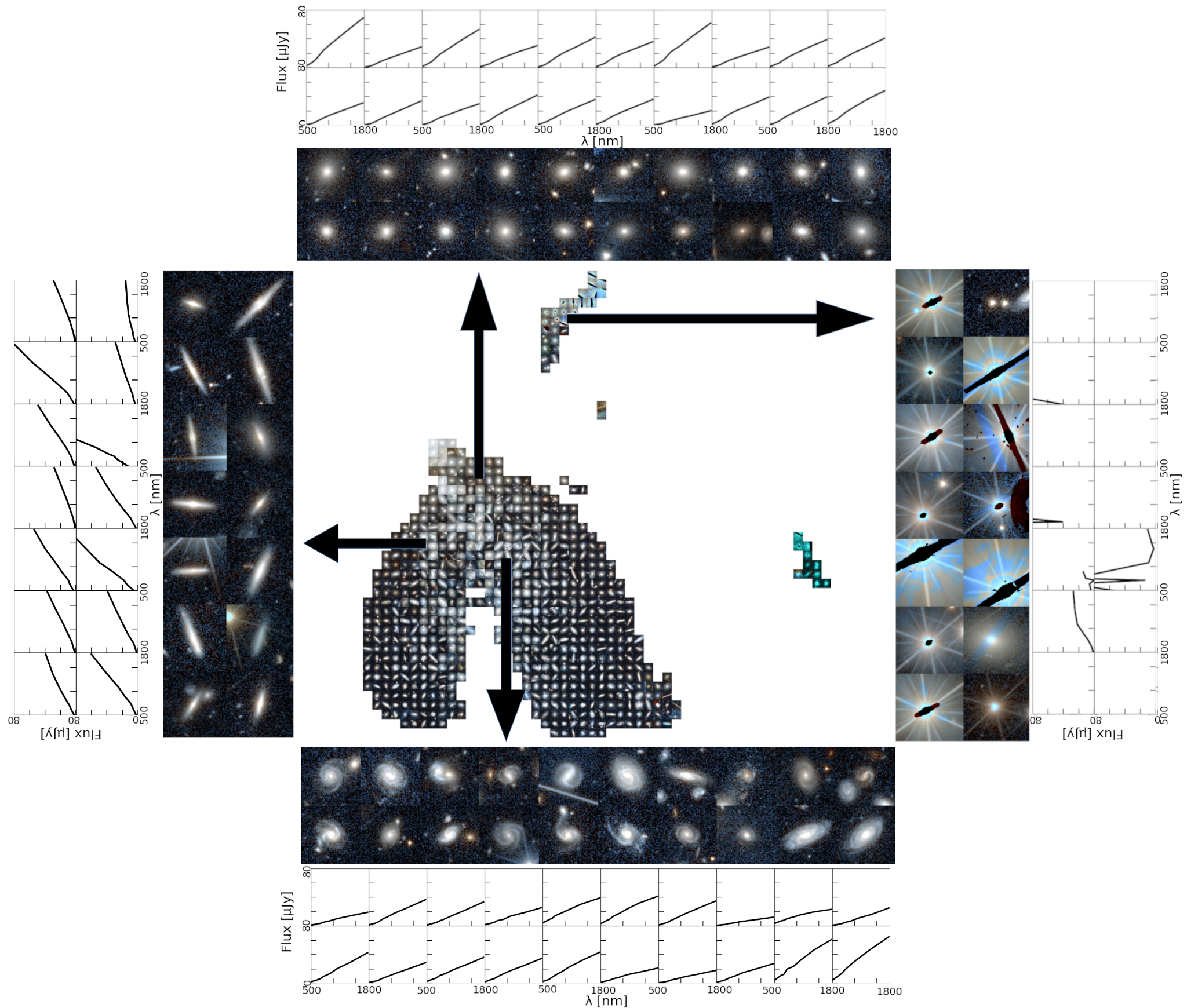} 
\caption{UMAP visualisations of the embeddings from {\tt AstroPT} trained on VIS+NISP+SEDs with example cutouts and SEDs, see Sect.~\ref{sc:embeddings} for more details. } 
\label{fig:UMAP_visualisation_cutout} 
\end{figure*}

A visualisation of the self-supervised embeddings trained on the VIS+NISP+SEDs images with example cutouts, and SEDs is shown in Fig.~\ref{fig:UMAP_visualisation_cutout}. The figure is created by binning the embedding space into a $\rm 50\times50$ grid and randomly selecting a sample from each cell, plotting its corresponding colour-mapped galaxy image at that location. The images are colour-mapped based on the VIS and NISP band images.  The RGB image is generated only for visualisation purposes by averaging the \YE-, \JE-, and \HE-band images to form an NISP channel, combining the VIS channel with this NISP channel to form a composite, and then stacking these three channels: VIS, NISP, and the composite. To preserve luminance, the luminosity channel of the resulting RGB image is replaced with the VIS channel, maintaining the brightness information while allowing for distinct colour contrasts. Zoom-in panels around the edges show that galaxies with similar morphologies tend to be clustered in different areas allowing for a separation of edge-on  (left zoom-in panel), elliptical (top zoom-in panel), and spiral (bottom zoom-in panel) galaxies, while stars and low-quality images are separated in distinct cluster (right zoom-in panel). 
 The zoom-in panels of spiral, and elliptical galaxies are selected within a photo-$z$ range of 0.5 to 0.7, ensuring a fair comparison of their SEDs. While no striking differences are observed, the SEDs appear to broadly reflect morphological trends. Specifically, edge-on disc galaxies and ellipticals tend to be redder. However, this distinction is subtle and subject to interpretation. Additionally, the SEDs of stars and sources with lower data quality show spurious features.  
A similar plot, but generated using the {\tt AstroPT} embeddings trained on VIS, and VIS+NISP data, is provided in Appendix~\ref{app:VIS_NISP_visualization}. 

\begin{figure*}[ht] 
\centering 
\includegraphics[width=0.99\textwidth]{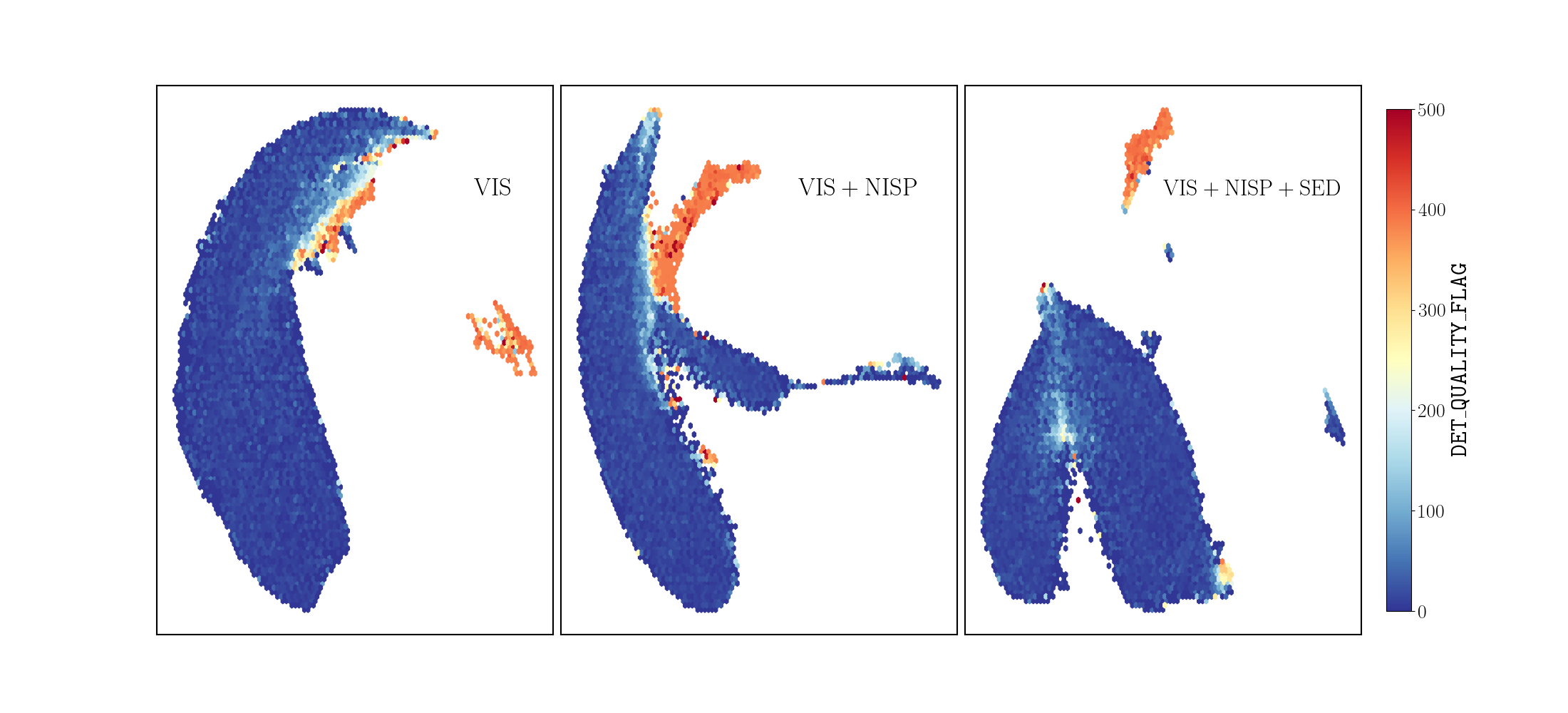}
\includegraphics[width=0.8\textwidth]{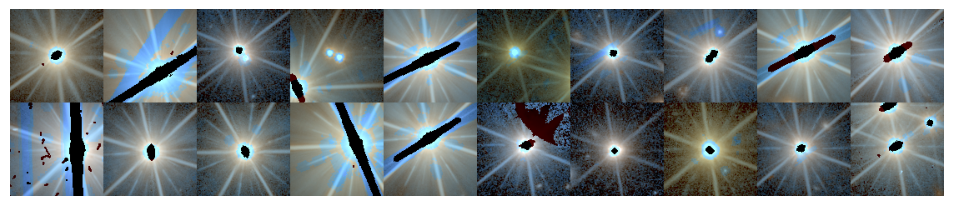} 
\caption{$Top$: UMAP visualisations of the self-supervised embeddings from {\tt AstroPT} trained on VIS (left), VIS+NISP (centre), and VIS+NISP+SED (right) images coloured by the detection quality flag.  This flag allows to identify sources with contaminants, e.g., proximity star ({\tt DET\_QUALITY\_FLAG}~$\geq 4$). $Bottom$: Example VIS+NISP cutouts of the outliers in the representative space. }
\label{fig:UMAP_visualisation_qualilty_flag} 
\end{figure*}

The visual similarity of sources located close to each other in the representative space is also confirmed by their properties. The outlier regions are clearly separated from the main representations and are mostly attributed to bad-quality imaging and stars. 
Figure~\ref{fig:UMAP_visualisation_qualilty_flag} shows the relationship between the embeddings trained on VIS images only, on VIS+NISP images, and VIS+NISP+SED, and the detection quality flag. The detection quality flag ({\tt DET\_QUALITY\_FLAG}) is a flag indicating specific issues with sources, such as contamination, blending, or proximity to bright stars~\citep{Merlin2015, Merlin2016, Merlin2019, Kummel2022}. 
The UMAP projections reveal significant structure within the embedding space, reflecting the quality of the data: sources with high-quality images ({\tt DET\_QUALITY\_FLAG}~$<4$) cluster tightly, whereas sources with contamination (e.g., blending or proximity to bright stars) form distinct outlier clusters. 
Including NISP data enhances the separation of poorly flagged sources, suggesting that NIR data provides complementary morphological information compared to VIS data.  This improvement may arise because NIR observations are less affected by dust extinction and better trace the underlying stellar mass distribution, which can help distinguish extended or blended sources that are challenging to classify in VIS alone. 
In particular, embeddings trained on VIS+NISP images reveal two primary branches: one dominated by sources flagged as unknown (i.e. not fitting galaxy, star, or QSO templates); and another branch associated with sources of lower quality. However, the separation between galaxies, stars, and QSOs remains less defined (see details in Appendix~\ref{app:star_galaxy_qso_separation}). 
 When SED information is incorporated into the training (VIS+NISP+SED), the embeddings show a bimodal structure, which is driven by the spatial distribution: the EDF-N and EDF-S are targeted by different telescopes (see Sect.~\ref{sc:data}). Besides these two main branches, stars are also more distinctly separated, benefiting from the additional spectral information provided by SEDs. Furthermore, the unknown sources form a more coherent and distinguishable cluster, indicating that the combination of imaging and SED data aids in isolating these objects from other classes. This highlights the potential of SED data to complement imaging in disentangling complex or ambiguous cases, such as blended sources or objects with uncertain classifications.

\begin{figure*}[ht] 
\centering 
\includegraphics[width=0.95\textwidth]{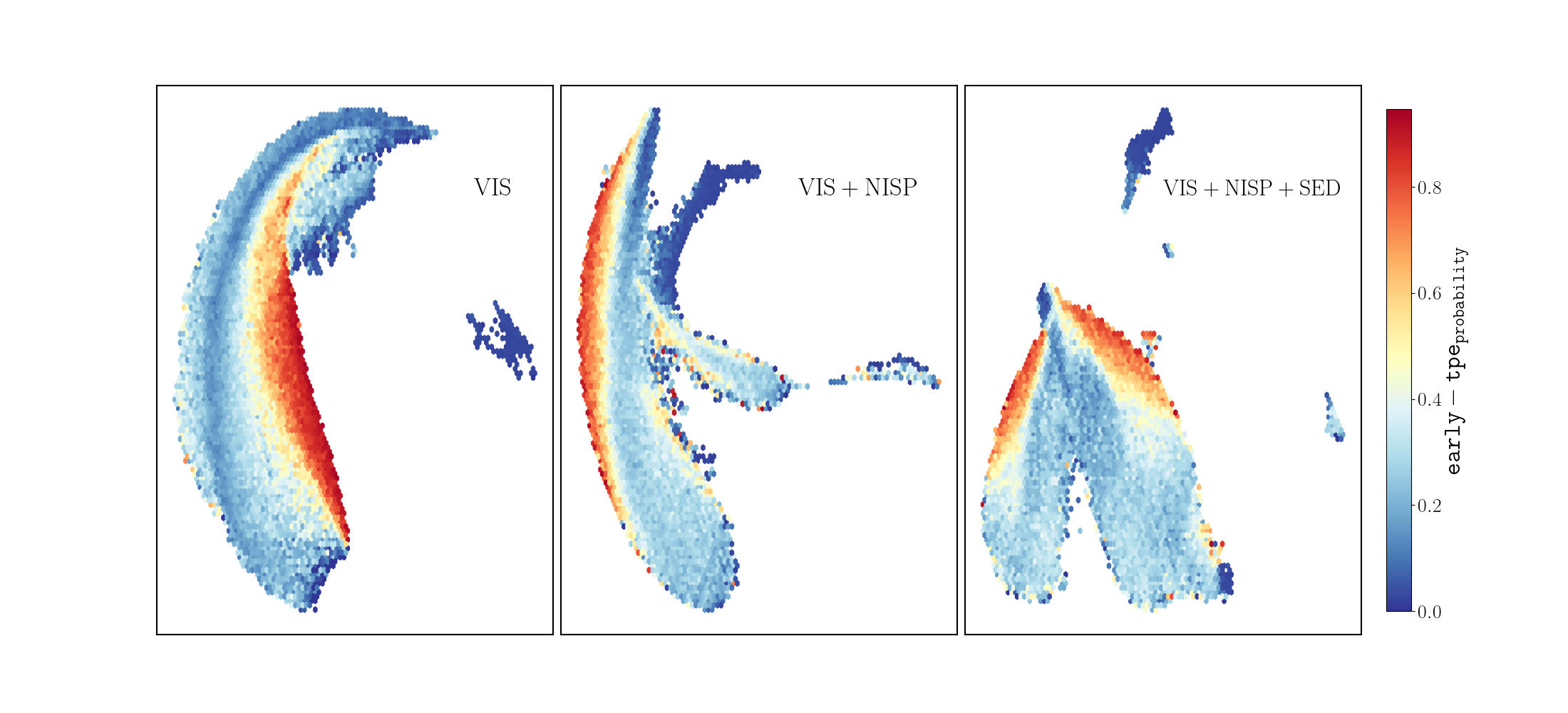} 
\includegraphics[width=0.95\textwidth]{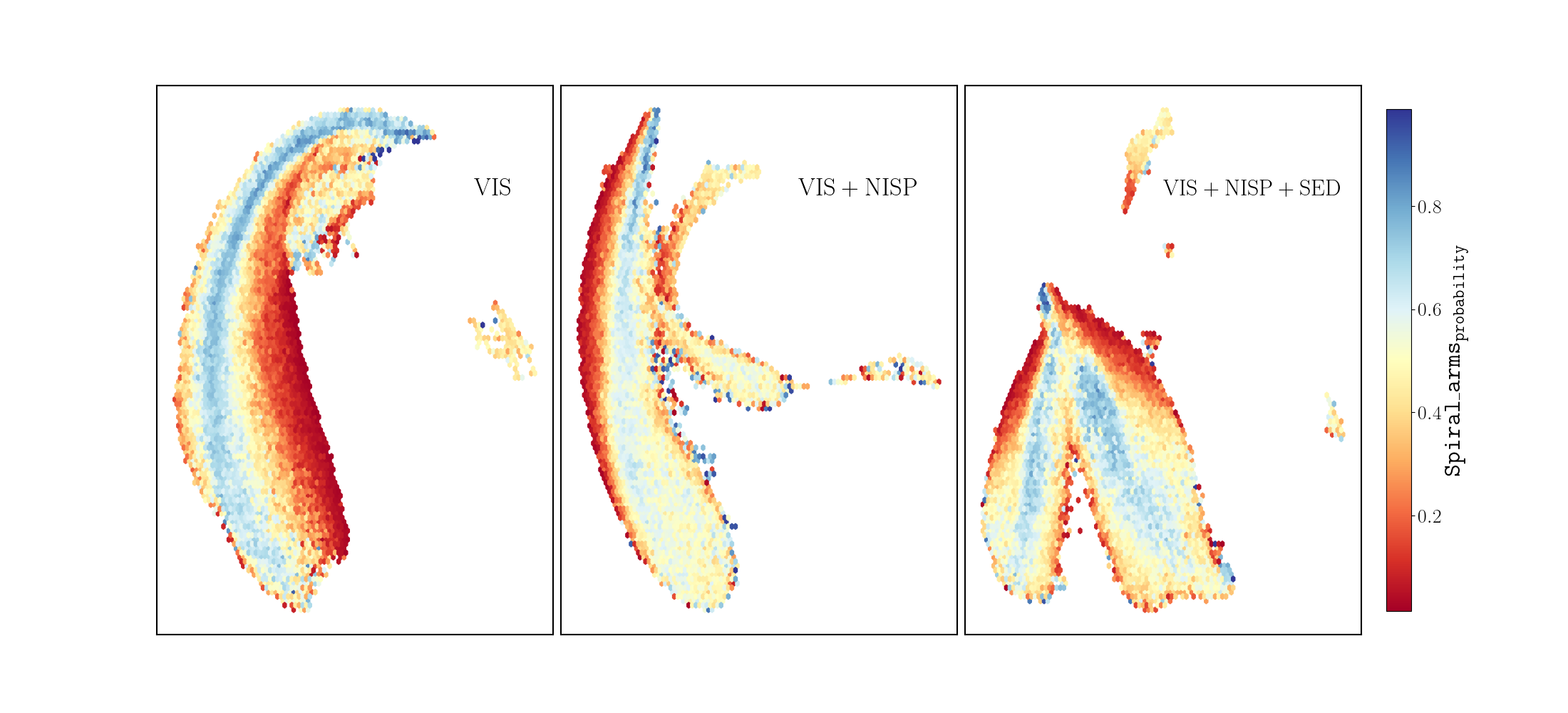} 
\caption{UMAP visualisations of the self-supervised embeddings from {\tt AstroPT} trained on VIS (left), VIS+NISP (centre), and VIS+NISP+SED (right) coloured by morphological parameters derived from the GZ catalogue~\citep{Q1-SP047}. Representations show clear clustering of smooth, non-spiral, and non-smooth, spiral galaxies.}
\label{fig:UMAP_visualisation_morphology_spiral} 
\end{figure*}

\begin{figure*}[ht] 
\centering 
\includegraphics[width=0.95\textwidth]{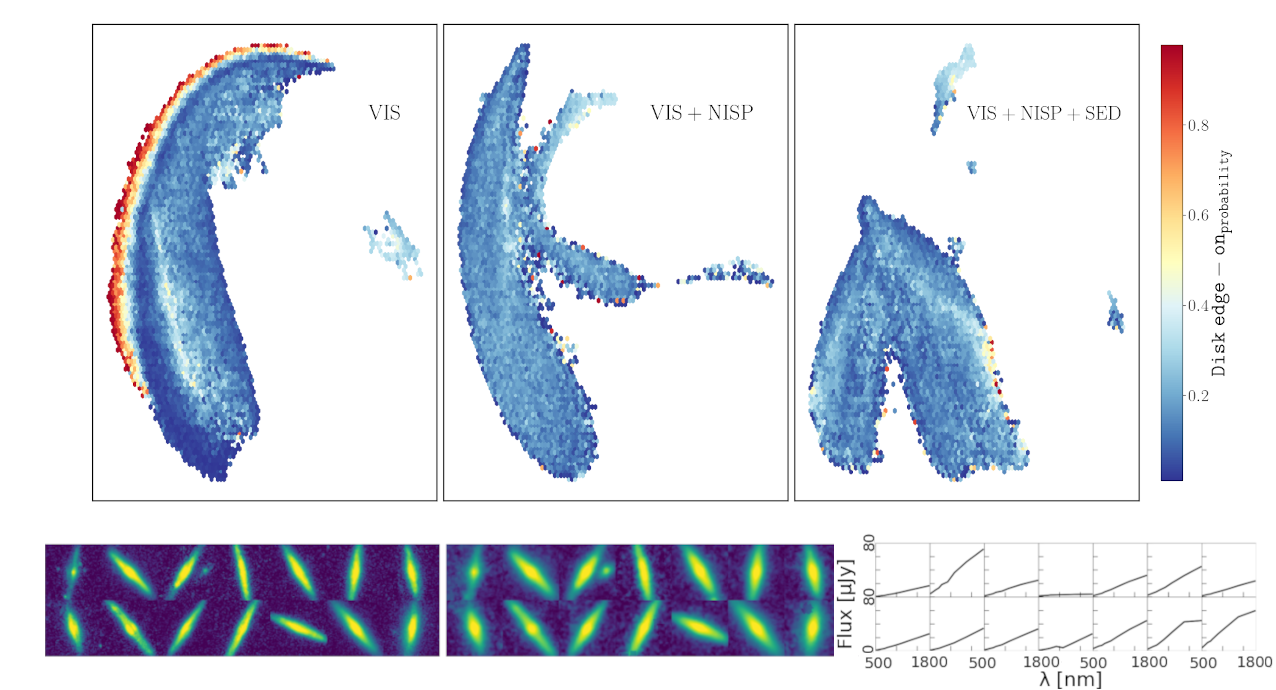}
\caption{UMAP visualisations of the self-supervised embeddings from {\tt AstroPt} coloured by morphological parameters derived from the GZ catalogue~\citep{Q1-SP047}. $Left$: Representations from VIS images alone, showing a distinct region for edge-on galaxies. $Middle$: Representations incorporating both VIS and NISP data, where the separation of edge-on galaxies becomes less distinct. $Right$: Representations incorporating VIS, NISP, and SED data. $Bottom$: Example cutouts and SEDs of the edge-on galaxies.} 
\label{fig:UMAP_visualisation_morphology_edge_on} 
\end{figure*}

We also examined the correlation of the self-supervised embeddings with morphological parameters derived from the GZ catalogue~\citep{Q1-SP047}. 
 Early-type galaxies typically exhibit smooth profiles, while late-type galaxies tend to show distinct features such as spiral arms or a bulging core, contributing to a more complex, less smooth morphology. Although not identical, the GZ classification between smooth and featured galaxies provided with the \Euclid Q1 release~\citep{Q1-SP047} closely approximates an early-type/late-type separation.
The UMAP visualisations reveal distinct clustering patterns that correspond to an early-type/late-type separation and the presence of spiral arms as shown in Fig.~\ref{fig:UMAP_visualisation_morphology_spiral}. 
 Across all embeddings -- whether {\tt AstroPT} is trained on VIS, VIS+NISP, or VIS+NISP+SED -- specific regions of the latent space are consistently associated with early-type galaxies, which also exhibit a low probability of hosting spiral arms. Conversely, other areas correspond to late-type galaxies with pronounced spiral structures. While the gradients within VIS+NISP+SED embeddings are clear, the separations between morphological types become harder to identify. This suggests that the embeddings prioritise spectral diversity over morphological clarity when SED data are included, highlighting a trade-off between the abstraction of spectral information and the preservation of morphological distinctions. This highlights the ability of automatic separation of elliptical and star-forming galaxy populations. 

Interestingly, edge-on galaxies form a clearly identifiable region in the embeddings trained on VIS images alone, highlighting the capability of {\tt AstroPT} to capture these specific morphological features (see the left panel in Fig.~\ref{fig:UMAP_visualisation_morphology_edge_on}). 
This morphological distinctiveness aligns with their SED properties: a rising flux toward longer wavelengths, indicative of more dust, supports expectations for edge-on star-forming disk galaxies. 
However, as additional data from the NISP bands, and SEDs are incorporated, this distinct separation becomes less pronounced in the VIS+NISP and VIS+NISP+SED embeddings, as seen in the middle, and right panels of Fig.~\ref{fig:UMAP_visualisation_morphology_edge_on}. While the inclusion of NISP and SED data enriches the latent space by providing additional spectral information, it does not necessarily enhance the clarity of morphological separations. Instead, the embeddings appear to encode a broader range of physical and morphological characteristics, becoming more abstract as additional data are integrated.
The reduction in morphological separation, particularly for edge-on galaxies, may also be attributed to the lower resolution of NISP images relative to VIS, which limits the preservation of finer structural details. 
This does not imply that {\tt AstroPT}'s performance degrades with lower-resolution data but instead highlights the trade-offs in representation fidelity when combining data sets of varying resolutions. These findings demonstrate the adaptability of {\tt AstroPT} to encode rich, multidimensional information, albeit at the cost of losing some morphological specificity in certain contexts. 

Similarly, we find that physical properties are correlated with the latent space and morphological parameters. Namely, the region associated with spiral galaxies is characterised by lower stellar masses, while the space occupied by elliptical galaxies tends to have higher masses. There is though no trend with the \Euclid spectroscopic redshift and only a subtle trend of higher redshift for more massive galaxies using photo-$z$s. More details are given in Appendix~\ref{app:quality_of_emb_PhysProp}.

These visualisations confirm that the self-supervised embeddings capture meaningful physical and morphological properties, with improved performance when both VIS and NISP data are utilised, except for the shape elongation (i.e., edge-on, ellipticity) which are caught by VIS embeddings but blended when using VIS+NISP imaging. The latent space's ability to distinguish between low- and high-quality data, as well as between source types, underscores the robustness of the {\tt AstroPT} framework for image analysis.

\subsection{Downstream tasks}\label{sec:downstream_tasks}

In our work, we evaluate the utility of self-supervised representations for different downstream tasks: morphology classification (see Sect.~\ref{sc:MorphologoyPredictions}),  photo-$z$ estimations (see Sect.~\ref{sc:Phtoz}), and stellar mass predictions (see Sect.~\ref{sc:physical_properties}).

\subsubsection{\label{sc:MorphologoyPredictions}Morphological properties}

\begin{figure*}[ht]
\centering
\includegraphics[width=0.95\textwidth]{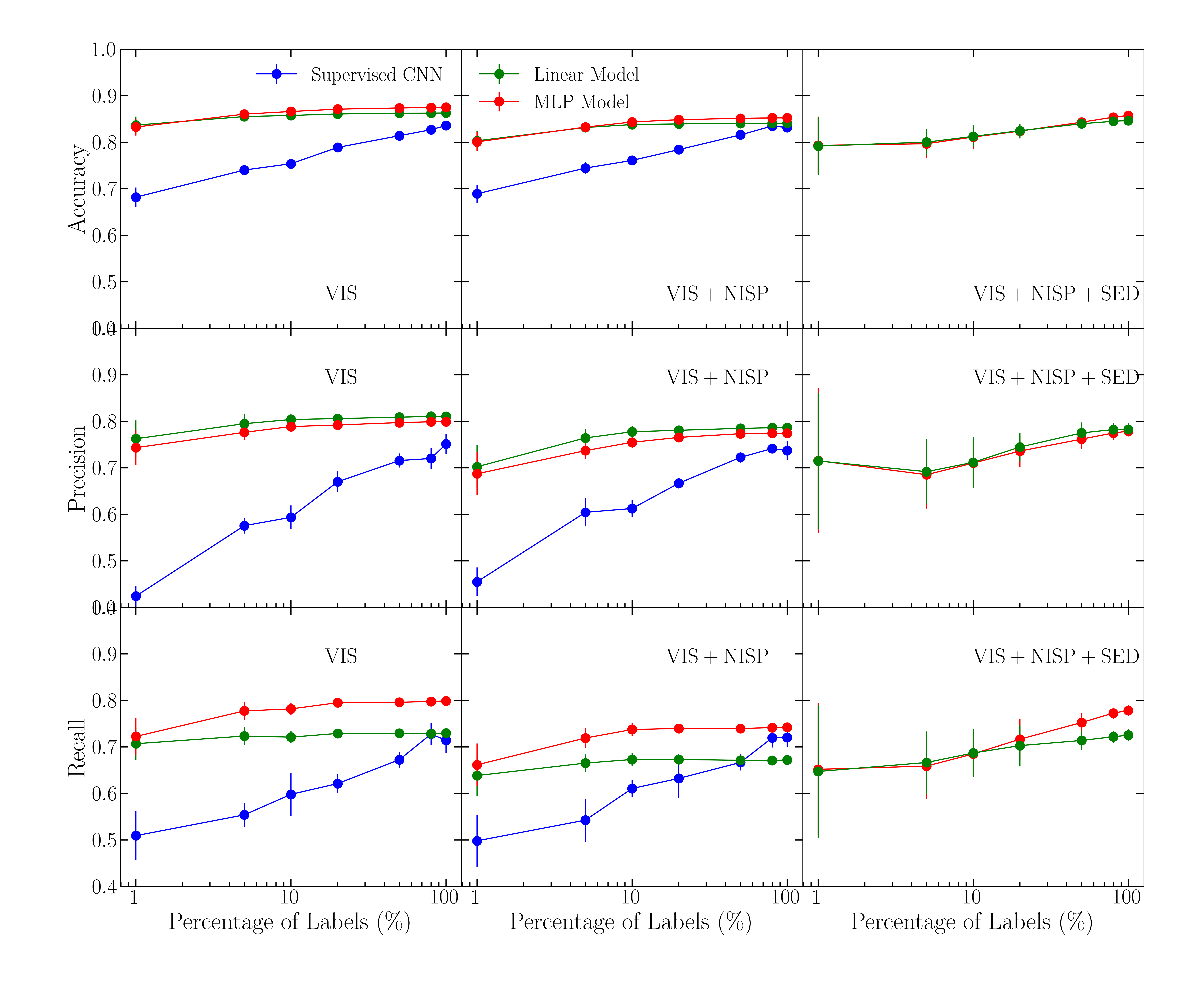} 
\caption{Comparison of the accuracy, precision, and recall as functions of the percentage of labels used for training, for early-type/late-type classification. Results are shown for the VIS (left), VIS+NISP (centre), and VIS+NISP+SED configurations. }
\label{fig:MorphologyPredTrue}
\end{figure*}

 The performance of three different methods -- a supervised CNN, a linear model, and an MLP model (see Sect.~\ref{sec:Downstreamtaks:methodology})--is evaluated for their ability to classify galaxy as early-type or late-type based on the morphological catalogue (see~\citealp{Q1-SP047}).  We define galaxies as early-type if their smooth probability exceeds 0.5 and as late-type otherwise. However, this threshold-based classification may introduce inaccuracies, as it relies on predicted continuous probabilities rather than an inherent binary distinction.
The accuracy, precision, and recall as a function of the percentage of labelled data are shown in Fig.~\ref{fig:MorphologyPredTrue}.  Table~\ref{tab:smooth_comparison} shows a statistical summary for models trained on 100\% of the labels, while Table~\ref{tab:smooth_comparison_extended} extends this comparison across all percentages of labelled data. 
As expected, the results suggest that performance varies with the percentage of the labelled data used for training, but this dependence is neither uniform nor consistent across methods. In particular, the linear and MLP models present relatively stable performance, with only minor fluctuations, suggesting that even modest amounts of labelled data are sufficient for robust classification. 
However, the supervised CNN shows a more pronounced dependence on labelled data, with a clear decline in accuracy, precision, and recall as the labelled data set size decreases.
Interestingly, incorporating NISP imaging does not enhance classification performance; in some cases, it even slightly lowers accuracy, precision, and recall, possibly due to the NISP spatial resolution (see also discussion about edge-on galaxies in Sect.~\ref{sc:embeddings}) or due to $z$-score normalisation (see Sect.~\ref{sc:Methodology}). Similarly, adding SEDs does not improve classification, likely because the SEDs contain little direct morphological information.
These findings suggest that while increasing the number of labelled examples provides some benefits, a self-supervised pre-training dramatically reduces the need for labelled samples. The self-supervised embeddings, trained on a relatively small labelled sample, already provide a robust foundation for morphological classification. Notably, the MLP model consistently achieves the highest performance across all metrics, particularly in distinguishing between early-type and late-type galaxies.

\begin{table*}[ht]
\centering
\caption{Comparison of early-type/late-type galaxy classification performance for different approaches. The model with the best (i.e., the highest) performance is bolded.}
\label{tab:smooth_comparison}
\begin{tabular}{lcccccc}
\hline\hline
Model & Data &  Accuracy & Precision & Recall & F1 score & FPR \\ 
\hline
Supervised CNN &  VIS & $0.84 \pm 0.01$ & $0.75 \pm 0.02$ & $0.71 \pm 0.03$ & $0.73 \pm 0.01$ & $0.11 \pm 0.02$ \\
Linear model & VIS & {$0.86 \pm 0.00$} & $0.81 \pm 0.01$ & $0.73 \pm 0.01$ & $0.77 \pm 0.00$ & $0.08 \pm 0.00$ \\
{\bf MLP model} & {\bf VIS} & $\mathbf{0.88 \pm 0.00}$ & $\mathbf{0.80 \pm 0.01}$ & $\mathbf{0.80 \pm 0.01}$ & $\mathbf{0.80 \pm 0.00}$ & $\mathbf{0.09 \pm 0.00}$ \\
\hline
Supervised CNN &  VIS+NISP & {$0.83 \pm 0.01$} & $0.74 \pm 0.02$ & $0.72 \pm 0.02$ & $0.73 \pm 0.00$ & $0.12 \pm 0.02$ \\
Linear model & VIS+NISP & {$0.84 \pm 0.00$} & $0.79 \pm 0.00$ & $0.67 \pm 0.01$ & $0.72 \pm 0.00$ & $0.08 \pm 0.00$ \\
MLP model & VIS+NISP & {$0.85 \pm 0.00$} & $0.77 \pm 0.00$ & $0.74 \pm 0.01$ & $0.76 \pm 0.00$ & $0.10 \pm 0.00$ \\
\midrule
Linear model & VIS+NISP+SED & {$0.85 \pm 0.01$} & $0.78 \pm 0.01$ & $0.73 \pm 0.01$ & $0.75 \pm 0.01$ & $0.10 \pm 0.01$ \\
MLP model & VIS+NISP+SED & {$0.86 \pm 0.01$} & $0.78 \pm 0.01$ & $0.78 \pm 0.01$ & $0.78 \pm 0.01$ & $0.10 \pm 0.01$ \\
\hline
\end{tabular}
\end{table*}

\subsubsection{\label{sc:Phtoz}Photometric redshifts}

\begin{figure}[ht] 
\centering 
\includegraphics[width=0.45\textwidth]{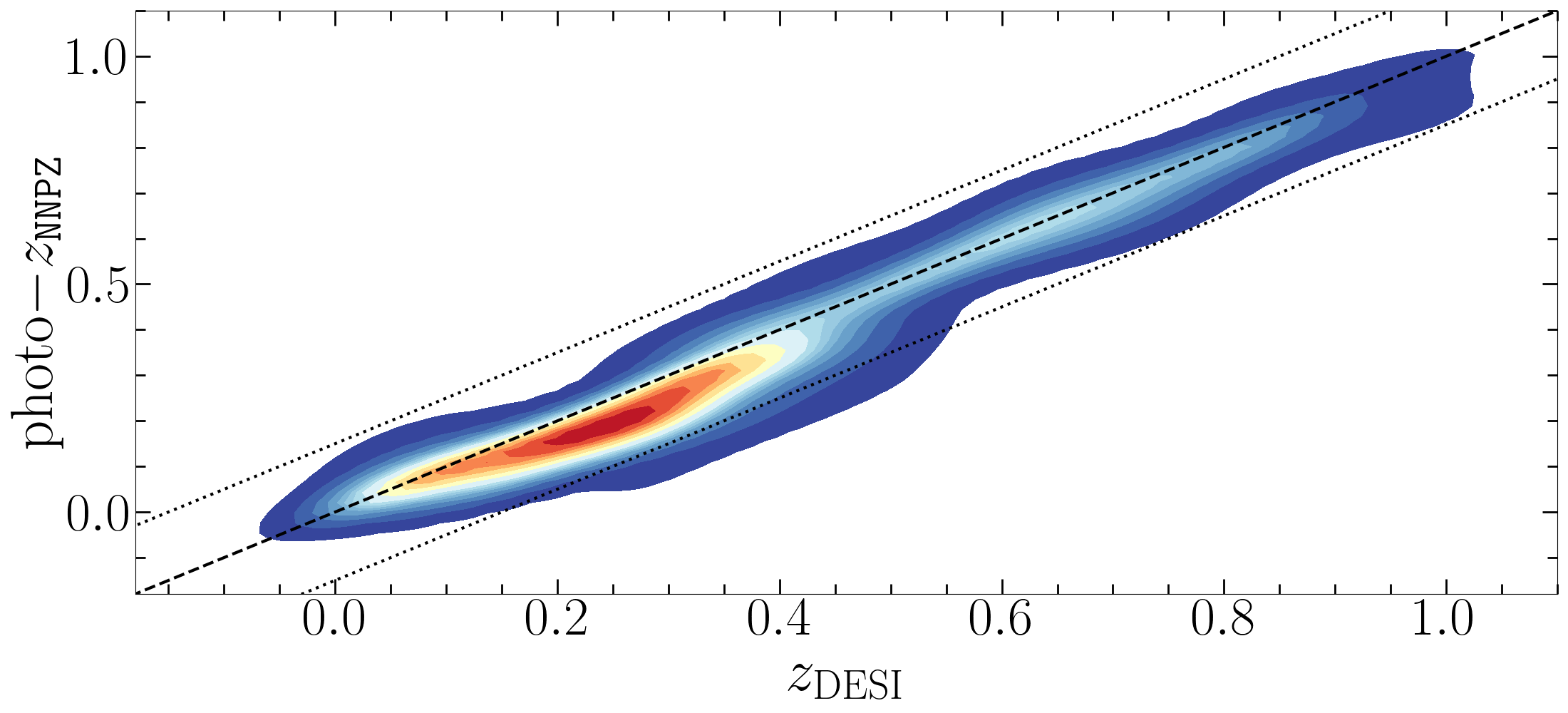} 
\includegraphics[width=0.45\textwidth]{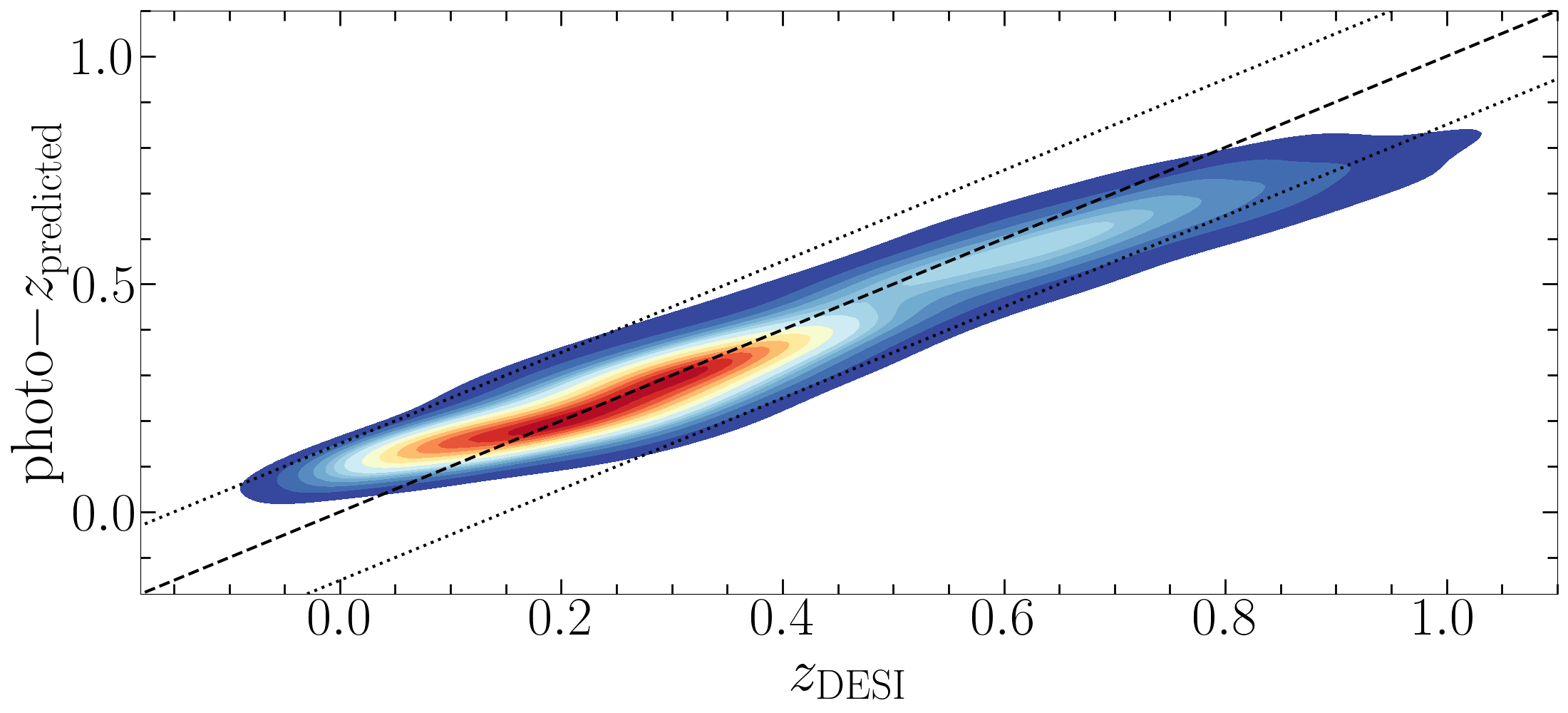} 
\caption{Comparison of photo-$z$ from \Euclid official {\tt NNPZ} pipeline (top) and predictions from the MLP model trained on {\tt AstroPT} embeddings (bottom) versus DESI  spec-$z$. A solid black line represents the 1:1 relation, while dotted lines indicate the outlier thresholds of 0.15.} 
\label{fig:Euclid_photoz} 
\end{figure}

\begin{figure*}[ht] 
\centering 
\includegraphics[width=0.95\textwidth]{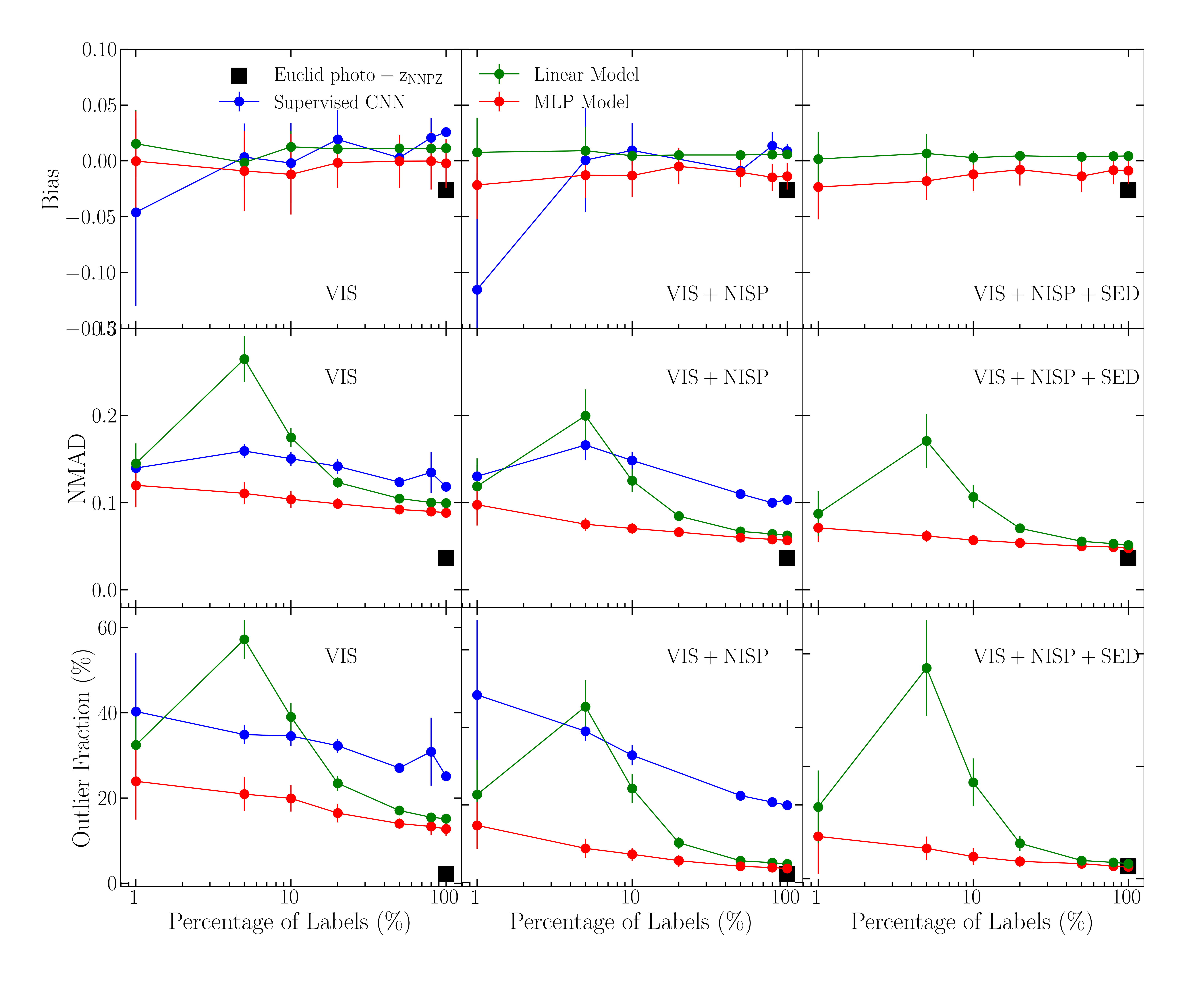} 
\caption{Comparison of the bias, NMAD, and outlier fraction as functions of the percentage of labels used for training, for true redshift (DESI spec-$z$)  versus \Euclid photo-$z$ predictions. Results are shown for the VIS (left), VIS+NISP (middle), and VIS+NISP+SED (right) configurations. } 
\label{fig:PhotZ} 
\end{figure*}

 We verify the performance of different methods in predicting photo-$z$ using the metrics described in Sect.~\ref{sec:Downstreamtaks:methodology}.
 The spec-$z$s from DESI, available for about $5\%$ of our sample, are assumed to be the true redshifts. The accuracy of \Euclid photo-$z$ derived with {\tt NNPZ} as part of the Q1 release is shown in the top panel in Fig.~\ref{fig:Euclid_photoz}.  
Figure~\ref{fig:PhotZ} shows the bias, NMAD, and outlier fraction as a function of the percentage of labels used for training for photo-$z$ obtained through these methods.  A statistical summary is provided in Table~\ref{tab:photoz_comparison} for a 100\% of labels used for training, for the full table see Table~\ref{tab:photoz_comparison_extended}. 
The results highlight the impact of adding NISP imaging data, and SED information, and the effectiveness of different training approaches.
The inclusion of NISP imaging improves photo-$z$ accuracy, reducing metrics to levels comparable to those achieved by the official \Euclid pipeline coming from the {\tt NNPZ} template fitting algorithm~\citep{Q1-TP005}. 
These improvements are consistent with previous findings, such as those by~\cite{Salvato2019}, who demonstrated the importance of including NIR bands to constrain galaxy redshifts, particularly at higher redshifts. 
The performance is further improved when the SED information is included in the training. The bottom panel in Fig.~\ref{fig:Euclid_photoz} illustrates the ability of the MLP model to predict photo-$z$s, showing that the VIS+NISP+SED configuration achieves tighter clustering of predictions around the ideal 1:1 line, while the {\tt NNPZ} photo-$z$s exhibit a more significant offset. 
 Notably, the MLP model that catches the non-linearity tends to outperform other methods across all metrics, particularly for low percentages of labelled data. This approach achieves lower NMAD, and outlier fractions demonstrating its ability to adaptively leverage the added NISP, and SED information. For instance, with 100\% labelled data, the MLP model achieves an $\rm NMAD = 0.048 \pm 0.002$ and an outlier fraction of $\rm 2.13\% \pm 0.38\%$, for training on VIS+NISP+SEDs approaching the performance of the official \Euclid photo-$z$ pipeline ($\rm NMAD = 0.037$ and an outlier fraction of $\rm 2.21\%$). 
Linear models, while simpler, deliver competitive results for higher percentages of labelled data, where they approach the performance of \Euclid pipeline, and more complex models. We note, that we observe a non-monotonic behaviour in model performance as a function of training sample size. Specifically, the NMAD and outlier fraction peak at 5\% training data, while performance improves at both smaller (1\%) and larger (20\%) sample sizes. This behaviour suggests the combination of the presence of the double descent effect~\citep[e.g.,][]{Schaeffer2023} in our regression model, and sample size balance. The double descent phenomenon occurs when test error initially decreases with increasing data size (classical regime), then spikes at an interpolation threshold, and finally decreases again as overparameterisation allows for better generalisation. In our case, with larger sample sizes ($>10\%$), the model moves into the overparameterised regime, where it can better capture meaningful patterns. On the other hand, when training on extremely limited data, the model may underfit, but given the low parameter count in linear regression this underfitting is less pronounced compared to the interpolation threshold. The fact that 1\% performs comparably to 10\% suggests that at extremely low sample sizes, the model is unable to overfit noise significantly, leading to stable results. 
In comparison, supervised CNN models exhibit modest improvements with increased labelled data, but generally underperform compared to other methods. 

Overall, an MLP model trained with SED (VIS+NISP+SED data), not only achieves the best overall metrics, with improvements in bias and outlier fraction values in comparison to photo-$z$s from {\tt NNPZ} for higher labelled data percentages, but also maintains superior results under low-data scenarios, demonstrating its robustness and adaptability for photo-$z$ estimation tasks requiring only the SED information. These results underscore the effectiveness of leveraging advanced training strategies for accurate photometric redshift estimation.

\begin{table*}[ht]
\centering
\caption{Comparison of photo-$z$ prediction performance for different approaches. The model with the best (i.e., the lowest) performance is bolded.}\label{tab:photoz_comparison}
\begin{tabular}{lrrrr}
\hline\hline
Model & Data & Bias & NMAD & Outlier fraction (\%) \\ 
\hline 
Supervised CNN & VIS & $0.026\pm 0.004$ & $0.118 \pm 0.002$ & $25.1 \pm 0.7$ \\
Linear model & VIS & $0.011\pm 0.003$ & $0.100\pm 0.002$ & $15.1\pm 0.6$ \\
MLP model & VIS & $-0.002 \pm 0.022$ & $0.089\pm 0.004$ & $12.7\pm 1.7$ \\
\hline 
Supervised CNN & VIS+NISP & $0.009 \pm 0.006$ & $0.103 \pm 0.003$ & $19.9 \pm 0.3$ \\
Linear model & VIS+NISP & $0.006\pm 0.002$ & $0.063\pm 0.001$ & $4.8\pm 0.4$ \\
MLP Model & VIS+NISP & $-0.014\pm 0.012$ & $0.057\pm 0.004$ & $3.6\pm 0.5$ \\
\hline 
Linear model & VIS+NISP+SED & $0.004\pm 0.001$ & $0.052\pm 0.001$ & $2.7\pm 0.3$ \\
{\bf MLP model} & {\bf VIS+NISP+SED} & $\mathbf{-0.009\pm 0.012}$ & $\mathbf{0.048\pm 0.002}$ & $\mathbf{2.1\pm 0.4}$ \\
\hline 
\multicolumn{2}{c}{\Euclid photo-$z$ } & --0.027 & 0.037 & 2.21 \\
\hline 
\end{tabular}
\end{table*}

\subsubsection{Stellar masses}\label{sc:physical_properties}

\begin{figure}[ht] 
\centering 
\includegraphics[width=0.45\textwidth]{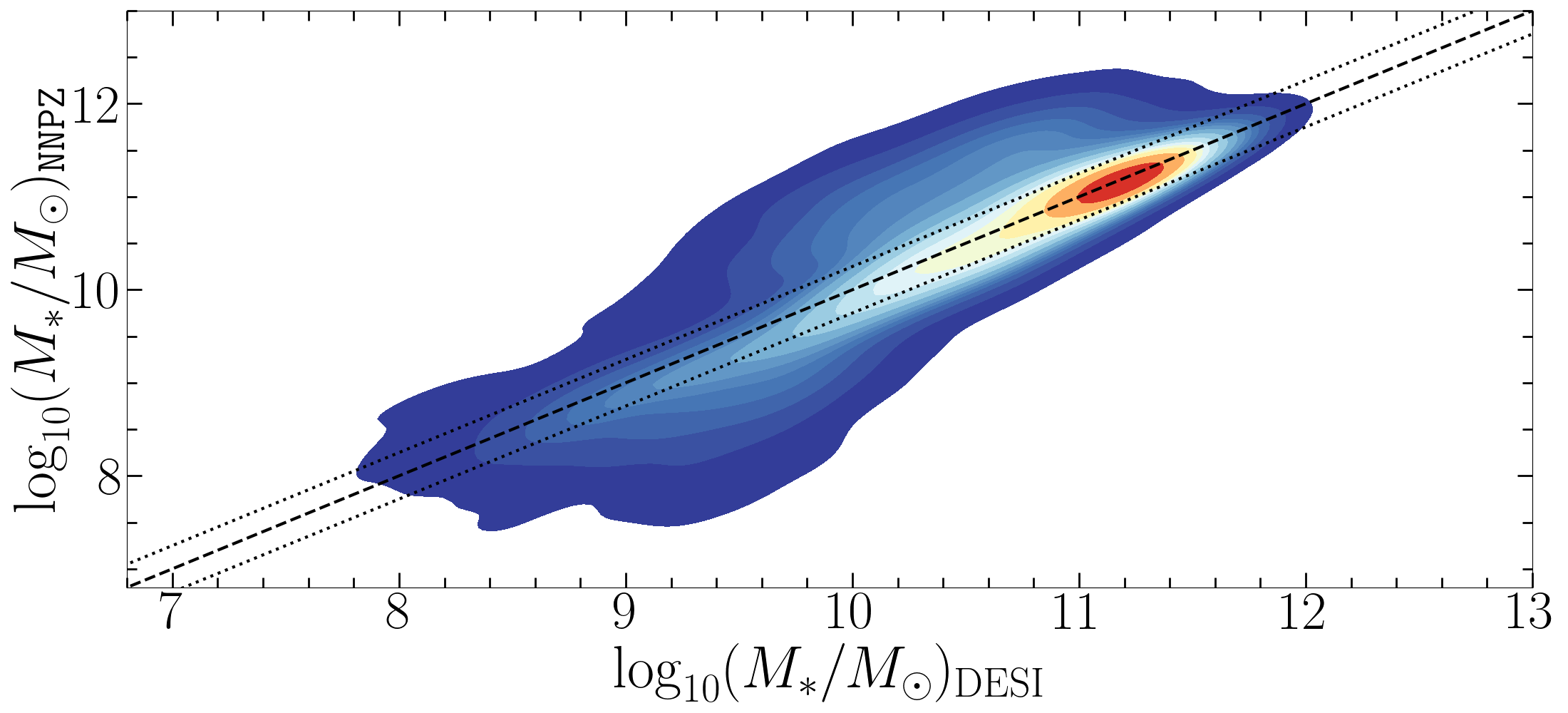} 
\includegraphics[width=0.45\textwidth]{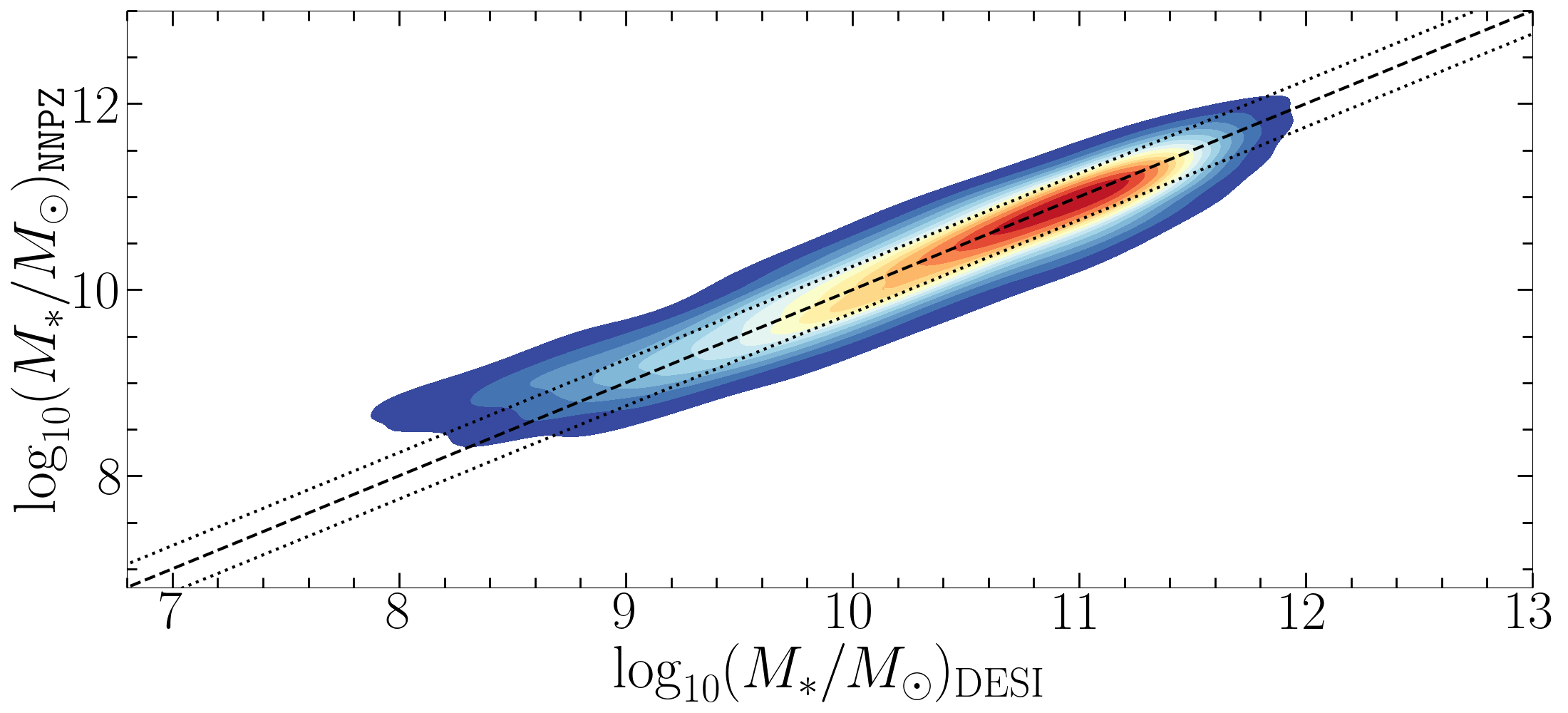} 
\caption{Comparison of stellar mass from \Euclid official {\tt NNPZ} pipeline (top) and predictions from MLP model trained on {\tt AstroPT} embeddings (bottom) versus DESI  stellar masses from~\cite{Siudek2024}. A solid black line represents the 1:1 relation, while dotted lines indicate the outlier thresholds of 0.25 dex.} 
\label{fig:Euclid_logM} 
\end{figure}

In this Section, we evaluate how {\tt AstroPT} model is able to predict the stellar masses in comparison to more traditional approaches.  As the ground truth, we assume the stellar masses of the DESI Value-Added Catalogue~\citep{Siudek2024}, which provides a robust data set derived from the SED fitting of optical and infrared photometry. First, we compare the stellar masses from \Euclid official pipeline~\citep{Q1-TP005} with the ones from DESI (see Fig.~\ref{fig:Euclid_logM}). The stellar masses are in agreement, as shown by the following 1:1 relation, but there is also a wide spread. 

\begin{figure*}[ht] 
\centering 
\includegraphics[width=0.95\textwidth]{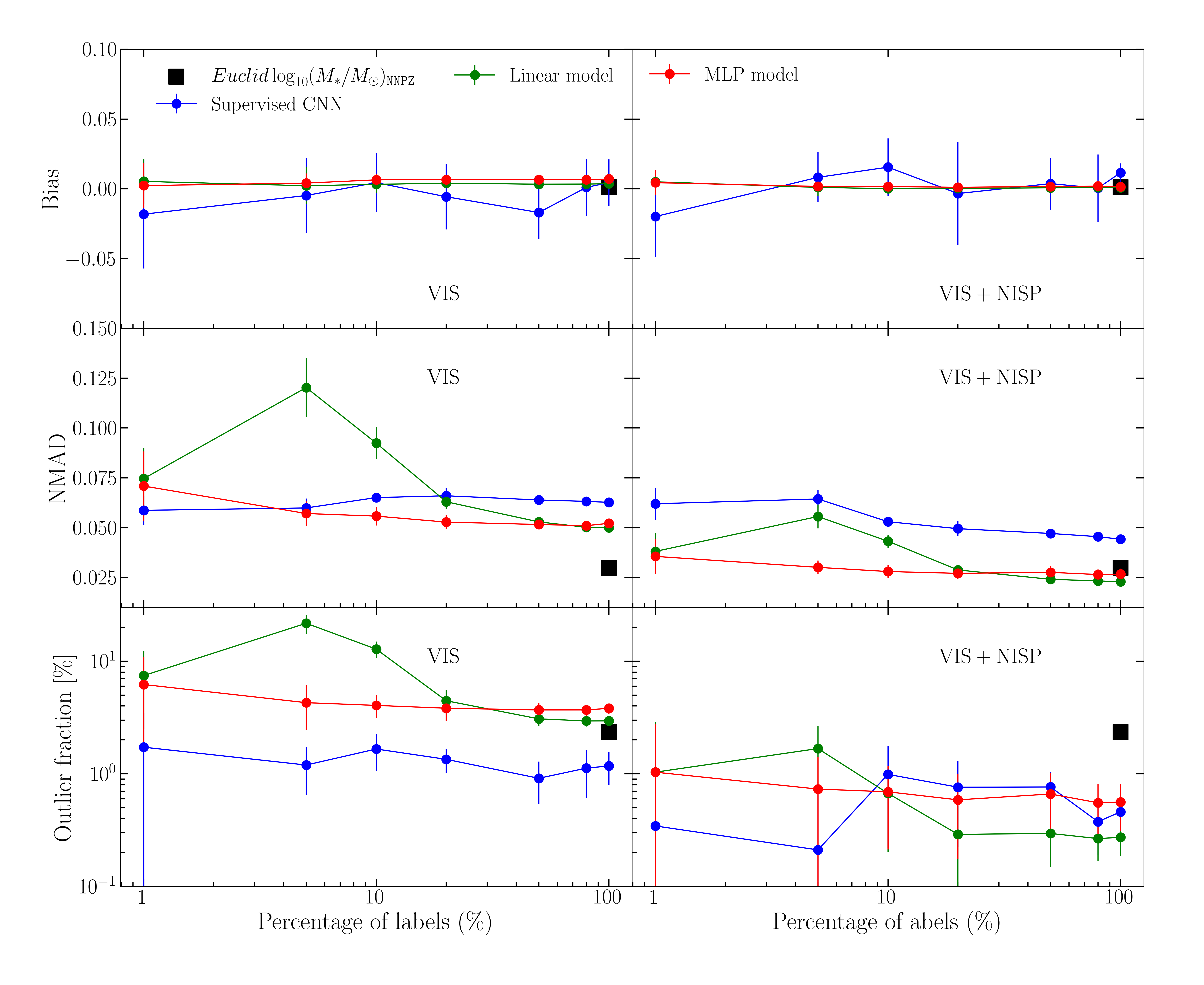} 
\caption{Comparison of the NMAD and outlier fraction as functions of the percentage of labels used for training, for DESI stellar masses from~\cite{Siudek2024}  versus \Euclid predictions. Results are shown for the VIS (left) and VIS+NISP (right) configurations. } 
\label{fig:logM_comparison} 
\end{figure*}

Similarly, as for the morphological parameters and photo-$z$s, we predict stellar mass using different methods and metrics introduced in Sect.~\ref{sec:Downstreamtaks:methodology}. 
We compare this property for both VIS and VIS+NISP models, but do not compare for our VIS+NISP+SED model as the $z$-score normalisation across the SEDs removes the additional information needed to predict this property. 
The bias, NMAD, and outlier fraction as a function of the percentage of labelled data for the stellar mass are shown in Fig.~\ref{fig:logM_comparison}. A statistical summary is given in Table~\ref{tab:logM_comparison}. 
The MLP model, and linear model consistently achieve the lowest NMAD across, but only MLP is independent on the number of labels used for training, demonstrating its robustness even with limited labelled data. The linear method, similarly as for the photo-$z$ estimates shows the non-homogenous behaviour due to sample size, and double descent effect (see Sect.~\ref{sc:Phtoz} for details). Nevertheless, for all labels used in training, both linear and MLP methods outperform estimates from \Euclid official pipeline when the {\tt AstroPT} is trained on both VIS and NISP data.  In particular, stellar mass predictions with the linear model are statistically even closer to DESI stellar masses than official {\tt NNPZ} estimates (with NMAD $0.023\pm 0.001$, and $0.027$, and outlier fraction $0.3\pm 0.1\%$, $2.34\%$ for MLP, and {\tt NNPZ}, respectively. 
 The performance degrades when solely VIS images are used for training. This highlights the added value of NISP's complementary wavelength coverage in constraining stellar mass. The distributions of the MLP stellar mass predictions is shown in the bottom panel in Fig.~\ref{fig:Euclid_logM}. 

\begin{table*}[ht]
\centering
\caption{Comparison of stellar mass prediction performance for different approaches. The model with the best (i.e., the lowest) performance is bolded.}
\label{tab:logM_comparison}
\begin{tabular}{lrrrr}
\hline\hline
Model & Data & Bias & NMAD & Outlier fraction (\%) \\ 
\hline
Supervised CNN & VIS  & $0.004 \pm 0.017$ & $0.063 \pm 0.001$ & $1.2\pm 0.4$ \\
Linear model & VIS & $0.004\pm 0.001$ & $0.050\pm 0.001$ & $3.0\pm 0.3$ \\
MLP model & VIS & $0.007\pm 0.002$ & $0.052\pm 0.002$ & $3.8\pm 0.4$ \\
\hline
Supervised CNN & VIS+NISP & $0.011 \pm 0.007$ & $0.044 \pm 0.001$ & $0.5\pm 0.1$ \\
{\bf Linear model}& {\bf VIS+NISP}  & $\mathbf{0.001\pm 0.001}$ & $\mathbf{0.023\pm 0.001}$ & $\mathbf{0.3\pm 0.1}$ \\
MLP model & VIS+NISP & $0.002\pm 0.001$ & $0.027\pm 0.003$ & $0.6\pm 0.3$ \\
\hline
\multicolumn{2}{c}{\Euclid $\Mstarsun$ } & 0.0011 & 0.0299 & 2.34 \\
\hline
\end{tabular}
\end{table*}

\subsubsection{Anomaly search}\label{sec:anomalies}

 To identify peculiar objects within the learned feature space, we apply two unsupervised anomaly detection algorithms -- isolation Forest (IF) and local outlier factor (LOF) -- to the {\tt AstroPT} embeddings. Both methods are implemented using the {\tt Scikit-learn} Python package \citep{pedregosa2011scikit}.
Before running the algorithms, we filter out spurious detections by removing sources with {\tt DET\_QUALITY\_FLAG} $> 0$. To ensure robustness against sensitivity to parameter choices, we execute each algorithm multiple times with varying hyperparameters. Specifically, IF is tested with different numbers of base estimators, ranging from 50 to 200 in increments of 50, and the maximum number of features for splitting is varied between 1 and 20 in steps of 5. Additionally, we set the contamination parameter to an estimated fraction of outliers in the data set and do not use bootstrap sampling. For LOF, we explore the effect of different neighbourhood sizes by varying the number of neighbours from 5 to 50 in steps of 5. We use the 'ball-tree'~\citep{Dolatshah2015} algorithm for nearest neighbour searches and employ the cosine distance metric, as it has been shown to better capture similarities in high-dimensional spaces. The contamination parameter is set to reflect the expected proportion of outliers.

\begin{figure}[ht] 
\centering 
\includegraphics[width=0.5\textwidth]{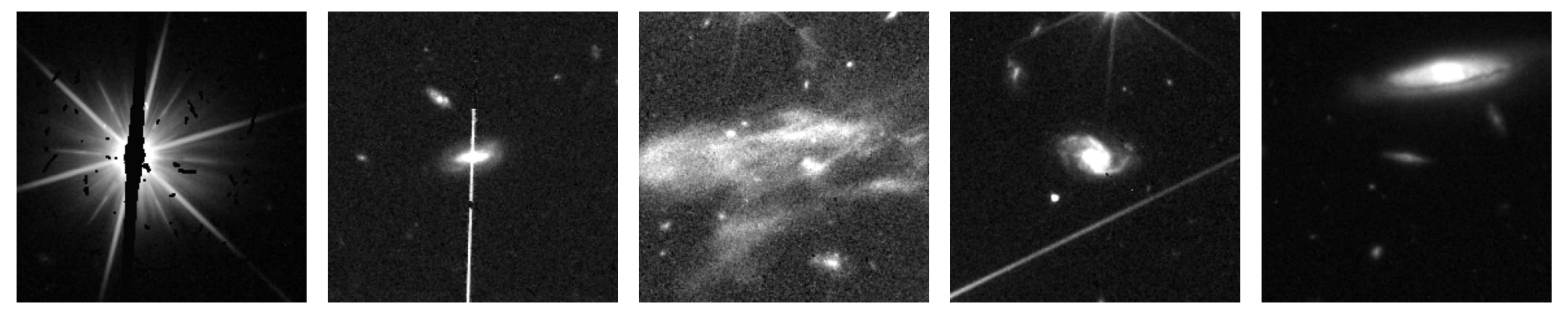} 
\includegraphics[width=0.5\textwidth]{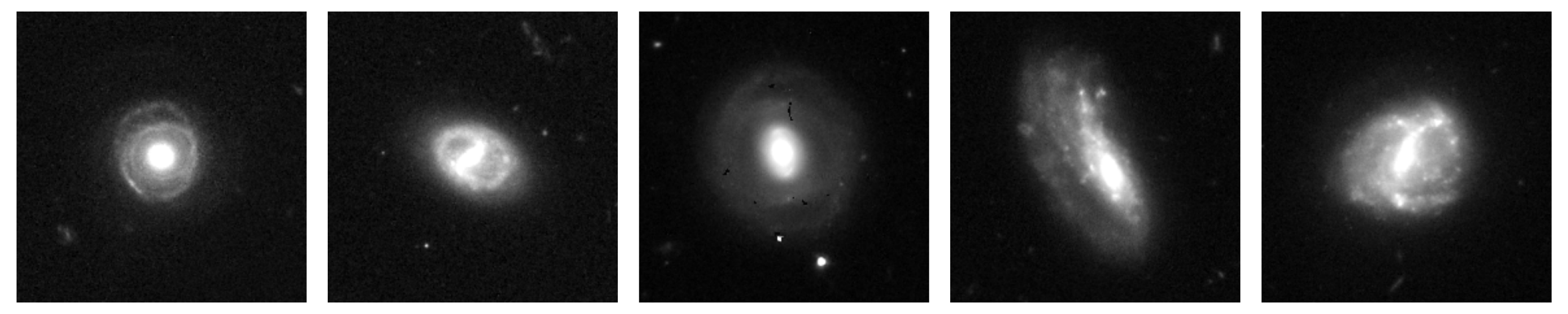} 
\caption{Example VIS cutouts of anomaly detections found with LOF (upper panel) and IF (bottom panel) on VIS embeddings.  } 
\label{fig:anomaly_vis} 
\end{figure}

\begin{figure}[ht] 
\centering 
\includegraphics[width=0.45\textwidth]{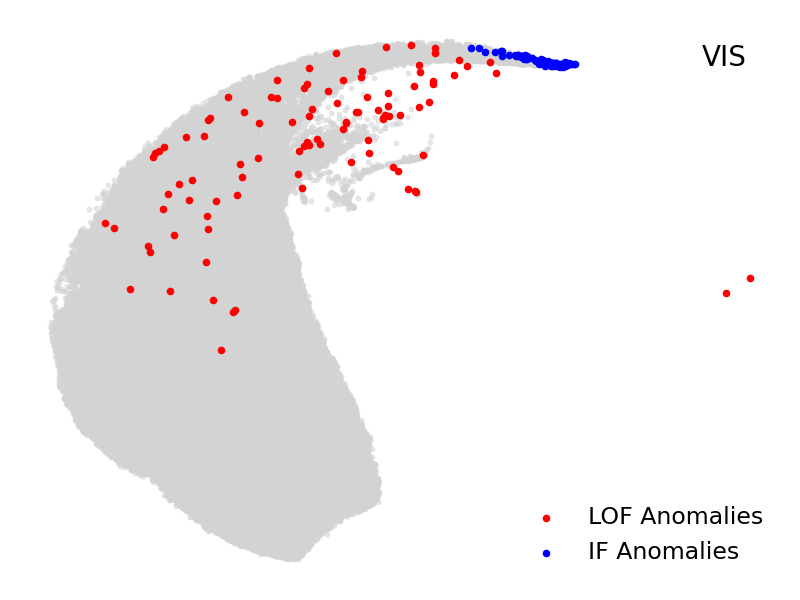} 
\caption{UMAP visualisations of the self-supervised embeddings trained on VIS images with anomalies detected with LOF, and IF algorithms.   } 
\label{fig:anomaly_vis_embeddings} 
\end{figure}

In the embeddings trained solely with VIS images, LOF predominantly detects anomalies often associated with spurious features, such as cosmic rays, bright stars in close proximity, or ghost artefacts (see upper panel in Fig.~\ref{fig:anomaly_vis}). We note that these objects were not flagged by standard quality-control metrics, indicating LOF's ability to identify subtle irregularities that might otherwise be overlooked. Additionally, LOF identifies extended objects in the outskirts of the cutout region, sometimes galaxies located far from image centres. These galaxies tend to be brighter and more extended than the central source, though such cases are relatively rare since most relevant features are concentrated near the image centre. In contrast, IF predominantly identifies large galaxies with rich morphological structures, including those with prominent rings indicative of recent merger events (see lower panel in Fig.~\ref{fig:anomaly_vis}). A  key distinction between the two methods emerges in the spatial distribution of detected anomalies within the embedding space: LOF-identified outliers are scattered throughout, whereas IF-detected anomalies cluster closely together in the latent space (see Fig.\ref{fig:anomaly_vis_embeddings}). These differences highlight the sensitivity of each method to different types of anomalies, with LOF favouring local density variations, and IF is adept at detecting sources with complex morphological structures.

When incorporating both VIS and NISP embeddings, the nature of the detected anomalies shifts. Both with IF and LOF algorithms, the VIS+NISP embedding highlights objects with multiple sources appearing within the field of view, potentially identifying interacting galaxies (see Fig.~\ref{fig:anomaly_vis_nisp} in Appendix~\ref{app:VIS_NISP_visualization}). However, this embedding is also prone to contamination by non-interacting systems, as the algorithms do not inherently account for redshift differences and instead focus on structural features within the images. Including redshift information as an additional constraint within the anomaly detection process could help disentangle interacting systems from unrelated projections of nearby and distant galaxies. Alternatively, utilising smaller image cutouts centred on each source might reduce confusion from neighbouring objects, improving both the accuracy and physical interpretability of detected anomalies. These refinements would enable more reliable identification of astrophysically meaningful outliers within the embedding space.

\begin{figure}[ht] 
\centering 
\includegraphics[width=0.5\textwidth]{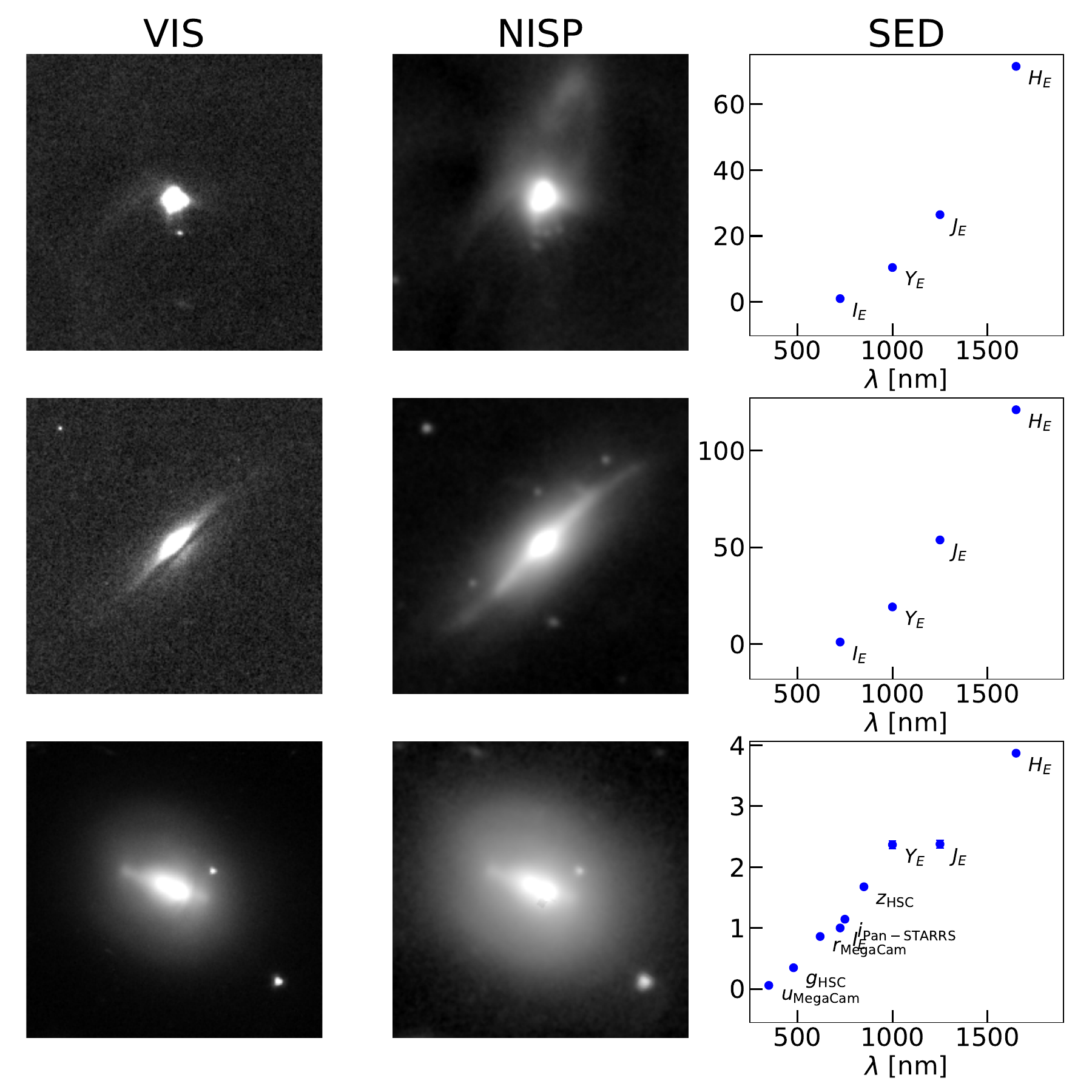} 
\caption{Anomalies detected with the IF algorithm in the VIS+NISP+SED embedding. For each galaxy we show VIS and NSIP images and the SED. Fluxes are normalised to the \Euclid VIS value.} 
\label{fig:anomaly_vis_nisp_sed} 
\end{figure}

When SED information is included, the nature of the detected anomalies changes once again. This exemplifies how complementary information can alter the structure of the embedding space and, consequently, the anomaly detection results. As more information is incorporated, anomalies become potentially more interesting. When using the IF algorithm, we detect objects with significant dust presence, predominantly in the \HE band (see Fig.~\ref{fig:anomaly_vis_nisp_sed}). The inclusion of SED data provides a measure of the relative importance of emission across different wavelengths, encoding valuable physical information. Interestingly, although the LOF algorithm still tends to detect anomalies driven by spurious features, it is nonetheless able to identify the two most prominent anomalies found with IF.

\subsubsection{Similarity search}\label{sec:similarity_search}

\begin{figure}[ht] 
\centering 
\includegraphics[width=0.5\textwidth]{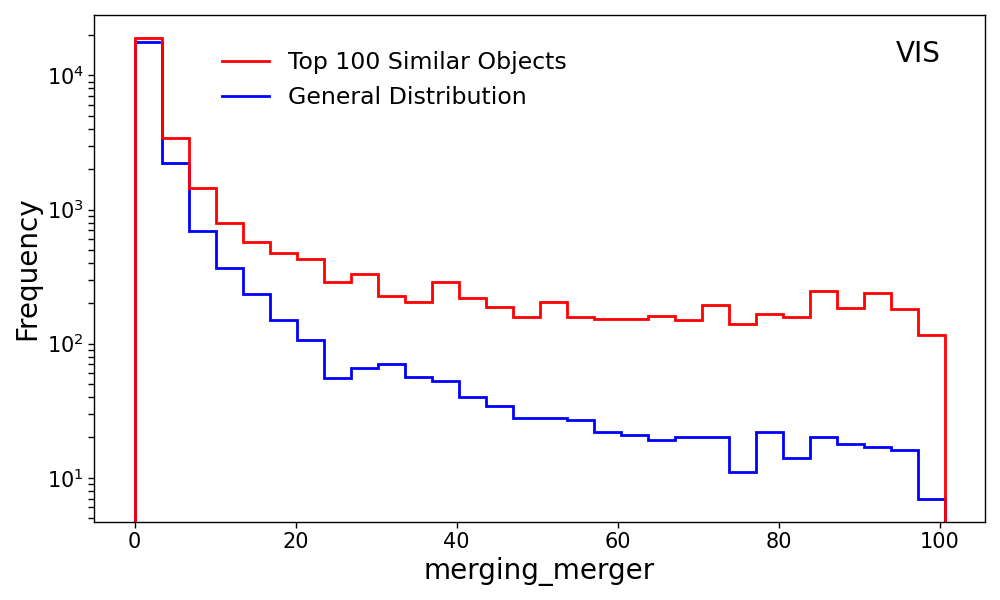} 
\caption{Distribution of $\tt merging\_merger$ parameter for similar objects computed with VIS embedding and general galaxy distribution.} 
\label{fig:interacting_simialrity} 
\end{figure}

{\tt AstroPT} projects objects with comparable features into the same regions of the embedding space, facilitating efficient similarity searches to identify objects of interest. This approach has shown great promise in various applications, particularly in identifying interacting galaxies. By using an interacting galaxy as a query, one can probe the embedding space to assess how many of its neighbouring galaxies also exhibit signs of interaction. This method allows for the rapid identification of potential interacting galaxies, contributing to large-scale surveys and studies of galaxy dynamics.
In this work, we explore the potential of similarity search for identifying interacting galaxies and investigate how this approach can be adapted to astronomical data sets. As shown in Fig.~\ref{fig:interacting_simialrity}, a significant fraction of the neighbouring galaxies to a query galaxy -- chosen as an interacting galaxy -- hereafter referred to as 'similar objects' -- are indeed interacting. 
In particular, the distribution of the $\tt merging\_merger$ parameter, which quantifies the probability of galaxies being in a merger process  \citep{Q1-SP048}, shows a higher fraction of merger candidates among the similar objects compared to the general galaxy population. This suggests that the similarity search effectively identifies galaxies involved in interactions. 
However, we also observe that the distribution of similar objects remains elongated, extending towards non-interacting galaxies as indicated by $\tt merging\_merger$ $\approx 0$, similar to the general distribution. This implies that, while the method is generally effective, it is not without contamination. In particular, the similarity search identifies some galaxies that are not truly interacting, highlighting the need for further refinement of the approach to reduce such false positives.

The presence of non-interacting neighbours, or 'false positives', in the identified clusters can be explained by the way {\tt AstroPT} was trained and operates. During the training process, a fixed image size was chosen to ensure that the largest galaxies were fully captured within the image frame. This approach works well for galaxies at lower redshifts, where the central regions of galaxies can be clearly separated. However, for higher-redshift galaxies, the fixed image size may include multiple objects or extended outskirts of galaxies within the same frame. As a result, the algorithm may erroneously prioritise similarities in the outskirts of these galaxies, where there may be multiple objects or extended features, rather than focusing solely on the central regions where interactions are more likely to be detected. This can lead to the identification of objects as interacting when they are not, especially if the objects in question lie at different redshifts or are not truly interacting.

Despite these limitations, the approach shows great potential in identifying interacting galaxies. The contamination in the results, while non-negligible, offers insight into the nuances of the similarity search method and highlights the need for further refinement, particularly in distinguishing objects that are at different redshifts or in disentangling the central regions of interacting galaxies from their outskirts. These findings contribute to the broader understanding of similarity search applications in astrophysics and suggest areas for improvement in future models to enhance the accuracy of interaction detection.

\section{\label{sc:Summary}Summary}

In this paper, we have explored {\tt AstroPT}, an autoregressive foundation model pre-trained on approximately $300\,000$ optical and infrared Q1 imaging and spectral energy distributions for deriving galaxy properties. By learning a unified and informative representation from multi-modal data, {\tt AstroPT} provides a scalable solution for analysing large, unlabelled data sets, reducing dependence on human-labelled training samples while maintaining high performance for a variety of downstream tasks (such as photometric redshift estimation, morphology classification, anomaly detection). 

The paper's highlights are as follows.
\begin{itemize}
    \item Advancing foundation models in astronomy: {\tt AstroPT} showcases the potential of multi-modal foundation models for a scalable and efficient resolution of many crucial astrophysical tasks, such as morphology classification, galaxy characterisation, and the search for peculiar sources. Unlike traditional task-specific models, {\tt AstroPT} provides a generalisable feature space applicable across multiple domains, demonstrating the feasibility of self-supervised learning for astronomy at scale. Our results provide critical insight into the synergy between optical, infrared, and SED data, guiding future strategies for galaxy characterisation and cosmological analysis with \Euclid. 
    \item Multi-modal synergies: The integration of optical, and NIR photometry highlights the advantages of multi-modal approaches. Pre-trained embeddings capture the inherent diversity of galaxy morphologies (see Figs.~\ref{fig:UMAP_visualisation_cutout} and~\ref{fig:UMAP_visualisation_morphology_spiral}), clearly isolating galaxy classes hardly distinguishable with standard methods, such as edge-on galaxies (see Fig.~\ref{fig:UMAP_visualisation_morphology_edge_on}) and outliers, such as spurious objects and stars missed by standard techniques (see Figs.~\ref{fig:UMAP_visualisation_qualilty_flag} and~\ref{fig:UMAP_visualisation_source_type}). 
    \item Unified representations for multiple tasks: Self-supervised embeddings provide a consistent representation for various downstream tasks, including photo-$z$ and stellar mass estimates, anomalies, and similarity searches.  This cross-task applicability underscores the flexibility of {\tt AstroPT}'s learned representations, which remain effective across different astrophysical domains. By leveraging the task-agnostic nature of self-supervised embeddings, we reduce dependence on large labelled data sets while maintaining high performance. This scalability is crucial for addressing the challenges of analysing large multi-modal data sets in future surveys.
    \item Morphological classification: Using solely optical images, our models achieve robust performance in predicting galaxy types. Surprisingly, incorporating NIR data and SEDs does not improve classification results, indicating that the additional spectral coverage does not introduce significant morphological information. The MLP and linear models demonstrate consistent performance across different data set sizes, maintaining higher robustness compared to traditional supervised approaches, even with a limited number of labels (see Fig.~\ref{fig:MorphologyPredTrue} and Table~\ref{tab:smooth_comparison}).  This suggests that deep-learning-based embeddings effectively capture morphological characteristics using only optical imaging, potentially simplifying future survey strategies. 
    \item Photo-$z$ estimations: By integrating NIR, optical images, and SEDs, our models achieve competitive accuracy in photo-$z$ estimates compared to the standard \Euclid pipeline relying on traditional methods.  Notably, the MLP model shows reduced bias and a lower fraction of catastrophic outliers, demonstrating that self-supervised embeddings can enhance photometric redshift predictions while being less reliant on large labelled training samples (see Fig.~\ref{fig:PhotZ}, and Table~\ref{tab:photoz_comparison}). This highlights the potential of deep-learning models adapted for standard tasks such as photo-$z$ estimations, and the importance of multi-band photometry for capturing diverse galaxy properties and achieving higher precision. 
    \item Stellar mass estimations: We explore the ability of {\tt AstroPT} to reconstruct fundamental galaxy properties, such as stellar mass, showing that MLP-based models achieve high accuracy even in low-label regimes. The linear model performs comparably well, showcasing the flexibility of the learned representations (see Fig.~\ref{fig:logM_comparison} and Table~\ref{tab:logM_comparison}). This result highlights the robustness of self-supervised learning in deriving astrophysical parameters, even when traditional template-fitting approaches require extensive labelled training sets. 
    \item Outlier and similarity search: {\tt AstroPT}’s anomaly and similarity search capabilities allow the identification of rare astrophysical objects, including galaxies with rings, interacting systems, and other unusual morphologies. Including SEDs in training, we can also identify very dusty sources. This demonstrates the model’s utility in detecting previously overlooked populations in large-scale galaxy surveys, paving the way for new discoveries in \Euclid and future astronomical data sets.
    \item  Limitations, and future perspectives: While {\tt AstroPT} demonstrates strong performance across multiple tasks, some limitations remain. The reliance on pre-training with Q1 data means that domain adaptation might be required for different survey depths or observational conditions. Additionally, the current classification thresholds, and sample selection, could introduce selection biases, necessitating further refinement. Future work should explore strategies to mitigate these biases and improve generalisation to different data sets.

\end{itemize}

This work underscores the transformative potential of combining multi-modal data sets, self-supervised learning, and foundation models for tackling large-scale challenges in galaxy analysis.  

%
%

\begin{acknowledgements}
\AckEC  
\AckQone

Based on data from UNIONS, a scientific collaboration using
three Hawaii-based telescopes: CFHT, Pan-STARRS, and Subaru
\url{www.skysurvey.cc}\,.

Based on data from the Dark Energy Camera (DECam) on the Blanco 4-m Telescope
at CTIO in Chile \url{https://www.darkenergysurvey.org}\,.


DESI construction and operations is managed by the Lawrence Berkeley National Laboratory. This research is supported by the U.S. Department of Energy, Office of Science, Office of High-Energy Physics, under Contract No. DE–AC02–05CH11231, and by the National Energy Research Scientific Computing Center, a DOE Office of Science User Facility under the same contract. Additional support for DESI is provided by the U.S. National Science Foundation, Division of Astronomical Sciences under Contract No. AST-0950945 to the NSF’s National Optical-Infrared Astronomy Research Laboratory; the Science and Technology Facilities Council of the United Kingdom; the Gordon and Betty Moore Foundation; the Heising-Simons Foundation; the French Alternative Energies and Atomic Energy Commission (CEA); the National Council of Science and Technology of Mexico (CONACYT); the Ministry of Science and Innovation of Spain, and by the DESI Member Institutions. The DESI collaboration is honored to be permitted to conduct astronomical research on Iolkam Du’ag (Kitt Peak), a mountain with particular significance to the Tohono O’odham Nation.

MHC and MS acknowledge support from the State Research
Agency (AEIMCINN) of the Spanish Ministry of Science and Innovation under the grants “Galaxy Evolution with Artificial Intelligence" with reference
PGC2018-100852-A-I00 and "BASALT" with reference PID2021-126838NBI00. MS acknowledges also support by the Polish National Agency for Academic Exchange (Bekker grant BPN/BEK/2021/1/00298/DEC/1). HDS acknowledges financial support by RyC2022-030469-I grant, funded by MCIN/AEI/10.13039/501100011033  and FSE+.
\end{acknowledgements}

%
%

\bibliography{Euclid_AstroPT, Euclid, Q1}

%

\begin{appendix}
  \onecolumn 

\section{Hyperparameters, compute, and carbon emissions}\label{app:hyperparameters}

\begin{table*}[h!t]
\centering
\caption{Hyperparameters used in pre-training {\tt AstroPT-90M}.}
\label{tab_hparams}
\begin{tabular}{ll}
\toprule
\hline
Parameter & {\tt AstroPT} pre-training \\
\midrule
\multicolumn{2}{l}{Model architecture} \\
\midrule
Active parameters & 90 millions \\
Number of layers & 12 \\
Model dimension & 768 \\
Number of heads & 12 \\
\midrule
\multicolumn{2}{l}{Training configuration} \\
\midrule
Batch size & 125\,440 \\
Learning rate & 0.0006 \\
Weight decay & 0.1 \\
Warmup steps & 2000 \\
Training steps & 5000 \\
Optimiser & AdamW \\
\bottomrule
\end{tabular}
\end{table*}

\noindent The training of deep-learning models requires considerable energy, which can contribute to carbon
emissions. To counteract further emission from unnecessary retraining we follow the recommendations of \citep{ref_strubell2019} and make available our model weights and inference
codein the Hugging Face repository\footnote{\url{https://huggingface.co/collections/msiudek/}}.
While this study made use of the 100\% renewable energy-powered ITER Teide HPC cluster \citep{ref_mampaso2022} for pre-training, for the sake of transparency we estimate and show our energy
usage during the pre-training of {\tt AstroPT} here: {\tt AstroPT-90M} takes 2 hours to train on 4xA100-40G
NVIDIA GPUs, which corresponds to 2\,kWh of energy use according to the Machine Learning CO2
Impact Calculator (\url{mlco2.github.io}).

\section{{\tt AstroPT} embeddings for VIS and VIS+NISP imaging}\label{app:VIS_NISP_visualization}

\begin{figure}[ht] 
\centering 
\includegraphics[width=0.49\textwidth]{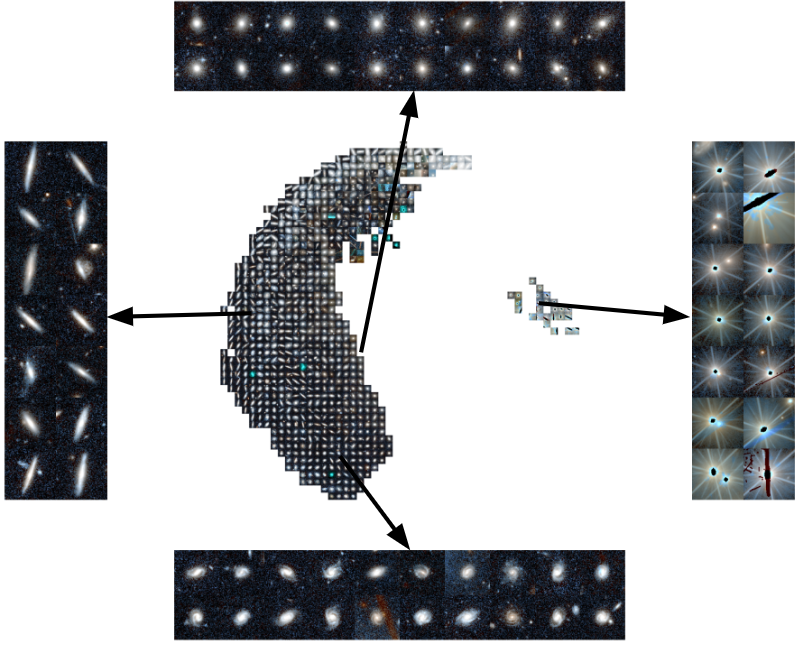} 
\includegraphics[width=0.49\textwidth]{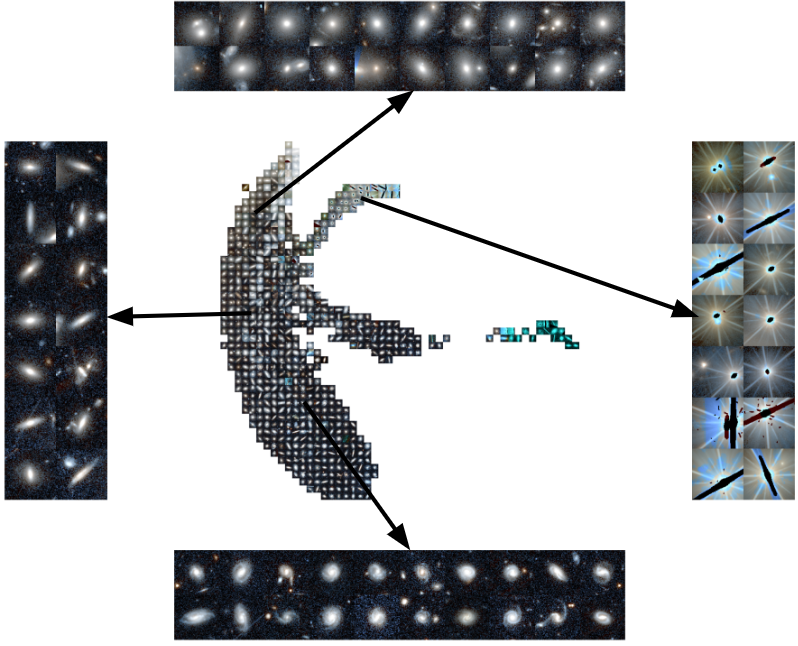} 
\caption{UMAP visualisations of the embeddings from {\tt AstroPT} trained on VIS (left) and VIS+NISP (right) images with example cutouts,  see text for more details. } 
\label{fig:UMAP_visualisation_cutout_VIS_NISP} 
\end{figure}

Figure~\ref{fig:UMAP_visualisation_cutout_VIS_NISP} presents a visualisation of the {\tt AstroPT} embeddings trained on the VIS (left) and combined VIS and NISP data (right). This plot demonstrates how the inclusion of both optical and infrared wavelengths influences the organisation and clustering of different galaxy types and stars in the embedding space.

\begin{figure}[ht] 
\centering 
\includegraphics[width=\textwidth]{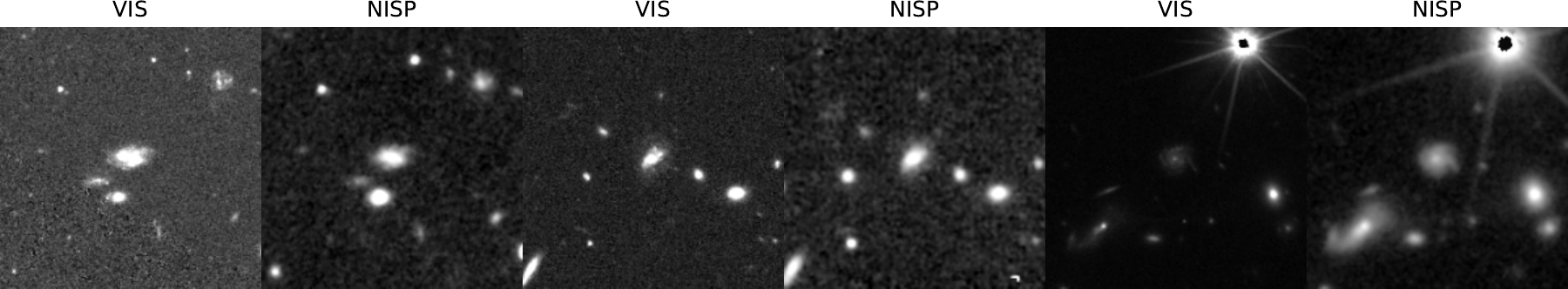} 
\includegraphics[width=\textwidth]{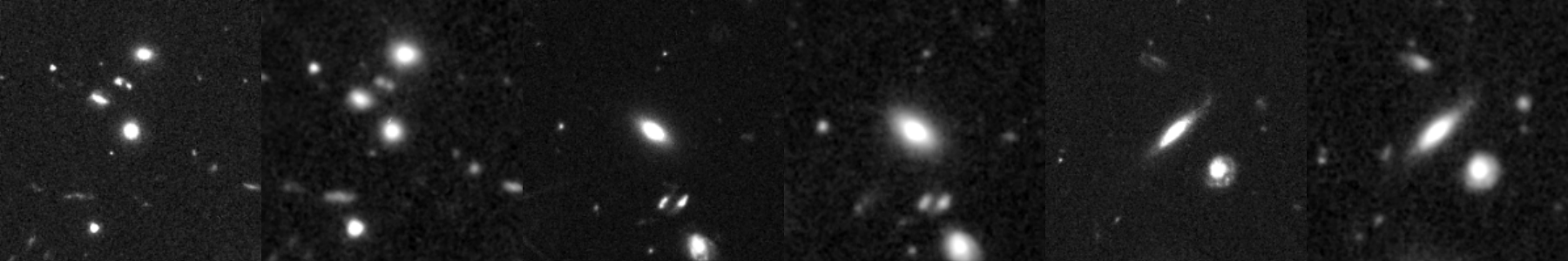} 

\caption{Example VIS and NISP cutouts of anomaly detections found on VIS+NISP embeddings with IF (upper panel) and LOF (bottom panel). } 
\label{fig:anomaly_vis_nisp} 
\end{figure}

In Fig.~\ref{fig:anomaly_vis_nisp}, we show typical anomalies found with LOF (upper panel) and IF (bottom panel) for the VIS+NISP embedding. As can be observed, anomalies consist of images that include multiple objects within the cutouts.

\section{Star-QSO-galaxy separation}\label{app:star_galaxy_qso_separation}

\begin{figure*}[ht] 
\centering 
\includegraphics[width=0.95\textwidth]{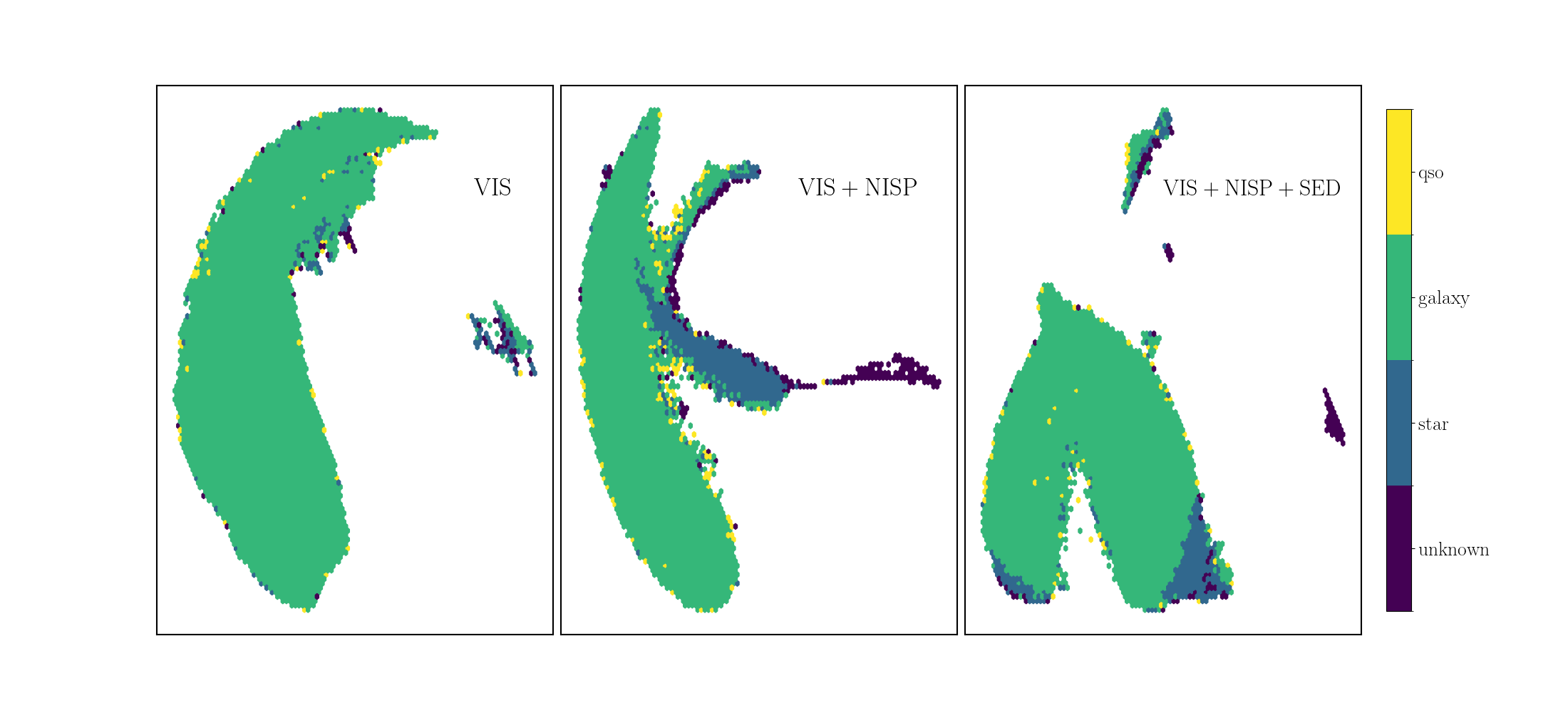} 
\caption{UMAP visualisations of the self-supervised embeddings from {\tt AstroPT} coloured by the source type (QSO, galaxy, star, or unknown). The left panel corresponds to VIS-only embeddings, the centre panel includes both VIS and NISP data, and the right panel includes VIS, NISP, and SED data.  } 
\label{fig:UMAP_visualisation_source_type} 
\end{figure*}
Figure~\ref{fig:UMAP_visualisation_source_type} shows the correlation of the embeddings with the source type (unknown, galaxy, star, QSO) derived from the spectroscopic pipeline. Clearly, the outlier regions of embeddings are occupied by stars and unknown sources, but stars are also found over all regions (see Fig.~\ref{fig:UMAP_visualisation_source_type}). Galaxies, QSOs, and stars tend to occupy a broad distribution. 
Adding NISP data sharpens the separation of unknown objects where a long tail gathers the invalid NISP data. 
 Incorporating SEDs further enhance the separation of unknown sources from the main space. 
\begin{figure*}[ht] 
\centering
\includegraphics[width=0.95\textwidth]{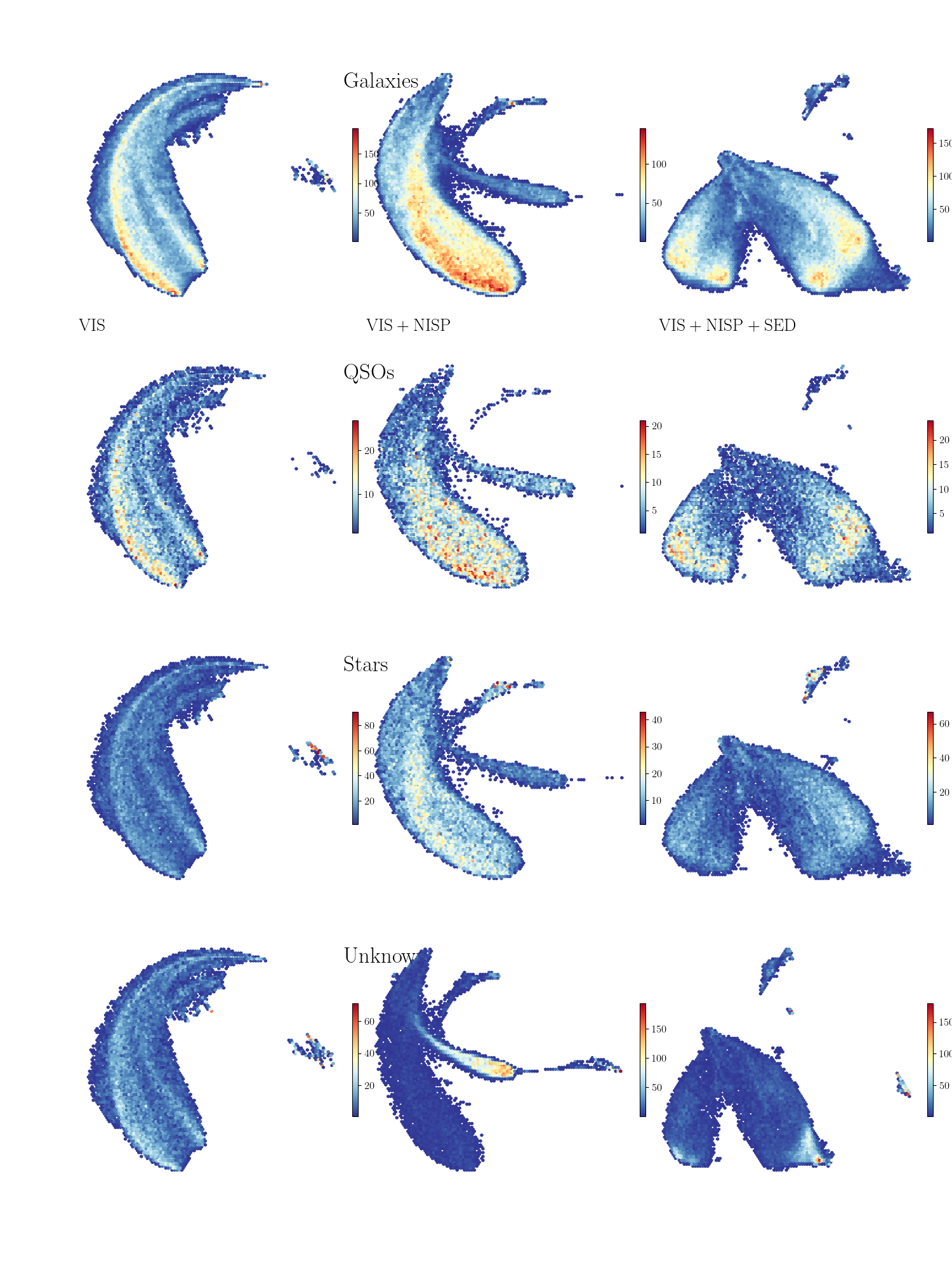}  
\caption{UMAP visualisations of the self-supervised embeddings from {\tt AstroPT}. $Top$: Coloured by the stars. $Bottom$: Colorued by QSO. Left panels correspond to VIS-only embeddings,  centre panels include both VIS, and NISP data, and the right panels include VIS, NISP, and SED data.} 
\label{fig:UMAPvisualisation_source_detail} 
\end{figure*}

\section{Quality of the embeddings: Physical properties}\label{app:quality_of_emb_PhysProp}
\begin{figure*}[ht] 
\centering 
\includegraphics[width=0.92\textwidth]{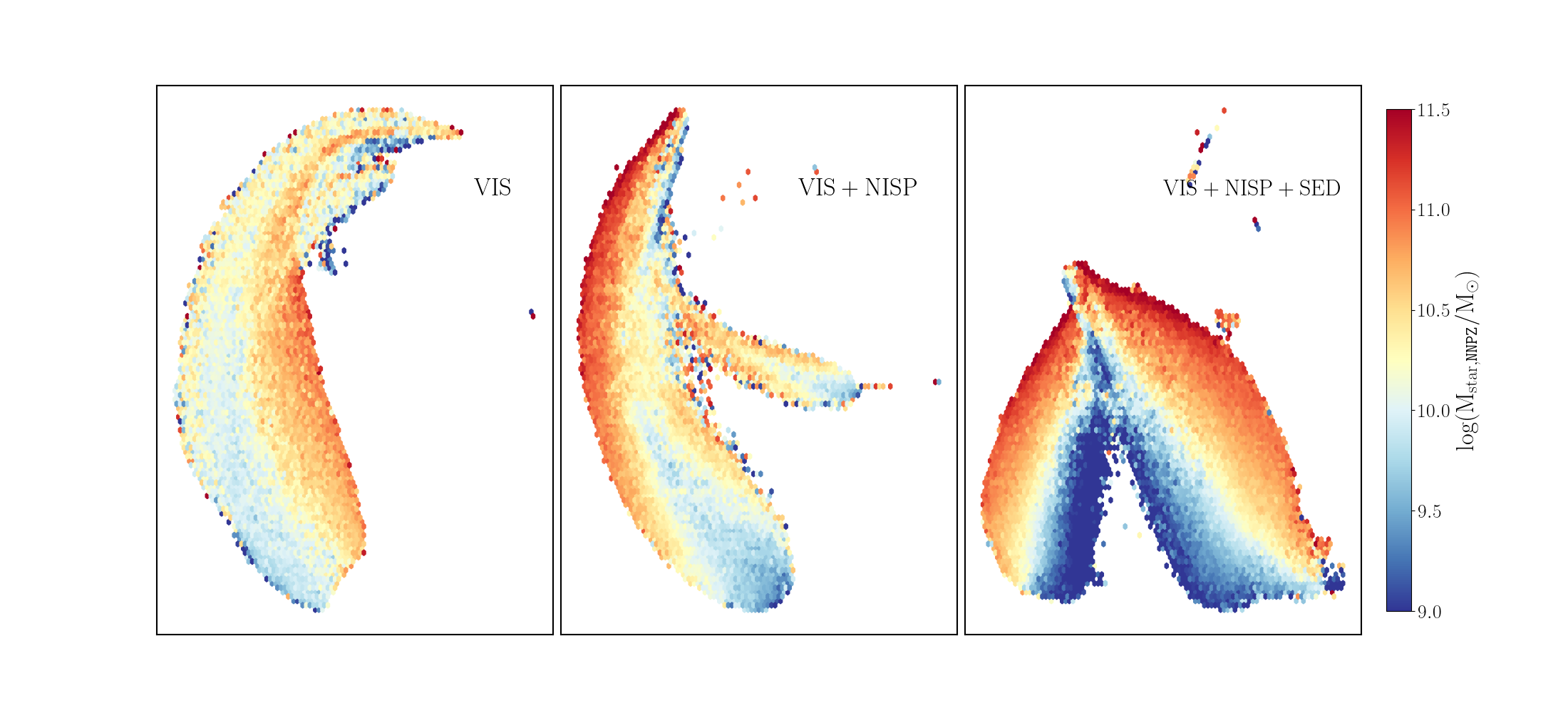} 
\includegraphics[width=0.92\textwidth]{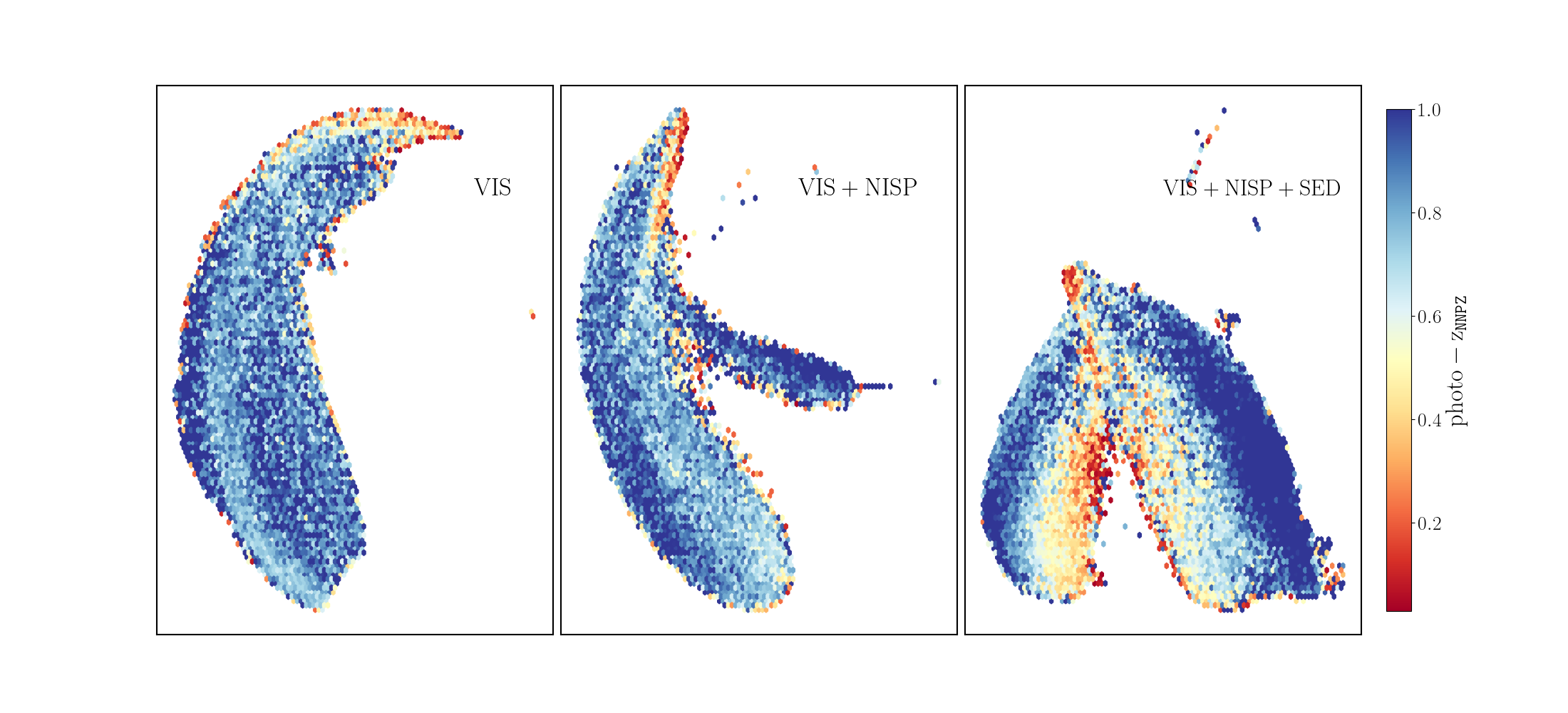} 

\caption{UMAP visualisation of self-supervised embeddings, highlighting correlations with stellar mass and photometric redshift. The left panels show the distribution of galaxies when {\tt AstroPT} is trained on VIS images, while the centre panels represent the training with VIS+NISP images, and the right panels correspond to training on VIS, NISP images, and SEDs. The top UMAP visualisations are colour-coded by stellar mass from \Euclid official pipeline (top). The bottom UMAP visualisations are colour-coded by \Euclid official photometric redshifts. } 
\label{fig:UMAP_visualisation_PhysProp_Euclid} 
\end{figure*}

\begin{figure*}[ht] 
\centering 
\includegraphics[width=0.92\textwidth]{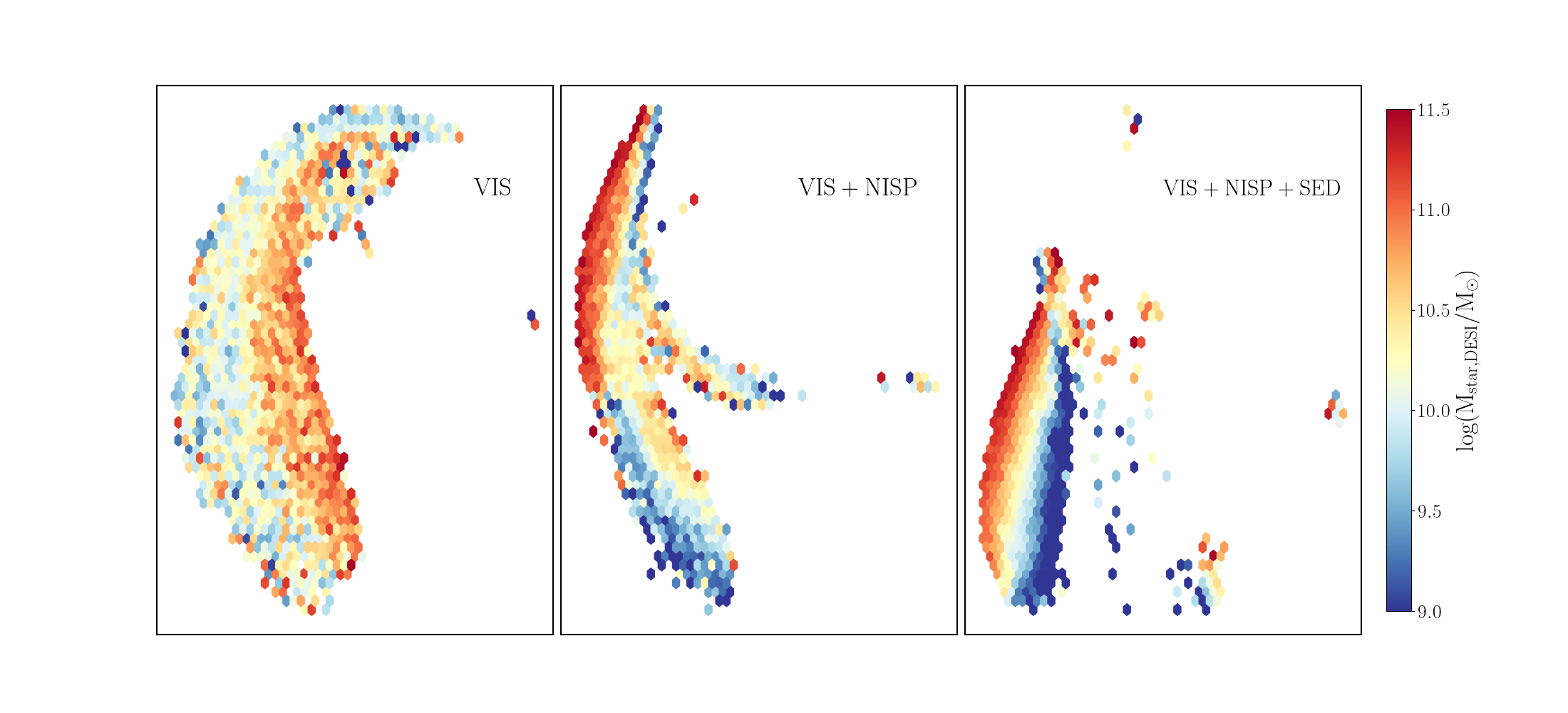} 
\includegraphics[width=0.92\textwidth]{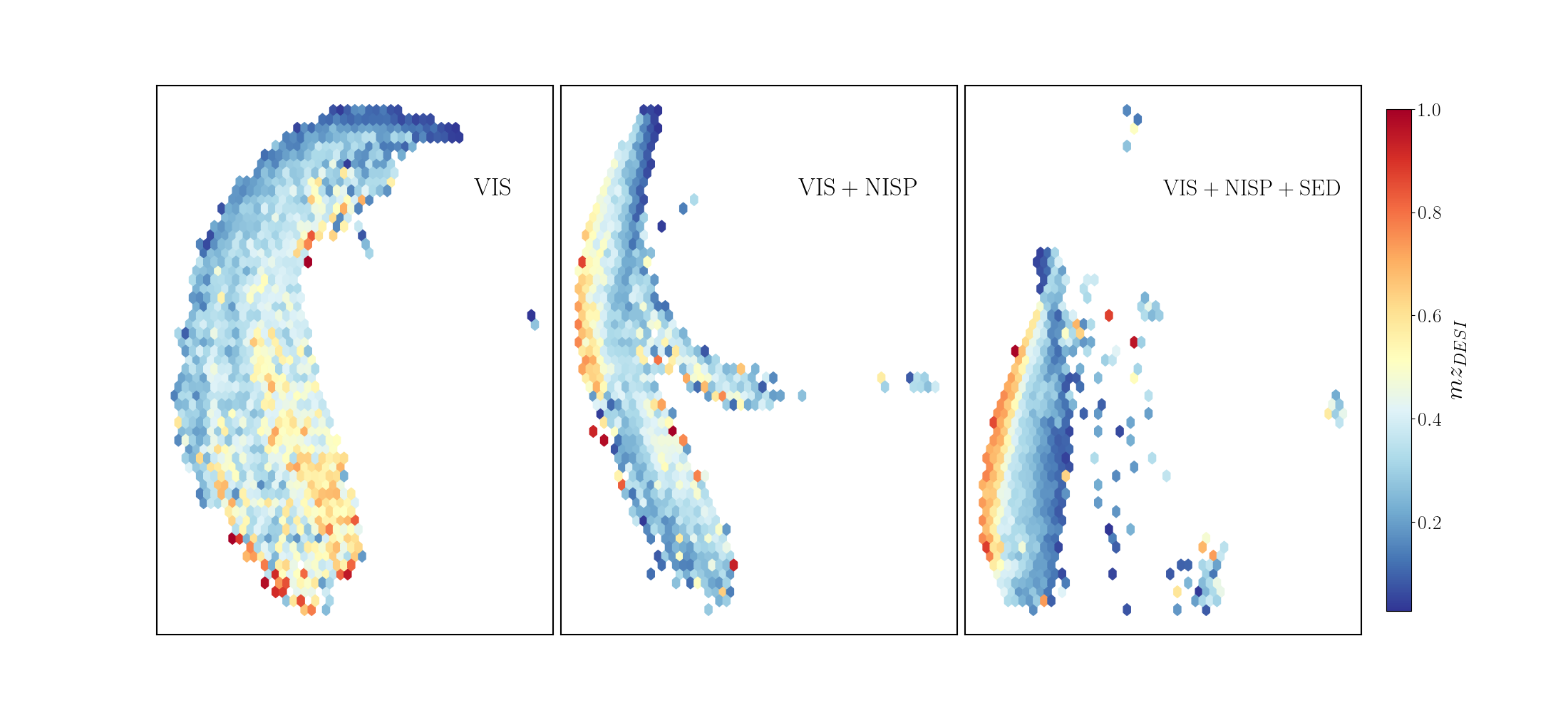} 

\caption{UMAP visualisation of self-supervised embeddings, highlighting correlations with stellar mass and photometric redshift. The left panels show the distribution of galaxies when {\tt AstroPT} is trained on VIS images, while the centre panels represent the training with VIS+NISP images, and the right panels correspond to training on VIS, NISP images, and SEDs. The left UMAP visualisations are colour-coded by stellar mass from DESI VAC~\citep{Siudek2024}. The right UMAP visualisations are colour-coded by the DESI spectroscopic redshift~\citep{DESI2024_EDR} available for about $5 \%$ of the sample. } 
\label{fig:UMAP_visualisation_PhysProp_DESI} 
\end{figure*}

Figures~\ref{fig:UMAP_visualisation_PhysProp_Euclid}, and~\ref{fig:UMAP_visualisation_PhysProp_DESI} highlight the relationship between the latent space and physical properties, such as stellar mass, and redshift. There is a clear connection between morphological trends (see Fig.~\ref{fig:UMAP_visualisation_morphology_spiral}) and stellar masses. Early-type galaxies, which are typically more massive, occupy regions of the UMAP space associated with higher stellar masses. In contrast, spiral galaxies, known for their lower stellar masses, cluster in distinct, separate regions. This differentiation is especially pronounced when embeddings are trained with VIS+NISP data, suggesting that the additional NISP features help to better constrain stellar mass distributions. This is further enhanced when the SEDs are included in the training. The \Euclid photometric redshift shows no strong correlation with the embeddings, although there is a subtle trend of higher stellar masses being linked to slightly higher redshifts, especially for VIS+NIS+SED, consistent with the evolutionary connection between galaxy mass and formation epoch. The trend with the redshift is clearer when the spectroscopic redshift from DESI is used. These results underscore the utility of self-supervised representations in capturing key physical galaxy properties, particularly with multi-band data.

\section{Accuracy of predictions\label{app:statistics}}
 
Table~\ref{tab:smooth_comparison_extended},~\ref{tab:photoz_comparison_extended}, and ~\ref{tab:logM_comparison_extended} provide a detailed summary of the performance metrics for morphology classification, photo-$z$, and stellar mass predictions, respectively, across different models and data configurations (see Sect.~\ref{sec:Downstreamtaks:methodology}). Metrics are reported for various percentages of labelled data used during training.
These tables complements Figs.~\ref{fig:MorphologyPredTrue},~\ref{fig:PhotZ},~\ref{fig:logM_comparison} by offering a more detailed view of the performance differences between the models. MLP model consistently achieves the best overall metrics, particularly when leveraging self-supervised embeddings with even a minimal set of labels. The linear model also demonstrates competitive performance, and serves as a robust baseline for comparison.

\begin{table*}[ht]
\centering
\caption{Comparison of early-type/late-type galaxy classification performance for different approaches. Results are reported for different percentages of labelled data used for training (1\%, 5\%, 10\%, 50\%, 80\%, and 100\%).}
\label{tab:smooth_comparison_extended}
\begin{tabular}{lrrrrrrr}
\hline\hline
Model & Data & Percentage &  Accuracy & Precision & Recall & F1 score & FPR \\ 
\hline
\noalign{\vskip 2pt}
Supervised CNN &  VIS & 1  & \textbf{$0.68 \pm 0.02$} & $0.42 \pm 0.02$ & $0.51 \pm 0.05$ & $0.46 \pm 0.02$ & $0.25 \pm 0.04$ \\
Supervised CNN &  VIS & 5  & \textbf{$0.74 \pm 0.01$} & $0.58 \pm 0.02$ & $0.55 \pm 0.03$ & $0.56 \pm 0.01$ & $0.18 \pm 0.02$ \\
Supervised CNN &  VIS & 10  & \textbf{$0.75 \pm 0.01$} & $0.59 \pm 0.03$ & $0.60 \pm 0.05$ & $0.59 \pm 0.01$ & $0.18 \pm 0.03$ \\
Supervised CNN &  VIS & 20  & \textbf{$0.79 \pm 0.01$} & $0.67 \pm 0.02$ & $0.62 \pm 0.02$ & $0.64 \pm 0.01$ & $0.14 \pm 0.02$ \\
Supervised CNN &  VIS & 50  & \textbf{$0.81 \pm 0.01$} & $0.72 \pm 0.01$ & $0.67 \pm 0.02$ & $0.69 \pm 0.01$ & $0.12 \pm 0.01$ \\
Supervised CNN &  VIS  & 80 & $0.83 \pm 0.01$ & $0.72 \pm 0.02$ & $0.73 \pm 0.02$ & $0.72 \pm 0.00$ & $0.13 \pm 0.02$ \\
Supervised CNN &  VIS & 100  & \textbf{$0.84 \pm 0.01$} & $0.75 \pm 0.02$ & $0.71 \pm 0.03$ & $0.73 \pm 0.01$ & $0.11 \pm 0.02$ \\
\hline 
MLP model & VIS & 1  & \textbf{$0.83 \pm 0.02$} & $0.74 \pm 0.04$ & $0.72 \pm 0.04$ & $0.73 \pm 0.03$ & $0.12 \pm 0.02$ \\
MLP model & VIS & 5  & \textbf{$0.86 \pm 0.01$} & $0.78 \pm 0.02$ & $0.78 \pm 0.02$ & $0.78 \pm 0.01$ & $0.10 \pm 0.01$ \\
MLP model & VIS & 10  & \textbf{$0.87 \pm 0.01$} & $0.79 \pm 0.01$ & $0.78 \pm 0.01$ & $0.79 \pm 0.01$ & $0.10 \pm 0.01$ \\
MLP model & VIS & 20  & \textbf{$0.87 \pm 0.00$} & $0.79 \pm 0.01$ & $0.80 \pm 0.01$ & $0.79 \pm 0.01$ & $0.09 \pm 0.00$ \\
MLP model & VIS & 50  & \textbf{$0.87 \pm 0.00$} & $0.80 \pm 0.01$ & $0.80 \pm 0.01$ & $0.80 \pm 0.00$ & $0.09 \pm 0.00$ \\
MLP model & VIS & 80 & $0.87 \pm 0.00$ & $0.80 \pm 0.01$ & $0.80 \pm 0.01$ & $0.80 \pm 0.00$ & $0.09 \pm 0.00$ \\
MLP model & VIS & 100  & \textbf{$0.88 \pm 0.00$} & $0.80 \pm 0.01$ & $0.80 \pm 0.01$ & $0.80 \pm 0.00$ & $0.09 \pm 0.00$ \\
\hline 
Linear model & VIS & 1  & \textbf{$0.84 \pm 0.02$} & $0.76 \pm 0.04$ & $0.71 \pm 0.03$ & $0.73 \pm 0.03$ & $0.10 \pm 0.02$ \\
Linear model & VIS & 5  & \textbf{$0.86 \pm 0.01$} & $0.80 \pm 0.02$ & $0.72 \pm 0.02$ & $0.76 \pm 0.01$ & $0.08 \pm 0.01$ \\
Linear model & VIS & 10  & \textbf{$0.86 \pm 0.01$} & $0.80 \pm 0.01$ & $0.72 \pm 0.01$ & $0.76 \pm 0.01$ & $0.08 \pm 0.01$ \\
Linear model & VIS & 20  & \textbf{$0.86 \pm 0.00$} & $0.81 \pm 0.01$ & $0.73 \pm 0.01$ & $0.77 \pm 0.01$ & $0.08 \pm 0.00$ \\
Linear model & VIS & 50  & \textbf{$0.86 \pm 0.00$} & $0.81 \pm 0.00$ & $0.73 \pm 0.01$ & $0.77 \pm 0.00$ & $0.08 \pm 0.00$ \\
Linear model & VIS & 80 & $0.86 \pm 0.00$ & $0.81 \pm 0.01$ & $0.73 \pm 0.01$ & $0.77 \pm 0.00$ & $0.08 \pm 0.00$ \\
Linear model & VIS & 100  & \textbf{$0.86 \pm 0.00$} & $0.81 \pm 0.01$ & $0.73 \pm 0.01$ & $0.77 \pm 0.00$ & $0.08 \pm 0.00$ \\
\hline 
Supervised CNN &  VIS+NISP & 1  & \textbf{$0.69 \pm 0.02$} & $0.46 \pm 0.03$ & $0.50 \pm 0.06$ & $0.47 \pm 0.04$ & $0.24 \pm 0.03$ \\
Supervised CNN &  VIS+NISP & 5  & \textbf{$0.74 \pm 0.01$} & $0.60 \pm 0.03$ & $0.54 \pm 0.05$ & $0.57 \pm 0.02$ & $0.16 \pm 0.03$ \\
Supervised CNN &  VIS+NISP & 10  & \textbf{$0.76 \pm 0.01$} & $0.61 \pm 0.02$ & $0.61 \pm 0.02$ & $0.61 \pm 0.01$ & $0.17 \pm 0.02$ \\
Supervised CNN &  VIS+NISP & 20  & \textbf{$0.78 \pm 0.01$} & $0.67 \pm 0.01$ & $0.63 \pm 0.04$ & $0.65 \pm 0.02$ & $0.15 \pm 0.02$ \\
Supervised CNN &  VIS+NISP & 50  & \textbf{$0.82 \pm 0.01$} & $0.72 \pm 0.01$ & $0.67 \pm 0.02$ & $0.69 \pm 0.01$ & $0.12 \pm 0.01$ \\
Supervised CNN & VIS+NISP & 80 & $0.83 \pm 0.00$ & $0.74 \pm 0.01$ & $0.72 \pm 0.02$ & $0.73 \pm 0.01$ & $0.11 \pm 0.01$ \\
Supervised CNN &  VIS+NISP & 100  & \textbf{$0.83 \pm 0.01$} & $0.74 \pm 0.02$ & $0.72 \pm 0.02$ & $0.73 \pm 0.00$ & $0.12 \pm 0.02$ \\
\hline 
MLP model & VIS+NISP & 1  & \textbf{$0.80 \pm 0.02$} & $0.69 \pm 0.05$ & $0.66 \pm 0.05$ & $0.67 \pm 0.04$ & $0.14 \pm 0.02$ \\
MLP model & VIS+NISP & 5  & \textbf{$0.83 \pm 0.01$} & $0.74 \pm 0.02$ & $0.72 \pm 0.02$ & $0.73 \pm 0.02$ & $0.12 \pm 0.01$ \\
MLP model & VIS+NISP & 10  & \textbf{$0.84 \pm 0.01$} & $0.76 \pm 0.01$ & $0.74 \pm 0.01$ & $0.75 \pm 0.01$ & $0.11 \pm 0.01$ \\
MLP model & VIS+NISP & 20  & \textbf{$0.85 \pm 0.00$} & $0.77 \pm 0.01$ & $0.74 \pm 0.01$ & $0.75 \pm 0.01$ & $0.10 \pm 0.00$ \\
MLP model & VIS+NISP & 50  & \textbf{$0.85 \pm 0.00$} & $0.77 \pm 0.01$ & $0.74 \pm 0.01$ & $0.76 \pm 0.00$ & $0.10 \pm 0.00$ \\
MLP model &VIS+NISP & 80 & $0.85 \pm 0.00$ & $0.77 \pm 0.01$ & $0.74 \pm 0.01$ & $0.76 \pm 0.00$ & $0.10 \pm 0.00$ \\
MLP model & VIS+NISP & 100  & \textbf{$0.85 \pm 0.00$} & $0.77 \pm 0.00$ & $0.74 \pm 0.01$ & $0.76 \pm 0.00$ & $0.10 \pm 0.00$ \\
\hline 
Linear model & VIS+NISP & 1  & \textbf{$0.80 \pm 0.02$} & $0.70 \pm 0.05$ & $0.64 \pm 0.04$ & $0.67 \pm 0.04$ & $0.12 \pm 0.02$ \\
Linear model & VIS+NISP & 5  & \textbf{$0.83 \pm 0.01$} & $0.76 \pm 0.02$ & $0.67 \pm 0.02$ & $0.71 \pm 0.01$ & $0.09 \pm 0.01$ \\
Linear model & VIS+NISP & 10  & \textbf{$0.84 \pm 0.01$} & $0.78 \pm 0.01$ & $0.67 \pm 0.01$ & $0.72 \pm 0.01$ & $0.09 \pm 0.01$ \\
Linear model & VIS+NISP & 20  & \textbf{$0.84 \pm 0.00$} & $0.78 \pm 0.01$ & $0.67 \pm 0.01$ & $0.72 \pm 0.01$ & $0.09 \pm 0.01$ \\
Linear model & VIS+NISP & 50  & \textbf{$0.84 \pm 0.00$} & $0.79 \pm 0.01$ & $0.67 \pm 0.01$ & $0.72 \pm 0.00$ & $0.08 \pm 0.00$ \\
Linear model & VIS+NISP & 80 & $0.84 \pm 0.00$ & $0.79 \pm 0.01$ & $0.67 \pm 0.01$ & $0.72 \pm 0.00$ & $0.08 \pm 0.00$ \\
Linear model & VIS+NISP & 100  & \textbf{$0.84 \pm 0.00$} & $0.79 \pm 0.00$ & $0.67 \pm 0.01$ & $0.72 \pm 0.00$ & $0.08 \pm 0.00$ \\
\hline 
MLP model & VIS+NISP+SED & 1  & \textbf{$0.79 \pm 0.06$} & $0.72 \pm 0.16$ & $0.65 \pm 0.14$ & $0.67 \pm 0.11$ & $0.13 \pm 0.08$ \\
MLP model & VIS+NISP+SED & 5  & \textbf{$0.80 \pm 0.03$} & $0.69 \pm 0.07$ & $0.66 \pm 0.07$ & $0.67 \pm 0.06$ & $0.14 \pm 0.03$ \\
MLP model & VIS+NISP+SED & 10  & \textbf{$0.81 \pm 0.03$} & $0.71 \pm 0.05$ & $0.69 \pm 0.05$ & $0.70 \pm 0.04$ & $0.13 \pm 0.03$ \\
MLP model & VIS+NISP+SED & 20  & \textbf{$0.82 \pm 0.02$} & $0.74 \pm 0.03$ & $0.72 \pm 0.04$ & $0.73 \pm 0.03$ & $0.12 \pm 0.02$ \\
MLP model & VIS+NISP+SED & 50  & \textbf{$0.84 \pm 0.01$} & $0.76 \pm 0.02$ & $0.75 \pm 0.02$ & $0.76 \pm 0.02$ & $0.11 \pm 0.01$ \\
MLP Model & VIS+NISP+SED & 80 & $0.85 \pm 0.01$ & $0.78 \pm 0.01$ & $0.77 \pm 0.01$ & $0.77 \pm 0.01$ & $0.11 \pm 0.01$ \\
MLP model & VIS+NISP+SED & 100  & \textbf{$0.86 \pm 0.01$} & $0.78 \pm 0.01$ & $0.78 \pm 0.01$ & $0.78 \pm 0.01$ & $0.10 \pm 0.01$ \\
\hline 
Linear model & VIS+NISP+SED & 1  & \textbf{$0.79 \pm 0.06$} & $0.71 \pm 0.15$ & $0.65 \pm 0.14$ & $0.66 \pm 0.11$ & $0.13 \pm 0.07$ \\
Linear model & VIS+NISP+SED & 5  & \textbf{$0.80 \pm 0.03$} & $0.69 \pm 0.07$ & $0.67 \pm 0.07$ & $0.68 \pm 0.05$ & $0.14 \pm 0.04$ \\
Linear model & VIS+NISP+SED & 10  & \textbf{$0.81 \pm 0.02$} & $0.71 \pm 0.05$ & $0.69 \pm 0.05$ & $0.70 \pm 0.04$ & $0.13 \pm 0.03$ \\
Linear model & VIS+NISP+SED & 20  & \textbf{$0.82 \pm 0.01$} & $0.74 \pm 0.03$ & $0.70 \pm 0.04$ & $0.72 \pm 0.03$ & $0.12 \pm 0.02$ \\
Linear model & VIS+NISP+SED & 50  & \textbf{$0.84 \pm 0.01$} & $0.78 \pm 0.02$ & $0.71 \pm 0.02$ & $0.74 \pm 0.02$ & $0.10 \pm 0.01$ \\
Linear model & VIS+NISP+SED & 80 & $0.85 \pm 0.01$ & $0.78 \pm 0.01$ & $0.72 \pm 0.01$ & $0.75 \pm 0.01$ & $0.10 \pm 0.01$ \\
Linear model & VIS+NISP+SED & 100  & \textbf{$0.85 \pm 0.01$} & $0.78 \pm 0.01$ & $0.73 \pm 0.01$ & $0.75 \pm 0.01$ & $0.10 \pm 0.01$ \\
\hline
\end{tabular}
\end{table*}

\begin{table*}[ht]
\centering
\caption{Comparison of photo-$z$ prediction performance for different approaches. Results are reported for different percentages of labelled data used for training (1\%, 5\%, 10\%, 50\%, 80\%, and 100\%). }
\label{tab:photoz_comparison_extended}
\begin{tabular}{lrrrrrr}
\hline\hline
{Model} & {Data} & {Percentage} & {Bias} & {NMAD} & {Outlier fraction (\%)} \\ 
\hline
\noalign{\vskip 2pt}
Supervised CNN &VIS & 1 & $-0.046 \pm 0.084$ & $0.140 \pm 0.012$ & $40.3 \pm 13.6$ \\
Supervised CNN &VIS & 5 & $0.003 \pm 0.030$ & $0.159 \pm 0.008$ & $34.9 \pm 2.2$ \\
Supervised CNN &VIS & 10 & $-0.002 \pm 0.036$ & $0.150 \pm 0.008$ & $34.6 \pm 2.4$ \\
Supervised CNN &VIS & 20 & $0.019 \pm 0.026$ & $0.142 \pm 0.009$ & $32.3 \pm 1.6$ \\
Supervised CNN &VIS & 50 & $0.003 \pm 0.009$ & $0.124 \pm 0.004$ & $27.1 \pm 1.2$ \\
Supervised CNN &VIS & 80 & $0.021 \pm 0.018$ & $0.135 \pm 0.024$ & $30.9 \pm 8.0$ \\
Supervised CNN &VIS & 100 & $0.026 \pm 0.004$ & $0.118 \pm 0.002$ & $25.1 \pm 0.7$ \\
\hline 
MLP model & VIS & 1 & $-0.000 \pm 0.044$ & $0.120 \pm 0.025$ & $23.9 \pm 9.0$ \\
MLP model & VIS & 5 & $-0.009 \pm 0.036$ & $0.111 \pm 0.013$ & $20.9 \pm 4.1$ \\
MLP model & VIS & 10 & $-0.012 \pm 0.036$ & $0.104 \pm 0.010$ & $19.9 \pm 3.1$ \\
MLP model & VIS & 20 & $-0.002 \pm 0.022$ & $0.099 \pm 0.006$ & $16.5 \pm 2.2$ \\
MLP model & VIS & 50 & $-0.000 \pm 0.024$ & $0.092 \pm 0.005$ & $14.0 \pm 1.2$ \\
MLP model & VIS & 80 & $-0.000 \pm 0.026$ & $0.090 \pm 0.006$ & $13.3 \pm 2.0$ \\
MLP model & VIS & 100 & $-0.002 \pm 0.022$ & $0.089 \pm 0.004$ & $12.7 \pm 1.7$ \\
\hline 
Linear model & VIS & 1 & $0.015 \pm 0.030$ & $0.145 \pm 0.023$ & $32.5 \pm 7.1$ \\
Linear model & VIS & 5 & $-0.002 \pm 0.022$ & $0.265 \pm 0.027$ & $57.2 \pm 4.5$ \\
Linear model & VIS & 10 & $0.013 \pm 0.014$ & $0.175 \pm 0.011$ & $39.1 \pm 3.2$ \\
Linear model & VIS & 20 & $0.011 \pm 0.007$ & $0.123 \pm 0.006$ & $23.5 \pm 1.7$ \\
Linear model & VIS & 50 & $0.011 \pm 0.003$ & $0.105 \pm 0.003$ & $17.1 \pm 0.9$ \\
Linear model & VIS & 80 & $0.011 \pm 0.003$ & $0.100 \pm 0.002$ & $15.5 \pm 0.7$ \\
Linear model & VIS & 100 & $0.011 \pm 0.003$ & $0.100 \pm 0.002$ & $15.1 \pm 0.6$ \\
\hline 
Supervised  CNN &VIS+NISP & 1 & $-0.115 \pm 0.104$ & $0.130 \pm 0.013$ & $48.4 \pm 19.3$ \\
Supervised  CNN &VIS+NISP & 5 & $0.001 \pm 0.047$ & $0.166 \pm 0.017$ & $39.0 \pm 2.6$ \\
Supervised  CNN &VIS+NISP & 10 & $0.009 \pm 0.024$ & $0.148 \pm 0.010$ & $32.8 \pm 2.6$ \\
Supervised  CNN &VIS+NISP & 50 & $-0.009 \pm 0.005$ & $0.110 \pm 0.004$ & $22.4 \pm 0.6$ \\
Supervised  CNN &VIS+NISP & 80 & $0.013 \pm 0.012$ & $0.100 \pm 0.001$ & $20.7 \pm 0.9$ \\
Supervised  CNN &VIS+NISP & 100 & $0.009 \pm 0.006$ & $0.103 \pm 0.003$ & $19.9 \pm 0.3$ \\
MLP Model  CNN &VIS+NISP & 1 & $-0.022 \pm 0.030$ & $0.098 \pm 0.024$ & $14.7 \pm 6.1$ \\
\hline 
MLP model & VIS+NISP & 1 & $-0.022 \pm 0.030$ & $0.098 \pm 0.024$ & $14.7 \pm 6.1$ \\
MLP model & VIS+NISP & 5 & $-0.013 \pm 0.020$ & $0.075 \pm 0.008$ & $8.8 \pm 2.5$ \\
MLP model & VIS+NISP & 10 & $-0.013 \pm 0.019$ & $0.070 \pm 0.006$ & $7.3 \pm 1.7$ \\
MLP model & VIS+NISP & 20 & $-0.005 \pm 0.016$ & $0.066 \pm 0.005$ & $5.6 \pm 1.5$ \\
MLP model & VIS+NISP & 50 & $-0.010 \pm 0.013$ & $0.060 \pm 0.004$ & $4.2 \pm 0.7$ \\
MLP model & VIS+NISP & 80 & $-0.015 \pm 0.012$ & $0.058 \pm 0.004$ & $3.8 \pm 0.6$ \\
MLP model & VIS+NISP & 100 & $-0.014 \pm 0.012$ & $0.057 \pm 0.004$ & $3.6 \pm 0.5$ \\
\hline
Linear model & VIS+NISP & 1 & $0.008 \pm 0.031$ & $0.119 \pm 0.032$ & $22.6 \pm 8.9$ \\
Linear model & VIS+NISP & 5 & $0.009 \pm 0.022$ & $0.200 \pm 0.030$ & $45.4 \pm 6.8$ \\
Linear model & VIS+NISP & 10 & $0.005 \pm 0.008$ & $0.125 \pm 0.013$ & $24.2 \pm 3.7$ \\
Linear model & VIS+NISP & 20 & $0.005 \pm 0.004$ & $0.085 \pm 0.006$ & $10.2 \pm 1.5$ \\
Linear model & VIS+NISP & 50 & $0.005 \pm 0.002$ & $0.067 \pm 0.002$ & $5.6 \pm 0.6$ \\
Linear model & VIS+NISP & 80 & $0.006 \pm 0.002$ & $0.064 \pm 0.002$ & $5.1 \pm 0.4$ \\
Linear model & VIS+NISP & 100 & $0.006 \pm 0.002$ & $0.063 \pm 0.001$ & $4.8 \pm 0.4$ \\
\hline 
MLP model & VIS+NISP+SED & 1 & $-0.023 \pm 0.029$ & $0.071 \pm 0.016$ & $7.5 \pm 6.7$ \\
MLP model & VIS+NISP+SED & 5 & $-0.018 \pm 0.017$ & $0.062 \pm 0.007$ & $5.4 \pm 2.1$ \\
MLP model & VIS+NISP+SED & 10 & $-0.012 \pm 0.015$ & $0.057 \pm 0.005$ & $4.0 \pm 1.5$ \\
MLP model & VIS+NISP+SED & 20 & $-0.008 \pm 0.014$ & $0.054 \pm 0.004$ & $3.1 \pm 1.0$ \\
MLP model & VIS+NISP+SED & 50 & $-0.014 \pm 0.014$ & $0.050 \pm 0.003$ & $2.7 \pm 0.9$ \\
MLP model & VIS+NISP+SED & 80 & $-0.008 \pm 0.013$ & $0.049 \pm 0.003$ & $2.3 \pm 0.4$ \\
MLP model & VIS+NISP+SED & 100 & $-0.009 \pm 0.012$ & $0.048 \pm 0.002$ & $2.1 \pm 0.4$ \\
\hline
Linear model & VIS+NISP+SED & 1 & $0.002 \pm 0.025$ & $0.088 \pm 0.026$ & $12.8 \pm 6.5$ \\
Linear model & VIS+NISP+SED & 5 & $0.007 \pm 0.017$ & $0.171 \pm 0.031$ & $37.5 \pm 8.5$ \\
Linear model & VIS+NISP+SED & 10 & $0.003 \pm 0.006$ & $0.107 \pm 0.013$ & $17.2 \pm 4.2$ \\
Linear model & VIS+NISP+SED & 20 & $0.004 \pm 0.004$ & $0.071 \pm 0.005$ & $6.3 \pm 1.3$ \\
Linear model & VIS+NISP+SED & 50 & $0.004 \pm 0.002$ & $0.056 \pm 0.002$ & $3.3 \pm 0.5$ \\
Linear model & VIS+NISP+SED & 80 & $0.004 \pm 0.002$ & $0.053 \pm 0.001$ & $2.9 \pm 0.3$ \\
Linear model & VIS+NISP+SED & 100 & $0.004 \pm 0.001$ & $0.052 \pm 0.001$ & $2.7 \pm 0.3$ \\
\hline
\multicolumn{2}{c}{\Euclid photo-$z$ } &100 &  -0.027 & 0.037 & 2.21 \\
\hline
\end{tabular}
\end{table*}

\begin{table*}[ht]
\centering
\caption{Comparison of stellar mass prediction performance for different approaches. Results are reported for different percentages of labelled data used for training (1\%, 10\%, 50\%, 80\%, and 100\%).}
\label{tab:logM_comparison_extended}
\begin{tabular}{lrrrrrr}
\hline\hline
{Model} & {Data} & {Percentage} & {Bias} & {NMAD} & {Outlier fraction (\%)} \\ 
\hline
\noalign{\vskip 2pt}
Supervised CNN & VIS & 1 & $-0.018 \pm 0.039$ & $0.059 \pm 0.007$ & $1.7 \pm 1.7$ \\
Supervised CNN & VIS & 5 & $-0.005 \pm 0.027$ & $0.060 \pm 0.005$ & $1.2 \pm 0.6$ \\
Supervised CNN & VIS & 10 & $0.004 \pm 0.021$ & $0.065 \pm 0.002$ & $1.7 \pm 0.6$ \\
Supervised CNN & VIS & 20 & $-0.006 \pm 0.024$ & $0.066 \pm 0.004$ & $1.3 \pm 0.3$ \\
Supervised CNN & VIS & 50 & $-0.017 \pm 0.019$ & $0.064 \pm 0.002$ & $0.9 \pm 0.4$ \\
Supervised CNN & VIS & 80 & $0.001 \pm 0.021$ & $0.063 \pm 0.001$ & $1.1 \pm 0.5$ \\
Supervised CNN & VIS & 100 & $0.004 \pm 0.017$ & $0.063 \pm 0.001$ & $1.2 \pm 0.4$ \\
\hline
MLP model& VIS & 1 & $0.002 \pm 0.016$ & $0.071 \pm 0.017$ & $6.2 \pm 4.6$ \\
MLP model& VIS & 5 & $0.004 \pm 0.007$ & $0.057 \pm 0.006$ & $4.3 \pm 1.8$ \\
MLP model& VIS & 10 & $0.006 \pm 0.004$ & $0.056 \pm 0.005$ & $4.0 \pm 0.9$ \\
MLP model& VIS & 20 & $0.007 \pm 0.003$ & $0.053 \pm 0.003$ & $3.8 \pm 0.9$ \\
MLP model& VIS & 50 & $0.006 \pm 0.002$ & $0.052 \pm 0.002$ & $3.7 \pm 0.6$ \\
MLP model& VIS & 80 & $0.006 \pm 0.002$ & $0.051 \pm 0.002$ & $3.7 \pm 0.4$ \\
MLP model& VIS & 100 & $0.007 \pm 0.002$ & $0.052 \pm 0.002$ & $3.8 \pm 0.4$ \\
\hline
Linear model& VIS & 1 & $0.005 \pm 0.016$ & $0.075 \pm 0.015$ & $7.4 \pm 4.9$ \\
Linear model& VIS & 5 & $0.002 \pm 0.013$ & $0.120 \pm 0.015$ & $21.7 \pm 4.2$ \\
Linear model& VIS & 10 & $0.003 \pm 0.006$ & $0.092 \pm 0.008$ & $12.8 \pm 2.1$ \\
Linear model& VIS & 20 & $0.004 \pm 0.004$ & $0.063 \pm 0.004$ & $4.5 \pm 1.1$ \\
Linear model& VIS & 50 & $0.003 \pm 0.002$ & $0.053 \pm 0.002$ & $3.1 \pm 0.4$ \\
Linear model& VIS & 80 & $0.003 \pm 0.001$ & $0.050 \pm 0.001$ & $2.9 \pm 0.3$ \\
Linear model& VIS & 100 & $0.004 \pm 0.001$ & $0.050 \pm 0.001$ & $3.0 \pm 0.3$ \\
\hline
Supervised CNN& VIS+NISP & 1 & $-0.020 \pm 0.029$ & $0.062 \pm 0.008$ & $0.3 \pm 1.0$ \\
Supervised CNN& VIS+NISP & 5 & $0.008 \pm 0.018$ & $0.064 \pm 0.005$ & $0.2 \pm 0.5$ \\
Supervised CNN& VIS+NISP & 10 & $0.015 \pm 0.021$ & $0.053 \pm 0.002$ & $1.0 \pm 0.8$ \\
Supervised CNN& VIS+NISP & 20 & $-0.003 \pm 0.037$ & $0.050 \pm 0.004$ & $0.8 \pm 0.5$ \\
Supervised CNN& VIS+NISP & 50 & $0.004 \pm 0.019$ & $0.047 \pm 0.002$ & $0.8 \pm 0.3$ \\
Supervised CNN&VIS+NISP & 80 & $0.001 \pm 0.024$ & $0.045 \pm 0.001$ & $0.4 \pm 0.2$ \\
Supervised CNN& VIS+NISP & 100 & $0.011 \pm 0.007$ & $0.044 \pm 0.001$ & $0.5 \pm 0.1$ \\
\hline

MLP model& VIS+NISP & 1 & $0.004 \pm 0.009$ & $0.036 \pm 0.009$ & $1.0 \pm 1.7$ \\
MLP model& VIS+NISP & 5 & $0.002 \pm 0.003$ & $0.030 \pm 0.003$ & $0.7 \pm 0.7$ \\
MLP model& VIS+NISP & 10 & $0.002 \pm 0.002$ & $0.028 \pm 0.003$ & $0.7 \pm 0.5$ \\
MLP model& VIS+NISP & 20 & $0.001 \pm 0.002$ & $0.027 \pm 0.003$ & $0.6 \pm 0.4$ \\
MLP model& VIS+NISP & 50 & $0.002 \pm 0.002$ & $0.028 \pm 0.003$ & $0.7 \pm 0.4$ \\
MLP model& VIS+NISP & 80 & $0.002 \pm 0.001$ & $0.026 \pm 0.003$ & $0.6 \pm 0.3$ \\
MLP model& VIS+NISP & 100 & $0.002 \pm 0.001$ & $0.027 \pm 0.003$ & $0.6 \pm 0.3$ \\
\hline

Linear model& VIS+NISP & 1 & $0.005 \pm 0.008$ & $0.038 \pm 0.009$ & $1.0 \pm 1.9$ \\
Linear model& VIS+NISP & 5 & $0.001 \pm 0.006$ & $0.056 \pm 0.006$ & $1.7 \pm 1.0$ \\
Linear model& VIS+NISP & 10 & $0.000 \pm 0.003$ & $0.043 \pm 0.003$ & $0.7 \pm 0.5$ \\
Linear model& VIS+NISP & 20 & $0.000 \pm 0.002$ & $0.029 \pm 0.002$ & $0.3 \pm 0.2$ \\
Linear model& VIS+NISP & 50 & $0.001 \pm 0.001$ & $0.024 \pm 0.001$ & $0.3 \pm 0.1$ \\
Linear model& VIS+NISP & 80 & $0.001 \pm 0.001$ & $0.023 \pm 0.001$ & $0.3 \pm 0.1$ \\
Linear model& VIS+NISP & 100 & $0.001 \pm 0.001$ & $0.023 \pm 0.001$ & $0.3 \pm 0.1$ \\
\hline
\multicolumn{2}{c}{\Euclid $\Mstarsun$ } & 100 & 0.0011 & 0.0299 & 2.34 \\
\hline
\end{tabular}
\end{table*}

\end{appendix}

\end{document}